\documentclass[onecolumn,notitlepage,floats,floatfix,amssymb,aps,prd,showpacs,superscriptaddress,groupedaddress,nofootinbib]{revtex4-2}  

\usepackage[T1]{fontenc}
\usepackage{lmodern} 
\usepackage{graphicx}  
\usepackage{dcolumn}   
\usepackage{bm}        
\usepackage{amssymb}   
\usepackage{amsmath}
\usepackage{bbold}
\usepackage{array}
\usepackage{makecell}
\usepackage{braket}
\usepackage{longtable}
\usepackage{supertabular,booktabs}

\usepackage{titlesec} 
\usepackage[colorlinks,citecolor=blue,urlcolor=blue,hypertexnames=true]{hyperref}
\usepackage{subfigure}

\usepackage[usenames,dvipsnames,svgnames,table]{xcolor} 

\newcommand{\be}{\begin{equation}}
\newcommand{\ee}{\end{equation}}
\newcommand{\bea}{\begin{eqnarray}}
\newcommand{\eea}{\end{eqnarray}}
\newcommand{\bml}{\begin{subequations}}
\newcommand{\eml}{\end{subequations}}
\newcommand{\bfig}{\begin{figure}}
\newcommand{\efig}{\end{figure}}

\newcommand{\bmat}{\begin{pmatrix}}
\newcommand{\emat}{\end{pmatrix}}
\usepackage{graphicx,slashed,booktabs,xcolor,multirow,float,
amsfonts,bbold,mathtools,sidecap,tikz,bm,enumitem}
\usepackage{multirow}
\usepackage{bbding}
\usepackage{titlesec}
\usepackage{hyperref}
\usepackage{wasysym}
\usepackage{amssymb}
\usepackage{pifont}

\renewcommand{\leq}{\leqslant}
\renewcommand{\geq}{\geqslant}

\usepackage[dvipsnames, usenames]{xcolor}

\definecolor{linkcolor}{rgb}{0.55, 0.13, .32}

\definecolor{oucrimsonred}{rgb}{0.6, 0.0, 0.0}
\definecolor{persianblue}{rgb}{0.11, 0.22, 0.73}
\definecolor{forestgreen}{rgb}{0.13,0.35,0.13}
\definecolor{lightgray}{rgb}{0.83, 0.83, 0.83}
 \hypersetup{colorlinks, citecolor=oucrimsonred, linkcolor=persianblue, urlcolor=oucrimsonred}
\definecolor{cornellred}{rgb}{0.7, 0.11, 0.11}
\definecolor{navyblue}{rgb}{0.0, 0.0, 0.5}
\definecolor{amethyst}{rgb}{0.6, 0.4, 0.8}
\definecolor{yellow}{rgb}{1.0, 1.0, 0.0}
\definecolor{firebrick}{rgb}{0.7, 0.13, 0.13}
\definecolor{tangerineyellow}{rgb}{1.0, 0.8, 0.0}
\definecolor{deepfuchsia}{rgb}{0.76, 0.33, 0.76}
\definecolor{amber}{rgb}{1.0, 0.75, 0.0}
\definecolor{VioletRed4}{rgb}{0.55, 0.13, .32}
\definecolor{indiagreen}{rgb}{0.07, 0.53, 0.03}
\definecolor{VioletRed4}{rgb}{0.55, 0.13, .32}

\usepackage{hyperref}

\usepackage{graphics,appendix,afterpage,makecell}

\usepackage{bbold}
\usepackage{tikz}
\usepackage{adjustbox}
\usepackage{tikz-feynman}
\usepackage{tcolorbox}

\definecolor{oucrimsonred}{rgb}{0.6, 0.0, 0.0}
\definecolor{persianblue}{rgb}{0.11, 0.22, 0.73}
\definecolor{forestgreen}{rgb}{0.13,0.35,0.13}
\definecolor{lightgray}{rgb}{0.83, 0.83, 0.83}
 \hypersetup{colorlinks, citecolor=oucrimsonred, linkcolor=persianblue, urlcolor=oucrimsonred}
\definecolor{cornellred}{rgb}{0.7, 0.11, 0.11}
\definecolor{navyblue}{rgb}{0.0, 0.0, 0.5}
\definecolor{amethyst}{rgb}{0.6, 0.4, 0.8}
\definecolor{yellow}{rgb}{1.0, 1.0, 0.0}
\definecolor{firebrick}{rgb}{0.7, 0.13, 0.13}
\definecolor{tangerineyellow}{rgb}{1.0, 0.8, 0.0}
\definecolor{deepfuchsia}{rgb}{0.76, 0.33, 0.76}
\definecolor{amber}{rgb}{1.0, 0.75, 0.0}
\definecolor{VioletRed4}{rgb}{0.55, 0.13, .32}
\definecolor{indiagreen}{rgb}{0.07, 0.53, 0.03}
\definecolor{VioletRed4}{rgb}{0.55, 0.13, .32}

\definecolor{oucrimsonred}{rgb}{0.6, 0.0, 0.0}
\newcommand\vertarrowbox[3][6ex]{%
  \begin{array}[t]{@{}c@{}} #2 \\
  \left\uparrow\vcenter{\hrule height #1}\right.\kern-\nulldelimiterspace\\
  \makebox[0pt]{\scriptsize#3}
  \end{array}%
}

\definecolor{mtcolor}{rgb}{.8,.3,.1}

\definecolor{violachiaro}{rgb}{1,0.6,1}

\definecolor{gbcolor}{rgb}{.43,.22,.12}
 
\definecolor{gbcolor2}{rgb}{.9,.2,.6}
\definecolor{gbcolor3}{rgb}{.3,.2,.6}

\definecolor{verdechiaro}{rgb}{0.6,1,0.6}
\definecolor{giallochiaro}{rgb}{1,1,0.6}
\definecolor{bluscuro}{rgb}{0.15, 0.2, 0.9}
\definecolor{verdes}{rgb}{0.1, 0.5, 0.1}%
\definecolor{tangerineyellow}{rgb}{1.0, 0.8, 0.0}
\definecolor{smokyblack}{rgb}{0.06, 0.05, 0.03}

\definecolor{americanrose}{rgb}{1.0, 0.01, 0.24}
\definecolor{cobalt}{rgb}{0.0, 0.28, 0.67}
\definecolor{brandeisblue}{rgb}{0.0, 0.44, 1.0}
\definecolor{mycolor}{rgb}{0.0, 0.0, 0.5}
\definecolor{oxfordblue}{rgb}{0.0, 0.13, 0.28}
\definecolor{azure}{rgb}{0.0, 0.5, 1.0}
\definecolor{turquoiseblue}{rgb}{0.0, 1.0, 0.94}
\newtcolorbox{mynewbox}[1]{colback=white!5!white,colframe=azure!75!black,fonttitle=\bfseries,title=#1}
\newtcolorbox{mybox}{colback=mycolor!5!white,colframe=azure!75!black}
\newtcolorbox{mynamedbox}[1]{colback=mycolor!5!white,colframe=azure!75!black,title=#1}
\definecolor{venetianred}{rgb}{0.78, 0.03, 0.08}
\newtcolorbox{mynamedbox1}[1]{colback=venetianred!5!white,colframe=venetianred!80!black,title=#1}
\newtcolorbox{mynamedbox2}[1]{colback=azure!5!white,colframe=azure!80!black,title=#1}

\definecolor{rossocorsa}{rgb}{0.83, 0.0, 0.0}

\tikzset{->-/.style={decoration={
  markings,
  mark=at position #1 with {\arrow{>}}},postaction={decorate}}}
\tikzset{-<-/.style={decoration={
  markings,
  mark=at position #1 with {\arrow{<}}},postaction={decorate}}} 

\def\be{\begin{equation}}
\def\ee{\end{equation}}
\def\ba{\begin{eqnarray}}
\def\ea{\end{eqnarray}}

\def\L*{{\cal L}_*}
\def\L{\mathcal{L}}
\def\({\left(}
\def\){\right)}

\def\<{\langle}
\def\>{\rangle}

 \def\neq {\not\equiv}

\def\cs2{c_{s}^{2}}

 \def\be   {\begin{equation}}   \def\ee   {\end{equation}}
 \def\ba   {\begin{array}}      \def\ea   {\end{array}}
 \def\bea  {\begin{eqnarray}}   \def\eea  {\end{eqnarray}}
 \def\bean {\begin{eqnarray*}}  \def\eean {\end{eqnarray*}}

\titleclass{\subsubsubsection}{straight}[\subsection]

\newcounter{subsubsubsection}[subsubsection]
\renewcommand\thesubsubsubsection{\thesubsubsection.\arabic{subsubsubsection}}

\titleformat{\subsubsubsection}
  {\normalfont\normalsize\bfseries}{\thesubsubsubsection}{1em}{}
\titlespacing*{\subsubsubsection}
{0pt}{3.25ex plus 1ex minus .2ex}{1.5ex plus .2ex}

\makeatletter
\renewcommand\paragraph{\@startsection{paragraph}{5}{\z@}%
  {3.25ex \@plus1ex \@minus.2ex}%
  {-1em}%
  {\normalfont\normalsize\bfseries}}
\renewcommand\subparagraph{\@startsection{subparagraph}{6}{\parindent}%
  {3.25ex \@plus1ex \@minus .2ex}%
  {-1em}%
  {\normalfont\normalsize\bfseries}}
\def\toclevel@subsubsubsection{4}
\def\toclevel@paragraph{5}
\def\toclevel@paragraph{6}
\def\l@subsubsubsection{\@dottedtocline{4}{7em}{4em}}
\def\l@paragraph{\@dottedtocline{5}{10em}{5em}}
\def\l@subparagraph{\@dottedtocline{6}{14em}{6em}}
\makeatother

\usepackage{titlesec} 
\usepackage[colorlinks,citecolor=blue,urlcolor=blue,hypertexnames=true]{hyperref}
\usepackage{subfigure}

\usepackage[usenames,dvipsnames,svgnames,table]{xcolor}

\usepackage{tikzsymbols}
\usepackage{natbib}
\usepackage{float}

\usepackage{tikz,xcolor,hyperref}

\usepackage{slashed}

\definecolor{lime}{HTML}{A6CE39}
\DeclareRobustCommand{\orcidicon}{
	\begin{tikzpicture}
	\draw[lime, fill=lime] (0,0) 
	circle [radius=0.2] 
	node[white] {{\fontfamily{qag}\selectfont \tiny ID}};
	\draw[white, fill=white] (-0.0625,0.095) 
	circle [radius=0.007];
	\end{tikzpicture}
	\hspace{-2mm}
}

\foreach \x in {A, ..., Z}{\expandafter\xdef\csname orcid\x\endcsname{\noexpand\href{https://orcid.org/\csname orcidauthor\x\endcsname}
			{\noexpand\orcidicon}}
}

\usepackage{bbold}
\usepackage{tikz}
\usepackage{adjustbox}
\usepackage{tikz-feynman}
\usepackage{tcolorbox}
\usepackage{enumitem}
\usepackage{amsfonts}

\setlist[itemize,1]{label=$\times$}
\setlist[itemize,2]{label=$\checkmark$}
\setlist[itemize,3]{label=$\diamond$}
\setlist[itemize,4]{label=$\bullet$}

\begin{document}

\title{\Large \textcolor{Sepia}{
Reconstructing inflationary potential from NANOGrav 15-year Data: A
  robust study using Non-Bunch Davies initial condition}}

\author{{\large  Sayantan Choudhury\orcidA{}}}
\email{sayantan\_ccsp@sgtuniversity.org, \\sayanphysicsisi@gmail.com (Corresponding author)
}

\affiliation{Centre For Cosmology and Science Popularization (CCSP),\\
        SGT University, Gurugram, Delhi- NCR, Haryana- 122505, India.}

\begin{abstract}

We discuss the theoretical framework behind reconstruction of a generic class of inflationary potentials for canonical single-field slow-roll inflation in a model-independent fashion. The Non-Bunch Davies (NBD) initial condition is an essential choice to determine the structure of potential and to accommodate the blue-tilted tensor power spectrum feature recently observed in NANOGrav. Using the reconstruction technique we found the favoured parameter space which supports blue tilted tensor power spectrum. The validity of the EFT prescription in inflation is also maintained through the use of a new field excursion formula while keeping the necessary and sufficient conditions on the sub-Planckian field values in check. We find that the reconstructed potential display inflection point behaviour, which has deeper connection with high energy physics.

\end{abstract}

\maketitle
\tableofcontents
\newpage

 \section{Introduction}

Two important predictions of the primordial inflationary paradigm are that during the accelerated phase of expansion, there would be scalar density perturbations and tensor perturbations. For more details, see  refs. \cite{Kazanas:1980tx,Starobinsky:1980te,Sato:1981ds,Guth:1980zm,Mukhanov:1981xt,Linde:1981mu,Albrecht:1982wi,Baumann:2009ds,Baumann:2018muz,Senatore:2013roa,Choudhury:2011sq,Choudhury:2011jt,Choudhury:2012yh,Choudhury:2012whm,Choudhury:2013zna,Choudhury:2013jya,Choudhury:2013iaa,Choudhury:2013woa,Choudhury:2014sxa,Choudhury:2014uxa,Choudhury:2014hua,Choudhury:2014kma,Choudhury:2014sua,Choudhury:2014hja,Choudhury:2015pqa,Choudhury:2015hvr,Choudhury:2016wlj,Choudhury:2016cso,Choudhury:2017cos,Naskar:2017ekm,Choudhury:2017glj,Choudhury:2018glz,HosseiniMansoori:2023zop,Geng:2015fla,WaliHossain:2014usl,Hossain:2014coa,Hossain:2014xha}. The studies of the temperature anisotropy in the cosmic microwave background (CMB) radiation have now successfully verified one of the predictions, namely the temperature anisotropy owing to the scalar density fluctuations, and has raised the unanswered challenge of how to incorporate the inflationary paradigm into a particle theory \cite{Mazumdar:2010sa}. It is to be noted that the end of inflation must allow for the creation of the necessary Standard Model degrees of freedom required for the development of the Big Bang Nucleosynthesis (BBN) phenomenon \cite{Pospelov:2010hj}, without the inclusion of any additional relativistic degrees of freedom, i.e. dark radiation \cite{Planck:2018vyg}, 
because inflation diminishes all matter contents except for the inflationary quantum vacuum fluctuations generated by the tensor and scalar perturbations which are then extended outside the Hubble horizon before re-entering at a far later period of cosmic evolution.

A significant number of cosmological discoveries \cite{Martin:2013tda, Benetti:2013cja, Martin:2013nzq,Creminelli:2014oaa,Dai:2014jja,Benetti:2016tvm,Campista:2017ovq,Keeley:2020rmo,Vagnozzi:2020rcz,Vagnozzi:2020dfn,Vagnozzi:2023lwo,Cabass:2022wjy,Cabass:2022ymb} are compatible with the inflationary paradigm \cite{Baumann:2009ds,Baumann:2018muz,Senatore:2013roa,Choudhury:2011sq,Choudhury:2011jt,Choudhury:2012yh,Choudhury:2012whm,Choudhury:2013zna,Choudhury:2013jya,Choudhury:2013iaa,Choudhury:2013woa,Choudhury:2014sxa,Choudhury:2014uxa,Choudhury:2014hua,Choudhury:2014kma,Choudhury:2014sua,Choudhury:2014hja,Choudhury:2015pqa,Choudhury:2015hvr,Choudhury:2016wlj,Choudhury:2016cso,Choudhury:2017cos,Naskar:2017ekm,Choudhury:2017glj,Choudhury:2018glz,HosseiniMansoori:2023zop,Geng:2015fla,WaliHossain:2014usl,Hossain:2014coa,Hossain:2014xha}, which is still in great health. One of the key scientific components of upcoming observational investigations will be additional testing of the inflationary paradigm \cite{CMB-S4:2016ple,SimonsObservatory:2018koc,SimonsObservatory:2019qwx}. The inflationary SGWB, however, has not yet been found. Based on our current understanding of various inflationary models, this signal is desired at a very low frequency domain, $f\leq {\cal O}(10^{-15}{\rm Hz}-10^{-16}{\rm Hz})$ 
\cite{Kamionkowski:2015yta}. However, searches in the higher frequency regime, where the crucial frequency band ${\cal O}(10^{-9}{\rm Hz} -10^{-7}{\rm Hz})$ as shown by the recently observed NANOGrav 15 year GW signal also lies \cite{NANOGrav:2023gor,Antoniadis:2023ott,Reardon:2023gzh,Xu:2023wog,NANOGrav:2023hde,NANOGrav:2023ctt,NANOGrav:2023hvm,NANOGrav:2023hfp,NANOGrav:2023tcn,NANOGrav:2023pdq,NANOGrav:2023icp,Antoniadis:2023lym,Antoniadis:2023puu,Antoniadis:2023aac,Antoniadis:2023xlr,Smarra:2023ljf,Reardon:2023zen,Zic:2023gta}, can intuitively predict interesting features related to modified canonical single field inflation models.  \footnote{Some other sources are able to explain the GW signal from NANOGrav 15 from various perspectives. Some of the refs. \cite{Siemens:2006yp,Caprini:2010xv,Ramberg:2019dgi,Caprini:2019egz,Ellis:2020awk,Rajagopal:1994zj,Jaffe:2002rt,Wyithe:2002ep,Sesana:2004sp,Burke-Spolaor:2018bvk,Athron:2023mer,Kitajima:2023cek} are pointed here completeness.}.

Next, we arrive at the matter of describing the quantum mechanical fluctuations during inflation in the presence of the correct quantum initial condition choice, generally implemented for the initial vacuum state. To date, the well-known {\it Bunch Davies} (BD) initial condition is the most successful candidate to define the quantum vacuum state necessarily required to describe the fluctuations generated from scalar and tensor perturbations and up to the pivot scale, i.e. at $k_*\sim 0.05 {\rm Mpc}^{-1}$ this fact is completely consistent with the CMB observation from Planck. {\it Bunch Davies} quantum vacuum state is a Euclidean vacuum state in Quantum Field Theory of de Sitter space which respects the de Sitter isometry. Such an initial condition is represented by a point within the large parameter space allowed by the underlying  symmetry. In this connection, it is important to note that the BD initial condition successfully describes the existence of a red-tilted tensor power spectrum. However, this fact is in huge tension as far as the recent findings from the NANOGrav 15 signal for GW are concerned. \textcolor{black}{The new NANOGrav 15 findings suggest that the tensor power spectrum may be described by the blue tilted feature, which is almost impossible to achieve by making use of the BD initial condition}~\footnote{\textcolor{black}{The exact origin of the NANOGrav signal is not yet confirmed.}}. The prime reason for this problem stems from the use of BD initial condition which gives us a very strict consistency relation between tensor-to-scalar ratio ($r$) and the tensor spectral tilt ($n_t)$ is obtained, i.e. $r=-8n_t$, which can only accommodate red tilted tensor spectral tilt $n_t<0$ to have $r>0$. Hence from the theoretical perspective, a violation of the good old consistency relation derived in the presence of the BD initial condition is required. 

Among many possibilities \footnote{To know about the other possibilities consider the refs. \cite{Hossain:2014xha,Hossain:2014coa,Hossain:2014ova,WaliHossain:2014usl,Geng:2015fla,Piao:2004tq,Liu:2010dh,Brandenberger:2008nx,Brandenberger:2011et,Khoury:2001wf,Lehners:2008vx,Brandenberger:2016vhg,Brandenberger:2023wtd,Koehn:2015vvy,Lehners:2015mra,Ijjas:2018qbo,Bhargava:2020fhl,Agullo:2016tjh,Bojowald:2005epg,Bojowald:1999tr,Biswas:2005qr}, where the authors have studied the models of quintessential and phantom models of inflation, string gas cosmology, ekpyrosis, bouncing cosmology etc.}, recently in ref. \cite{Choudhury:2023kam} we have explicitly pointed that {\it Non-Bunch Davies} (NBD) initial condition is the most promising candidate using which one can provide a new consistency relation which in turn can successfully accommodate the blue-tilted tensor power spectrum $n_t>0$ to have tensor-to-scalar ratio $r>0$. For more details on NBD initial condition see the refs. \cite{Choudhury:2017glj,Choudhury:2020yaa,Choudhury:2021tuu,Adhikari:2021ked,Akama:2023jsb,Albayrak:2023hie,Choudhury:2022mch,Colas:2022kfu,Aalsma:2022eru,Chapman:2022mqd,Letey:2022hdp,Penna-Lima:2022dmx,Kanno:2022mkx,Fumagalli:2021mpc,Sleight:2021plv,Chen:2010bka,Wang:2014kqa,Ashoorioon:2014nta,Ashoorioon:2013eia}. \textcolor{black}{It is important to note that the signed coordinate
transformation has been recently introduced in refs \cite{Guendelman:2024ezk,Guendelman:2023spd,Guendelman:2023vso} which provides an interesting and innovative transformation that
changes the signature of the Jacobian if considered locally. It resembles the Bogoliubov transformations in a non-trivial
curved space background, where it mixes the positive and
negative frequency components. Here the non-trivial curved
background is important as in the case of the Minkowski
flat vacuum there is no mixing of the positive and negative
frequency components appearing. As an immediate consequence of this local signed coordinate transformation the
Fock space operators of the Non-Bunch Davies initial quantum state are written in terms of its Bunch Davies counterpart through Bogoliubov transformations in terms of positive
and negative frequency components of the creation and annihilation operators.} Now once we allow for the possibility of NBD vacuum, it will immediately show its impact in all the observables that we compute using the scalar and tensor perturbations, and the flow equations, related to Cosmological Beta functions and slow-roll hierarchy, will also be strongly affected, particularly the contributions directly related to the tensor perturbations. To maintain the observational constraints obtained from Planck we need to design the NBD initial condition in such a fashion that after accommodating the blue-tilted tensor power spectrum and violation of the existing consistency relation all the findings from the scalar perturbation sector gets mildly affected. From the theoretical perspective, a NBD initial state describes a bigger parameter space and also accommodates the information regarding the BD initial condition in a single point in the representative large parameter family and respects all the necessary requirements to preserve the de Sitter group. 
Since the inclusion of the NBD initial condition has a clear advantage in resolving the tensions in the known paradigm of inflation by accommodating both new NANOGrav 15 and CMB data together, from a realistic point of view, one should look into the bigger picture by exploring the new physics in the broad parameter space allowed by the NBD initial condition.

Further, we discuss the field excursion of the inflaton field which is commonly described in terms of the well-known {\it Lyth bound} which was derived in the presence of BD initial condition. However, this bound is strict against accommodating the Effective Field Theory (EFT) prescription within the framework of inflation to achieve the necessary requirement of the e-foldings $\Delta {\cal N}\sim 60$. The inclusion of the NBD initial condition helps us to violate the mentioned known bound and can accommodate very easily the EFT prescription for sub-Planckian models of inflation. NBD initial condition helps us to satisfy the necessary condition for the field excursion, $|\Delta \phi|=|\phi_{\rm end}-\phi_*|\ll M_{\rm pl}$ as well as the sufficient condition, $\phi_{\rm end}<M_{\rm pl}$ and $\phi_{*}<M_{\rm pl}$, where $\phi_{\rm end}$ and $\phi_{*}$ are respectively the inflaton field values at the end of inflation and at the CMB pivot scale $k_*=0.05{\rm Mpc}^{-1}$. Utilizing all the above-mentioned facts together we will reconstruct the inflationary potential structure using observational constraints obtained from CMB data and the NANOGrav 15-year signal for GWs. 

This paper is arranged as follows: In \ref{sec2}, we start the discussion by noting the power spectrum and the spectral tilt from both scalar and tensor perturbations using NBD initial condition. In \ref{sec3}, we discuss in detail the new consistency relation that arises due to the changes in tensor spectrum discussed in the last section and can accommodate the blue-tilted tensor power spectrum feature which is a prime component to reconstruct the structure of the inflationary potential. In \ref{sec4}, we discuss the validity of EFT in the sub-Planckian regime based on the newly proposed field excursion formula by strictly following the necessary and sufficient conditions throughout the analysis. In \ref{sec5a}, we comment on back-reaction with Non Bunch Davies vacuum in the light of Effective Field Theory. In \ref{sec5b}, we focus on the reconstruction of the inflationary potential by starting with a brief discussion on the theoretical background for the same, modified with the presence of the NBD initial condition, and then moving towards a detailed construction of the structure of the potential and analysis of the slow-roll parameters. In \ref{sec6}, we then consider a successful UV completion of our present potential reconstruction scenario by focusing on the framework of $N=1$ SUGRA theory and highlight the similarity of our model-independent analysis with the known inflection point models obtained from $N=1$ SUGRA MSSM inflationary framework. In \ref{sec7}, we present our results obtained from numerical analysis of the potential reconstruction and show that the outcomes follow by keeping the robustness of the theoretical analysis done in the previous sections intact. Finally, in \ref{sec8}, we highlight the main findings as a result of our analysis.

\section{Non-standard inputs for potential reconstruction}
\label{sec2}

For this section, our objective is to provide the inputs related to scalar and tensor perturbations derived using NBD initial condition. For the detailed derivation of these quantities see the ref. \cite{Choudhury:2023kam}. 

%
\begin{enumerate}
    \item \underline{\bf Power spectrum from scalar perturbation:}
    In the super-Hubble regime, through the use of the NBD initial condition, the expression for the scalar power spectrum can be written as follows:
 \bea \textcolor{black}{\Delta^{2}_{\zeta}(k)
=\Delta^{2}_{\zeta}(k_*)\left(\frac{k}{k_*}\right)^{n_s-1+\frac{1}{2}\alpha_s\ln\left(\frac{k}{k_*}\right)+\frac{1}{6}\kappa_s\ln^2\left(\frac{k}{k_*}\right)}}.\quad \eea
where the amplitude of the scalar perturbation at the pivot scale, $k=k_*=0.05{\rm Mpc}^{-1}$, is given by:
\bea \Delta^{2}_{\zeta}(k_*)
=\left(\frac{H^2}{8\pi^2M^2_{\rm pl}\epsilon_V}\right)_*2^{2\nu_{s}-3}\left|\frac{\Gamma(\nu_{s})}{\Gamma\left(\frac{3}{2}\right)}\right|^2 {\cal J}\quad {\rm where}\quad{\cal J}=\Bigg|{\alpha}^{(s)}_{{\bf k}}\exp\left(-\frac{i\pi}{2}\left(\nu_{s}+\frac{1}{2}\right)\right)-{\beta}^{(s)}_{{\bf k}}\exp\left(\frac{i\pi}{2}\left(\nu_{s}+\frac{1}{2}\right)\right)\Bigg|^2.\eea
Here $*$ notation is used to quantify the parameters of the theory in the pivot scale. 
Also, the factor $\nu_s$ is known as the mass parameter characterizing the quasi-de Sitter expansion of our universe and is given by the following expression:
   \bea \nu_s=\left(\frac{3}{2}+3\epsilon_V-\eta_V\right)\quad\quad\quad {\rm where}\quad\quad\epsilon_V=\frac{M^2_{\rm pl}}{2}\left(\frac{V^{'}(\phi)}{V(\phi)}\right)^2,\quad\quad \eta_V=M^2_{\rm pl}\left(\frac{V^{''}(\phi)}{V(\phi)}\right),\eea
   where in this description $\epsilon_V$ and $\eta_V$ represent the potential dependent slow-roll parameters. Here $M_{\rm pl}$ is the reduced Planck scale, the symbol $'$ represents derivative with respect to the inflaton field $\phi$ and $V(\phi)$ represents the inflationary potential. The NBD initial condition is described in the scalar sector of primordial perturbation by the two Bogoliubov coefficients, ${\alpha}^{(s)}_{{\bf k}}$ and ${\beta}^{(s)}_{{\bf k}}$ which must satisfy the normalization condition, $|{\alpha}^{(s)}_{{\bf k}}|^2-|{\beta}^{(s)}_{{\bf k}}|^2=1$.
   The aforementioned modified structure of the power spectrum for the scalar modes captures the details of the Cosmological Perturbation Theory and the computation of scalar modes from {\it Mukhanov Sasaki} equation in the presence of the NBD condition. Utilizing all of these facts together we have arrived at the above-mentioned result. For compactness, we have not provided such details in this paper, however from the basic understanding one can easily arrive at such a result as quoted above.

\item \underline{\bf Spectral tilt from scalar perturbation:}
From the previously mentioned result of the power spectrum for the scalar modes, the spectral tilt in the presence of the initial NBD condition can be computed as:
  \begin{equation}
    n_s-1=\left(\frac{d\ln \Delta^{2}_{\zeta}(k)}{d\ln k}\right)=3-2\nu_s+\Bigg(\frac{d {\cal J}}{d\ln k}\Bigg)\approx 2\eta_V-6\epsilon_V,
  \end{equation}
where we fix the initial NBD condition at the pivot scale in such a fashion that it satisfies the following condition:
\bea \Bigg(\frac{d {\cal J}}{d\ln k}\Bigg)\ll \epsilon_V\quad\quad{\rm and}\quad\quad  \Bigg(\frac{d {\cal J}}{d\ln k}\Bigg)\ll \eta_V.\eea
This is pointing toward the fact that the features of the scalar perturbations are not very sensitive toward the specific choices of the individual structures of the Bogoliubov coefficients describing the scalar sector. If we add this contribution, then this will be a very small correction in the final result of the scalar spectral tilt. For this reason, and for the sake of simplicity, we have currently not considered such contributions in our computation. For a better understanding let us remind ourselves that this above-mentioned justification that we have imposed is present only at the pivot scale where the CMB observation takes place.  Thus, it does not imply that the contribution will be completely flat throughout the evolution with respect to the wave number scale all over the region.

\item \underline{\bf Tensor power spectrum:}
In the super-Hubble region, the expression for the tensor power spectrum with NBD initial condition can be written as follows:
\bea\textcolor{black}{\Delta^{2}_{t}(k)=\Delta^{2}_{t}(k_*)\left(\frac{k}{k_*}\right)^{n_t+\frac{1}{2}\alpha_t\ln\left(\frac{k}{k_*}\right)+\frac{1}{6}\kappa_t\ln^2\left(\frac{k}{k_*}\right)}}.\quad\eea
where the amplitude of the tensor perturbation at the pivot scale $k=k_*=0.05{\rm Mpc}^{-1}$ is given by:
  \bea \Delta^{2}_{t}(k_*)=\left(\frac{2H^2}{\pi^2M^2_{\rm pl}}\right)_*\times {\cal Y},\quad\quad{\rm where}\quad{\cal Y}:=\sum_{\lambda=+,\times}\Bigg|{\alpha}^{(t)}_{{\bf k},\lambda}-{\beta}^{(t)}_{{\bf k},\lambda}\Bigg|^2=\Bigg\{\Bigg|{\alpha}^{(t)}_{{\bf k},+}-{\beta}^{(t)}_{{\bf k},+}\Bigg|^2+\Bigg|{\alpha}^{(t)}_{{\bf k},\times}-{\beta}^{(t)}_{{\bf k},\times}\Bigg|^2\Bigg\}.\eea
Here the symbol $+$ and $\times$ represent two helicities of the Primordial Gravitational Waves (PGWs), where the contributions from both of them have been taken into account. The NBD initial condition is described for the tensor sector of primordial perturbations by the two Bogoliubov coefficients, ${\alpha}^{(t)}_{{\bf k},+}$, ${\alpha}^{(t)}_{{\bf k},\times}$ and ${\beta}^{(t)}_{{\bf k},+}$, ${\beta}^{(t)}_{{\bf k},\times}$ which satisfy the normalization condition, 
\bea \sum_{\lambda=+,\times}\Bigg(|{\alpha}^{(t)}_{{\bf k},\lambda}|^2-|{\beta}^{(t)}_{{\bf k},\lambda}|^2\Bigg)=\Bigg\{\Bigg(|{\alpha}^{(t)}_{{\bf k},+}|^2-|{\beta}^{(t)}_{{\bf k},+}|^2\Bigg)+\Bigg(|{\alpha}^{(t)}_{{\bf k},\times}|^2-|{\beta}^{(t)}_{{\bf k},\times}|^2\Bigg)\Bigg\}=1,\eea 
for the present scenario. This modified structure of the tensor power spectrum captures the details of the Cosmological Perturbation Theory and the computation of corresponding two-helicity dependent tensor modes from {\it Mukhanov Sasaki} equation in the presence of the NBD quantum condition. Utilizing all of these facts together we have arrived at the above-mentioned result. \textcolor{black}{Further, it is important to point out that in the present context of the discussion, we have used different NBD initial conditions for scalar and tensor modes, which appear as different Bogoliubov coefficients. This type of construction is actually motivated by the studies performed in ref. \cite{Hui:2001ce,Ganc:2011dy,Brahma:2018hrd}. This type of construction allows large-scale dependence in the tensor spectrum compared to the scalar one due to its strong backreaction, which is necessarily required to produce large non-Gaussianities if one tries to probe it in various cosmological observations. Theoretically, these types of NBD states, which have different Bogoliubov coefficient structures in scalar and tensor modes, appear in the subgroup ${\rm ISO(1,4)}$ of the full de Sitter isommetry group ${\rm SO(1,4)}$. Within this subgroup, if we perform signed coordinate
transformation \cite{Guendelman:2024ezk,Guendelman:2023spd,Guendelman:2023vso} then it changes the structure of the Bogoliubov coefficients in the scalar and tensor modes. The prime reason is that due to having the helicity dependence in the tensor modes it transforms in a completely different way compared to the scalar modes, which do not depend on the helicity. As an immediate consequence of this local signed coordinate transformation in ${\rm ISO(1,4)}$ for the tensor modes (in the presence of additional helicity components comping from two types of polarizations), it induces a large amount of scale dependence compared to the scale modes. }

\item \underline{\bf Tensor spectral tilt:}
From the previously mentioned result of the power spectrum for the tensor modes the spectral tilt in the presence of NBD initial condition can be computed as:
  \bea n_t&=&\left(\frac{d\ln \Delta^{2}_{t}(k)}{d\ln k}\right)\approx -2\epsilon_V+{\cal G}\quad\quad{\rm where}\quad {\cal G}:=\Bigg(\frac{d{\cal Y}}{d\ln k}\Bigg),\eea
which further implies that the tilt of the tensor modes is highly dependent on the detailed structure of the Bogoliubov coefficients describing the two helicity-dependent sectors of PGW. \textcolor{black}{Due to having additional factor ${\cal G}$ appearing due to NBD vacuum the tensor spectral tilt $n_t$ will not be considered to be a constant. It will run with the momentum scale and most importantly the running and running of the running of the function ${\cal G}$ will fix the scale-dependent behaviour of the tensor spectral tilt $n_t$, which we have explicitly studied in detail in the context of cosmological beta functions in the later half of this paper. }

\item \underline{\textbf{Tensor-to-scalar ratio:}}
Next, the expression for the tensor-to-scalar ratio in the super-Hubble regime for the NBD initial condition can be written as:
 \bea \textcolor{black}{r(k)
=r(k_*)\left(\frac{k}{k_*}\right)^{(n_t-n_s+1)+\frac{1}{2}(\alpha_t-\alpha_s)\ln\left(\frac{k}{k_*}\right)+\frac{1}{6}(\kappa_t-\kappa_s)\ln^2\left(\frac{k}{k_*}\right)}}.\quad\eea
where at the pivot scale $k=k_*=0.05{\rm Mpc}^{-1}$, the tensor-to-scalar ratio is quantified as:
  \bea r(k_*) \label{tensortoscalar}
=\frac{\Delta^{2}_{t}(k_*)}{\Delta^{2}_{\zeta}(k_*)}=16\epsilon_V{\cal F},\quad{\rm where} \quad {\cal F}:=2^{3-2\nu_{s}}\left|\frac{\Gamma\left(\frac{3}{2}\right)}{\Gamma(\nu_{s})}\right|^2 \times\left(\frac{{\cal Y}}{{\cal J}}\right).\eea
\end{enumerate}

\section{Blue-tiltedness through new consistency relation}
\label{sec3}

It is anticipated that the old consistency relation, $r=-8n_t$, derived for the canonical single field models of inflation with BD initial condition, will be violated due to the modifications appearing in the amplitude and tilt of the tensor spectrum as a result of having NBD initial condition. Here, we discovered that the following shortened formula provides the new consistency connection \footnote{One crucial aspect of the aforementioned relation is that it yields the old consistency relation and can only explain red-tilted tensor spectrum in the case of the BD initial condition, where ${\cal F}=1$, ${\cal G}=0$ and ${\cal J}=1$. The NANOGrav 15-year Data Set ruled out this scenario and suggested the presence of a blue-tilted tensor power spectrum that is very hard to explain using an initial BD condition.
}:  

 \bea r=8{\cal F}\left({\cal G}-n_t\right),\eea
where the factors ${\cal F}$ and ${\cal G}$ are already defined before in terms of the Bogoliubov coefficients of the NBD vacuum. 
For the present purpose, to explain the existence of the blue-tilted tensor power spectrum from the perspective of the NANOGrav 15 signal \footnote{The tensor power spectrum from the recent NANOGrav 15-year Data Set has been found to be strongly blue-tilted and explained using the recommended fitting formula, $n_t=-0.14\log_{10}r+0.58$.}, we parameterize the factor ${\cal F}$ such that it is extremely small and treated to give a constant contribution while ${\cal G}$ is, in general, a specific function of the tensor tilt $n_t$. When taking into account all frequency domains, we recommend the following functional forms using the numerical approach that best fits the data \footnote{The conversion relationship between the frequency and the wave number is given below, which we are frequently using throughout the analysis performed in this paper:
\bea f=1.6\times 10^{-9}{\rm Hz}\times\left(\frac{k}{10^6 {\rm Mpc}^{-1}}\right).\eea}~~\footnote{\textcolor{black}{Disclaimer: It seems like the tensor special tilt $n_t$ for the parametrization of the factor ${\cal G}$ takes a constant value. However, in the present context of discussion, it is not at all true. Here $n_t$ acts as a parameter that takes different values as well as signatures in the three mentioned phases of the parameterization of the factor ${\cal G}$. Also, the scale dependence in ${\cal G}$ can be further computed in terms of the running of $n_t$ which we have computed later in this paper. Such running is extremely useful to reconstruct inflationary potential in terms of its higher derivatives, which are propagated in various types of slow-roll parameters $(\xi^2_V,\sigma^3_V)$, which is the prime objective of this paper. Though $n_t$ here acts as a parameter, rather than a constant number due to having running in ${\cal G}$, we keep the symbol as $n_t$, provided one should always remember that it is not a constant in this context. We are extremely thankful to the anonymous referee for pointing out this confusion.} }:
\bea  {\cal G}
&=& \displaystyle
\displaystyle\left\{
	\begin{array}{ll}
		\displaystyle 0& \mbox{for}\quad {\bf Phase\;I:}\;10^{-20}{\rm Hz}< f < 10^{-17}{\rm Hz} \;  \\  
			\displaystyle 
			\displaystyle 10^{28}\exp(-22n_t)+n_t \quad\quad\quad\quad\quad\quad\quad\quad& \mbox{for}\quad  {\bf Phase\;II:}\;10^{-17}{\rm Hz}< f < 10^{-7}{\rm Hz} \; \\
   \displaystyle 
			\displaystyle 10^{28+6.89n_t}\exp(22n_t)-n_t \quad\quad\quad\quad\quad\quad\quad\quad& \mbox{for}\quad  {\bf Phase\;III:}\;10^{-7}{\rm Hz}< f < 1{\rm Hz} \;. 
	\end{array}
\right. \eea
Here it is important to note that in our previous work \cite{Choudhury:2023kam}, to explain the NANOGrav 15 signal for GW we have fixed this parameter at ${\cal F}\sim{\cal O}(10^{-22})$, which is able to cover the range of the tensor-to-scalar ratio, $7.9\times 10^{-18}\leq r\leq 2.3\times 10^{-5}$ for the blue tilt $1.2\leq n_t\leq 2.5$. However, in a broader perspective, it is important to note the value for the parameter ${\cal F}$ can't be strictly fixed at previous mentioned value within the frequency domain $10^{-20}{\rm Hz}<f<1{\rm Hz}$. By performing the reconstruction of the inflationary potential in this paper, we will show that this mentioned value of the parameter ${\cal F}$ can be pushed to a larger value as far as the CMB constraint on tensor-to-scalar ratio, $r<0.06$ for the blue tensor tilt $n_t$ as favoured by NANOGrav 15 signal satisfies together. We will discuss this issue elaborately in the next section.

As a result, the following is how the new consistency relations in the three significant frequency domains indicated above may be represented mathematically:
\bea  r
&=& 8\times {\cal F}\times\displaystyle
\displaystyle\left\{
	\begin{array}{ll}
		\displaystyle n_t & \mbox{for}\quad {\bf Phase\;I:}\;10^{-20}{\rm Hz}< f < 10^{-17}{\rm Hz} \;  \\  
			\displaystyle 
			\displaystyle 10^{28}\exp(-22n_t) \quad\quad\quad\quad\quad\quad\quad\quad& \mbox{for}\quad  {\bf Phase\;II:}\;10^{-17}{\rm Hz}< f < 10^{-7}{\rm Hz} \; \\
   \displaystyle 
			\displaystyle 10^{28+6.89n_t}\exp(22n_t) \quad\quad\quad\quad\quad\quad\quad\quad& \mbox{for}\quad  {\bf Phase\;III:}\;10^{-7}{\rm Hz}< f < 1{\rm Hz} \;. 
	\end{array}
\right. \eea

There are a small number of theoretical frameworks which can accommodate the blue-tilted tensor spectrum. See refs. \cite{Hossain:2014xha,Hossain:2014coa,Hossain:2014ova,WaliHossain:2014usl,Geng:2015fla,Piao:2004tq,Liu:2010dh,Brandenberger:2008nx,Brandenberger:2011et,Khoury:2001wf,Lehners:2008vx,Brandenberger:2016vhg,Brandenberger:2023wtd,Koehn:2015vvy,Lehners:2015mra,Ijjas:2018qbo,Bhargava:2020fhl,Agullo:2016tjh,Bojowald:2005epg,Bojowald:1999tr} for more details on this issue. However, the NANOGrav 15-year Data Set suggests the presence of a significant spectral blue tilt that is incomprehensible from the aforementioned examples without taking into account the specifics of reheating phenomena.

In ref. \cite{Vagnozzi:2023lwo} the author demonstrated that a large amount of blue tilt in the tensor power spectrum supports very low magnitude reheating temperatures, which is impossible to generate from the conventional known models of inflation.  We are certain that some of the ambiguities and other issues linked with the detailed scenario of reheating may be addressed in a more straightforward manner once the microphysical structure of reheating is clearly understood,  which as of yet is something none of us are even somewhat familiar with.  See refs. \cite{Allahverdi:2010xz,Mazumdar:2010sa,Kohri:2009ka,Mazumdar:2003bs,Choudhury:2020yaa,Choudhury:2021tuu,Choudhury:2018rjl,Choudhury:2018bcf} for details.  For this reason in this paper, we only concentrate on NBD initial condition to solve the issue.

\section{Effective Field Theory validation with new field excursion formula}
\label{sec4}

Since the key component of this paper concerns with the NBD initial condition, the well-known field excursion formula for the inflaton field—also known as the {\it Lyth bound} —initially developed in the presence of the BD initial condition will also change as a result of the presence of different Bogoliubov coefficients that accurately explain the scalar and tensor perturbations. 

Here we have found after detailed computation that the new field excursion formula for NBD initial condition can be expressed as:
 \bea \frac{|\Delta\phi|}{M_{\rm pl}}&=&\sqrt{\frac{r(k_*)}{8{\cal F}}}\frac{2}{\left(n_t+n_s-1\right)}\Bigg|\Bigg\{1-\exp\left(-\Delta {\cal N}\left(\frac{n_t+n_s-1}{2}\right)\right)\Bigg\}\Bigg|\quad\quad{\rm where}\quad\Delta {\cal N}=\ln\left(\frac{k_*}{k_{\rm end}}\right). \eea
where the field excursion is defined as, $|\Delta\phi|:=|\phi_{\rm end}-\phi_*|$,
where $\phi_{\rm end}$ and $\phi_*$ represent respectively the values of the inflaton field at the end of inflation and the pivot scale\footnote{\textcolor{black}{It is critical to emphasise in this context that the {\it Lyth bound} affects whether inflation occurs in the super/sub-Planckian regime. The old {\it Lyth bound} frequently suggests a super-Planckian inflationary paradigm, which contradicts the basics of the EFT prescription.  Since great efforts have been made in the past to avoid such powerful effects of the mentioned bound, the EFT of inflation regulated by the BD initial condition is reliable.  See refs. \cite{Choudhury:2015hvr,Choudhury:2015pqa,Choudhury:2014sua,Choudhury:2014kma,Choudhury:2013iaa,Hossain:2014ova}. We revisit this age-old subject in our study because we believe that the NBD initial condition is one of the probable solutions which might help to circumvent the {\it Lyth bound} and its occurrence proves the inflationary paradigm.}}. Also, the total e-foldings is denoted by, $\Delta {\cal N}$. To validate the EFT prescription through the sub-Planckian inflationary paradigm we demand the strict conditions which need to be satisfied during the reconstruction of inflationary potential in the light of NANOGrav 15 data,
$|\Delta\phi|\ll M_{\rm pl}$, along with $\phi_*< M_{\rm pl}$ and $\phi_{\rm end}<M_{\rm pl}$.
A very crucial point to be noted here is that in the corresponding literature, there is confusion regarding the criteria to validate the EFT prescription in the sub-Planckian regime. In most of the cases, it is a myth that if $|\Delta\phi|\ll M_{\rm pl}$ condition is achieved then EFT is validated for inflation. However, more precisely this is the necessary condition to validate EFT, but not the sufficient one. In this context, $\phi_*< M_{\rm pl}$ and $\phi_e<M_{\rm pl}$ provides sufficient condition to validate EFT which in most of the cases either ignored or not taken seriously at all. In this paper, to reconstruct inflationary potential within the framework of EFT in the sub-Planckian regime, we have followed both the restriction strictly throughout our analysis.  For the estimation, let us now fix $n_t\sim 2$, $n_s\sim 0.96$, and $\Delta{\cal N}\sim 60$ for the numerical estimations which gives,
$\frac{|\Delta\phi|}{M_{\rm pl}}\sim {\cal O}(10^{-2}-10^{-1})$ for $\epsilon_V\sim 0.0044$ \cite{Planck:2018jri}, which means EFT prescription works well. After conducting this study, we discovered that the results of the NBD initial condition are more suitable than the BD initial condition-driven results.

\section{Comment on back-reaction with Non Bunch Davies vacuum in the light of Effective Field Theory }
\label{sec5a}

\textcolor{black}{In this section, our prime objective is to discuss the issue related to the back reactions that may appear after allowing the Non-Bunch Davies (NBD) vacuum within the framework of Effective Field Theory. To illustrate this issue in detail, let us start with the following expressions which belongs to the tensor sector of the perturbation theory}:
  \bea \textcolor{black}{\label{y}{\cal Y}=\Bigg\{\Bigg|{\alpha}^{(t)}_{{\bf k},+}-{\beta}^{(t)}_{{\bf k},+}\Bigg|^2+\Bigg|{\alpha}^{(t)}_{{\bf k},\times}-{\beta}^{(t)}_{{\bf k},\times}\Bigg|^2\Bigg\}}.\eea
  \textcolor{black}{Further using the following condition}:
  \bea \textcolor{black}{\Bigg\{\Bigg(|{\alpha}^{(t)}_{{\bf k},+}|^2-|{\beta}^{(t)}_{{\bf k},+}|^2\Bigg)+\Bigg(|{\alpha}^{(t)}_{{\bf k},\times}|^2-|{\beta}^{(t)}_{{\bf k},\times}|^2\Bigg)\Bigg\}=1},\eea 
  \textcolor{black}{one can recast the equation (\ref{y}) in the following form}:
  \bea \label{y1} \textcolor{black}{{\cal Y}}&=& \textcolor{black}{1+2|{\beta}^{(t)}_{{\bf k},+}|^2+2|{\beta}^{(t)}_{{\bf k},\times}|^2-\bigg({\alpha}^{(t)*}_{{\bf k},+}{\beta}^{(t)}_{{\bf k},+}+{\alpha}^{(t)}_{{\bf k},+}{\beta}^{(t)*}_{{\bf k},+}+{\alpha}^{(t)*}_{{\bf k},\times}{\beta}^{(t)}_{{\bf k},\times}+{\alpha}^{(t)}_{{\bf k},\times}{\beta}^{(t)*}_{{\bf k},\times}\bigg)}\nonumber\\
  &=& \textcolor{black}{1+2|{\beta}^{(t)}_{{\bf k},+}|^2+2|{\beta}^{(t)}_{{\bf k},\times}|^2-2\bigg(|{\alpha}^{(t)}_{{\bf k},+}| |{\beta}^{(t)}_{{\bf k},+}|\cos \theta^{(t)}_{+}+|{\alpha}^{(t)}_{{\bf k},\times}| |{\beta}^{(t)}_{{\bf k},\times}|\cos \theta^{(t)}_{\times}\bigg)}\nonumber\\
  &=& \textcolor{black}{1+2\bigg(|{\beta}^{(t)}_{{\bf k},+}|^2-|{\alpha}^{(t)}_{{\bf k},+}| |{\beta}^{(t)}_{{\bf k},+}|\cos \theta^{(t)}_{+}\bigg)+2\bigg(|{\beta}^{(t)}_{{\bf k},\times}|^2-|{\alpha}^{(t)}_{{\bf k},\times}| |{\beta}^{(t)}_{{\bf k},\times}|\cos \theta^{(t)}_{\times}\bigg).}\eea
  \textcolor{black}{Here $\theta^{(t)}_{+}$ and $\theta^{(t)}_{\times}$ are relative angles between $({\alpha}^{(t)}_{{\bf k},+},{\beta}^{(t)}_{{\bf k},+})$ and $({\alpha}^{(t)}_{{\bf k},\times},{\beta}^{(t)}_{{\bf k},\times})$ respectively. 
  In the above expression, we have explicitly used the fact that the two polarization modes of the gravitational waves appearing from the tensor perturbations behave in a different fashion. The prime reason for such a general choice will be more clear as we proceed with our discussion in this section. For the time being let us assume that the value of the parameter ${\cal Y}$ is extremely small, say for numerical purposes assume that ${\cal Y}\sim {\cal O}(10^{-20})$. In such a situation one can easily neglect the term ${\cal Y}$ compared to the unity as appearing in the RHS of the above equation. As a consequence, we are left with the following constraint condition}:
\bea\label{c} \textcolor{black}{\underbrace{\bigg(|{\beta}^{(t)}_{{\bf k},+}|^2-|{\alpha}^{(t)}_{{\bf k},+}| |{\beta}^{(t)}_{{\bf k},+}|\cos \theta^{(t)}_{+}\bigg)}=-\Bigg[\underbrace{\bigg(|{\beta}^{(t)}_{{\bf k},\times}|^2-|{\alpha}^{(t)}_{{\bf k},\times}| |{\beta}^{(t)}_{{\bf k},\times}|\cos \theta^{(t)}_{\times}\bigg)}+\frac{1}{2}\Bigg]}.\eea
\textcolor{black}{If there is any back-reaction associated with the tensor modes due to having large values of the Bogoliubov coefficients $({\alpha}^{(t)}_{{\bf k},+},{\beta}^{(t)}_{{\bf k},+})$ and $({\alpha}^{(t)}_{{\bf k},\times},{\beta}^{(t)}_{{\bf k},\times})$, then it nullifies by following the constraint equation (\ref{c}) and this is happening due to having distinctive behaviours of the two polarization modes. More precisely, due to having opposite signatures in the terms appearing in the parenthesis symbol $\underbrace{}$ the effect of the underlying back reaction becomes nullified. Though this situation is fine tuned, however, one should not mix up fine-tuning with back reaction in the present context of the discussion. This sort of situation is completely consistent with the outcomes obtained in this paper, which we have discussed in the later half with numerics, and completely at par with the observational constraint using which we have reconstructed the inflationary potential in the presence of NBD vacuum. Most importantly, such a prescription is completely consistent with the EFT constraints because in this specific case the UV scale associated with the excited tensor modes to the stress-energy tensor $M_T$ turns out to be $M_T>H_{inf}$ (for ${\cal Y}\sim {\cal O}(10^{-20})$) i.e. $M_T\sim 5.21\times 10^{-3}~M_{\rm pl}\sim 1.28\times 10^{17}~{\rm GeV}$ for the inflationary scale $H_{inf}\sim 1.4\times 10^{-3}~M_{\rm pl}\sim 3.43\times 10^{15}~{\rm GeV}$. From this discussion we have learnt that if we are very far away from the BD values of the Bogoliubov coefficients then due to having opposite distinctive behaviours of the two polarization modes of the gravitational waves one can able to compensate the back-reaction effects and maintain EFT requirements in this computation.}

\textcolor{black}{However, this might not always be true. At least we can discuss a pathological situation where the back reaction becomes large and the EFT constraints might get hampered. This is the special case where the two polarization modes of the gravitational waves exactly behave in a similar manner. To illustrate this issue in detail, let us start with the following expressions which belongs to the tensor sector of the perturbation theory:}
  \bea \label{ya} \textcolor{black}{{\cal Y}=2\Bigg|{\alpha}^{(t)}_{{\bf k}}-{\beta}^{(t)}_{{\bf k}}\Bigg|^2~~~~{\rm as}~~{\alpha}^{(t)}_{{\bf k},+}={\alpha}^{(t)}_{{\bf k},\times}={\alpha}^{(t)}_{{\bf k}}~~{\rm and}~~{\beta}^{(t)}_{{\bf k},+}={\beta}^{(t)}_{{\bf k},\times}={\beta}^{(t)}_{{\bf k}}.}\eea
  \textcolor{black}{Further using the following condition:}
  \bea \textcolor{black}{\Bigg(|{\alpha}^{(t)}_{{\bf k}}|^2-|{\beta}^{(t)}_{{\bf k}}|^2\Bigg)=1,}\eea 
  \textcolor{black}{one can recast the equation (\ref{ya}) in the following form:}
  \bea \label{y1a} \textcolor{black}{{\cal Y}=2\bigg( 1+2|{\beta}^{(t)}_{{\bf k}}|^2-2\sqrt{|{\beta}^{(t)}_{{\bf k}}|^2\left(1+|{\beta}^{(t)}_{{\bf k}}|^2\right)}\cos \theta^{(t)}\bigg).}\eea
  \textcolor{black}{Here $\theta^{(t)}$ is the relative angle between $({\alpha}^{(t)}_{{\bf k}},{\beta}^{(t)}_{{\bf k}})$. Here if we fix $\theta^{(t)}=(0,\pi)$ then we get the following acceptable solutions:}
  \bea\label{fgq1} \textcolor{black}{|{\beta}^{(t)}_{{\bf k}}|=\Bigg(\underbrace{\frac{2-{\cal Y}}{2\sqrt{2{\cal Y}}}}_{\bf Solution~I},\underbrace{-\frac{2-{\cal Y}}{2\sqrt{2{\cal Y}}}}_{\bf Solution~II}\Bigg),}\eea
  \textcolor{black}{using which $|{\alpha}^{(t)}_{{\bf k}}|$ can be further uniquely computed as:}
  \bea \textcolor{black}{|{\alpha}^{(t)}_{{\bf k}}|=\underbrace{\sqrt{1+\bigg(\frac{2-{\cal Y}}{2\sqrt{2{\cal Y}}}\bigg)^2}}_{\bf Both~for~ Solution~I~\&~II}.}\eea
  \textcolor{black}{Here it is important to note that the Solution I is valid for small ${\cal Y}$ and Solution II is valid for large value of ${\cal Y}$ as in both the cases we want $|{\beta}^{(t)}_{{\bf k}}|>0$. For this reason, equation (\ref{fgq1}) can be further approximated in the following simplified form for the better understanding purpose:}
   \bea\label{fgq2} \textcolor{black}{|{\beta}^{(t)}_{{\bf k}}|=\Bigg(\underbrace{\frac{1}{\sqrt{2{\cal Y}}}}_{\bf Solution~I},\underbrace{\frac{1}{2}\sqrt{\frac{{\cal Y}}{2}}}_{\bf Solution~II}\Bigg).}\eea
   \textcolor{black}{Now, as we are interested in very small value of the parameter ${\cal Y}$ which is ${\cal Y}\sim {\cal O}(10^{-20})$ we further consider Solution I using which we get an rough estimate $|{\beta}^{(t)}_{{\bf k}}|\sim {\cal O}(10^{10})$. This allows large back-reaction in the present computation and is that case one can further estimate that the UV scale associated with the excited tensor modes to the stress-energy tensor $M_T$ turns out to be $M_T\ll H_{inf}$, which further implies that the EFT prescription is violated. For more details on this issue see the ref \cite{Ashoorioon:2018sqb}. To resolve this issue one needs to consider either of the following proposals:}
   \begin{enumerate}
       \item\underline{\bf Option I:}\\ \\
       \textcolor{black}{To cancel the back reaction and validate EFT prescription one needs to consider different behaviour of the two polarization modes of the gravitational waves as we have discussed earlier in detail.}

       \item \underline{\bf Option II:}\\ \\
       \textcolor{black}{Instead of using two boundary values of the relative angle $\theta^{(t)}=(0,\pi)$ if we fix it in between i.e. if we consider $0<\theta^{(t)}<\pi$ then we get the following acceptable solutions for $|{\beta}^{(t)}_{{\bf k}}|>0$, which are given by:} 
       \bea \textcolor{black}{|{\beta}^{(t)}_{{\bf k}}|}
&=& \displaystyle
\displaystyle\left\{
	\begin{array}{ll}
		\displaystyle \textcolor{black}{\frac{1}{2} \sqrt{\left|\frac{\left(2-2 \cos ^2(\theta^{(t)})-{\cal Y}\right)+\sqrt{\cos ^2(\theta^{(t)}) \left(2 \cos (2 \theta^{(t)})+{\cal Y}^2-2\right)}}{\cos ^2(\theta^{(t)})-1}\right|}} & \mbox{\bf Solution~A}\;  \\  \\
			\displaystyle 
			\displaystyle \textcolor{black}{\frac{1}{2} \sqrt{\left|\frac{\left(2-2 \cos ^2(\theta^{(t)})-{\cal Y}\right)-\sqrt{\cos ^2(\theta^{(t)}) \left(2 \cos (2 \theta^{(t)})+{\cal Y}^2-2\right)}}{\cos ^2(\theta^{(t)})-1}\right|}} \quad\quad& \mbox{\bf Solution~B} \; 
	\end{array}
\right. \eea
\textcolor{black}{From these obtained solutions one can directly see that pathological situation appears for $\theta^{(t)}=(0,\pi)$ as for both the values of the relative angles, $|{\beta}^{(t)}_{{\bf k}}|$ becomes divergent because of the zero contribution coming from the denominators. So strictly both of these relative angles are redundant which is necessarily needed to avoid large back-reaction and validate the EFT prescription.}

\textcolor{black}{For the clear demonstration purpose let us further choose the relative angle $\theta^{(t)}=\pi/2$ which is lying within the window $0<\theta^{(t)}<\pi$ for which both the Solution A and Solution B coincides and finally gives:}
\bea \textcolor{black}{|{\beta}^{(t)}_{{\bf k}}|=\frac{1}{2}\sqrt{|2-{\cal Y}|}}.\eea
\textcolor{black}{Now as we demand $|{\beta}^{(t)}_{{\bf k}}|>0$ and real the above solution is valid for both the small and large values of the parameter ${\cal Y}$. Now, if we fix ${\cal Y}\sim {\cal O}(10^{-20})$ using the above expression for $\theta^{(t)}=\pi/2$ we get, $|{\beta}^{(t)}_{{\bf k}}|\sim \sqrt{2}\sim 1.414$, $|{\alpha}^{(t)}_{{\bf k}}|\sim \sqrt{3}\sim 1.732$ (such that $|{\alpha}^{(t)}_{{\bf k}}|^2-|{\beta}^{(t)}_{{\bf k}}|^2=1$) on which the back reaction is negligible and one can further show that the UV scale associated with the excited tensor modes to the stress-energy tensor $M_T$ turns out to be $M_T>H_{inf}$ (for ${\cal Y}\sim {\cal O}(10^{-20})$) i.e. $M_T\sim 5.21\times 10^{-3}~M_{\rm pl}\sim 1.28\times 10^{17}~{\rm GeV}$ for the inflationary scale $H_{inf}\sim 1.4\times 10^{-3}~M_{\rm pl}\sim 3.43\times 10^{15}~{\rm GeV}$.}
   \end{enumerate}

   \textcolor{black}{Now we extend this discussion for the scalar modes appearing from the perturbation. Since the scalar modes has no polarization, to avoid the back-reaction no such option I is appearing here just like tensor modes.  To illustrate this issue in detail, let us start with the following expressions which belongs to the scalar sector of the perturbation theory:}
   \bea \label{sc1} \textcolor{black}{{\cal J}=\Bigg|{\alpha}^{(s)}_{{\bf k}}\exp\left(-\frac{i\pi}{2}\left(\nu_{s}+\frac{1}{2}\right)\right)-{\beta}^{(s)}_{{\bf k}}\exp\left(\frac{i\pi}{2}\left(\nu_{s}+\frac{1}{2}\right)\right)\Bigg|^2~~~~~{\rm where}~~~~ \nu_s=\left(\frac{3}{2}+3\epsilon_V-\eta_V\right).}\eea
   \textcolor{black}{Further using the following condition:}
   \bea \textcolor{black}{|{\alpha}^{(s)}_{{\bf k}}|^2-|{\beta}^{(s)}_{{\bf k}}|^2=1,}\eea
   \textcolor{black}{one can recast equation (\ref{sc1}) in the following form:}
   \bea \label{sc2} \textcolor{black}{{\cal J}=\bigg[ 1+2|{\beta}^{(s)}_{{\bf k}}|^2-2\sqrt{|{\beta}^{(s)}_{{\bf k}}|^2\left(1+|{\beta}^{(s)}_{{\bf k}}|^2\right)}\cos \left(\theta^{(s)}+\left(\nu_s+\frac{1}{2}\right)\pi\right)\bigg].}\eea
    \textcolor{black}{Here $\theta^{(s)}$ is the relative angle between $({\alpha}^{(s)}_{{\bf k}},{\beta}^{(s)}_{{\bf k}})$. Here if we fix $\theta^{(s)}+\left(\nu_s+\frac{1}{2}\right)\pi=(0,\pi)$ then we get the following acceptable solutions:}
  \bea\label{fgq1a} \textcolor{black}{|{\beta}^{(s)}_{{\bf k}}|=\Bigg(\underbrace{\frac{1-{\cal J}}{2\sqrt{{\cal J}}}}_{\bf Solution~I},\underbrace{-\frac{1-{\cal J}}{2\sqrt{{\cal J}}}}_{\bf Solution~II}\Bigg)},\eea
  \textcolor{black}{using which $|{\alpha}^{(s)}_{{\bf k}}|$ can be further uniquely computed as:}
  \bea \textcolor{black}{|{\alpha}^{(t)}_{{\bf k}}|=\underbrace{\sqrt{1+\bigg(\frac{1-{\cal J}}{2\sqrt{{\cal J}}}\bigg)^2}}_{\bf Both~for~ Solution~I~\&~II}.}\eea
  \textcolor{black}{Here it is important to note that the Solution I is valid for small ${\cal J}$ and Solution II is valid for large value of ${\cal J}$ as in both the cases we want $|{\beta}^{(s)}_{{\bf k}}|>0$. For this reason, equation (\ref{fgq1a}) can be further approximated in the following simplified form for the better understanding purpose:}
   \bea\label{fgq2a} \textcolor{black}{|{\beta}^{(s)}_{{\bf k}}|=\Bigg(\underbrace{\frac{1}{2\sqrt{{\cal J}}}}_{\bf Solution~I},\underbrace{\frac{1}{2}\sqrt{{\cal J}}}_{\bf Solution~II}\Bigg).}\eea
   \textcolor{black}{Now, as we are interested in very small value of the parameter ${\cal J}$ which is ${\cal J}\sim {\cal O}(10^{-6})$ we further consider Solution I using which we get an rough estimate $|{\beta}^{(s)}_{{\bf k}}|\sim {\cal O}(10^{3})$. This allows large back-reaction in the present computation and is that case one can further estimate that the UV scale associated with the excited scalar modes to the stress-energy tensor $M_S$ turns out to be $M_S\ll H_{inf}$, which further implies that the EFT prescription is violated. For more details on this issue see the ref \cite{Ashoorioon:2018sqb}. To resolve this issue one needs to consider the following proposal.}

   \textcolor{black}{Instead of using two boundary values of the relative angle $\theta^{(s)}+\left(\nu_s+\frac{1}{2}\right)\pi=(0,\pi)$ if we fix it in between i.e. if we consider $0<\theta^{(s)}+\left(\nu_s+\frac{1}{2}\right)\pi<\pi$ then we get the following acceptable solutions for $|{\beta}^{(s)}_{{\bf k}}|>0$, which are given by: }
       \bea \textcolor{black}{|{\beta}^{(s)}_{{\bf k}}|}
&=& \footnotesize
\displaystyle\left\{
	\begin{array}{ll}
		\displaystyle \textcolor{black}{\frac{1}{2} \sqrt{\left|\frac{\left(2-2 \cos ^2\left(\theta^{(s)}+\left(\nu_s+\frac{1}{2}\right)\pi\right)-2{\cal J}\right)+\sqrt{2\cos ^2\left(\theta^{(s)}+\left(\nu_s+\frac{1}{2}\right)\pi\right) \left(2 \cos \left(2\theta^{(s)}+2\left(\nu_s+\frac{1}{2}\right)\pi\right)+2{\cal J}^2-1\right)}}{\cos ^2\left(\theta^{(s)}+\left(\nu_s+\frac{1}{2}\right)\pi\right)-1}\right|}} & \mbox{\bf Solution~A}\;  \\  \\
			\displaystyle 
			\displaystyle \textcolor{black}{\frac{1}{2} \sqrt{\left|\frac{\left(2-2 \cos ^2\left(\theta^{(s)}+\left(\nu_s+\frac{1}{2}\right)\pi\right)-2{\cal J}\right)-\sqrt{2\cos ^2\left(\theta^{(s)}+\left(\nu_s+\frac{1}{2}\right)\pi\right) \left(2 \cos \left(2\theta^{(s)}+2\left(\nu_s+\frac{1}{2}\right)\pi\right)+2{\cal J}^2-1\right)}}{\cos ^2\left(\theta^{(s)}+\left(\nu_s+\frac{1}{2}\right)\pi\right)-1}\right|}} \quad\quad& \mbox{\bf Solution~B} \; 
	\end{array}
\right. \eea
\textcolor{black}{From these obtained solutions one can directly see that pathological situation appears for $\theta^{(s)}+\left(\nu_s+\frac{1}{2}\right)\pi=(0,\pi)$ as for both the values of the relative angles, $|{\beta}^{(s)}_{{\bf k}}|$ becomes divergent because of the zero contribution coming from the denominators. So strictly both of these relative angles are redundant which is necessarily needed to avoid large back-reaction and validate the EFT prescription.}

\textcolor{black}{For the clear demonstration purpose let us further choose the relative angle $\theta^{(s)}+\left(\nu_s+\frac{1}{2}\right)\pi=\pi/2$ which is lying within the window $0<\theta^{(s)}+\left(\nu_s+\frac{1}{2}\right)\pi<\pi$ for which both the Solution A and Solution B coincides and finally gives:}
\bea \textcolor{black}{|{\beta}^{(s)}_{{\bf k}}|=\sqrt{2|1-{\cal J}|}.}\eea
\textcolor{black}{Now as we demand $|{\beta}^{(s)}_{{\bf k}}|>0$ and real the above solution is valid for both the small and large values of the parameter ${\cal J}$. Now, if we fix ${\cal J}\sim {\cal O}(10^{-6})$ using the above expression for $\theta^{(s)}+\left(\nu_s+\frac{1}{2}\right)\pi=\pi/2$ we get, $|{\beta}^{(s)}_{{\bf k}}|\sim \sqrt{2}\sim 1.414$, $|{\alpha}^{(s)}_{{\bf k}}|\sim \sqrt{3}\sim 1.732$ (such that $|{\alpha}^{(s)}_{{\bf k}}|^2-|{\beta}^{(s)}_{{\bf k}}|^2=1$) on which the back reaction is negligible and one can further show that the UV scale associated with the excited scalar modes to the stress-energy tensor $M_S$ turns out to be $M_S>H_{inf}$.}

\textcolor{black}{From this discussion we conclude in this section that the back-reaction issue in the scalar and tensor sector of the perturbation theory for the excited NBD modes can be avoided by the proper choice of the relative angles between the Bogoliubov coefficients. Strictly $(0,\pi)$ values of the relative angle needs to be avioded to suppress the effect of back-reaction and to validate the EFT prescription. } 

\section{Framework of non-standard inflaton potential reconstruction}\label{sec5b}

For this section, our main objective will be to reconstruct the inflationary potential in presence of NBD initial condition, which will fix the structure of canonical single-field inflation from the perspective of observational data, particularly using CMB observation and recently released NANOGrav 15-year data.

\subsection{Non-standard setup}

Here our goal is to write down the expression for the inflationary potential and its derivatives at the pivot scale $k=k_*\sim 0.05{\rm Mpc}^{-1}$. At $k=k_{*}$ the inflaton field value is given by $\phi_*$ in terms of which, and by making use of all the previously mentioned inputs written for inflationary observables, we write the following relations which will be helpful to construct the structure of the inflationary potential in the next subsection:
 \bea \textcolor{black}{V(\phi_*)}&=&\textcolor{black}{\frac{3}{2}\bigg(\frac{\pi^2 \Delta^2_{\zeta}(k_*)}{{\cal J}}\bigg)\times\frac{r}{{\cal F}}\;M^4_{\rm pl}=\frac{3}{2}\bigg(\frac{\pi^2 r\Delta^2_{\zeta}(k_*)}{{\cal Y}}\bigg)\;M^4_{\rm pl}},\\
\textcolor{black}{V^{'}(\phi_*)}&=&\textcolor{black}{\frac{3}{2}\bigg(\frac{\pi^2 \Delta^2_{\zeta}(k_*)}{{\cal J}}\bigg)\times\frac{r}{{\cal F}}\sqrt{\frac{r}{8{\cal F}}}\;M^3_{\rm pl}=\frac{3}{2}\bigg(\frac{\pi^2 r\Delta^2_{\zeta}(k_*)}{{\cal Y}}\bigg)\sqrt{\frac{r}{8{\cal F}}}\;M^3_{\rm pl}},\\
\textcolor{black}{V^{''}(\phi_*)}&=&\textcolor{black}{\frac{3}{4}\bigg(\frac{\pi^2 \Delta^2_{\zeta}(k_*)}{{\cal J}}\bigg)\times\frac{r}{{\cal F}}\bigg(\frac{3}{8}\frac{r}{{\cal F}}+n_s-1\bigg)\;M^2_{\rm pl}=\frac{3}{4}\bigg(\frac{\pi^2 r\Delta^2_{\zeta}(k_*)}{{\cal Y}}\bigg)\bigg(\frac{3}{8}\frac{r}{{\cal F}}+n_s-1\bigg)\;M^2_{\rm pl}}.\eea

The aforementioned factors are necessary and sufficient for reconstructing the inflationary potential structure, which we discuss in the next section in detail. Here ${\cal J}$ and ${\cal F}$ are the modification factors appearing from the detailed structure of the Bogoliubov coefficients of the scalar and tensor perturbations in the context of NBD initial condition. For a consistency check of the above-mentioned results, if we substitute ${\cal J}=1$ and ${\cal F}=1$ then the results for the BD initial condition can be easily visualized. However since BD initially supports red tilted tensor power spectrum, which is against the findings of the NANOGrav 15 signal, in this article we will mainly focus on the outcomes and further details of NBD initial condition. Further, it is important to note that the above expressions can be easily derived by making use of the new consistency relation which incorporates blue tilted tensor modes, its relationship with the scale of inflation, the amplitude of the scalar power spectrum, its spectral tilt, and its relationship with the potential dependent slow-roll parameters $\epsilon_V$ and $\eta_V$, which are previously defined using the first and second derivatives of the inflationary potential $V(\phi)$. In this construction, the symbol $'$ describes the derivatives with respect to the inflaton field.
To understand the features and structure of the inflationary potential in a more detailed fashion one can further introduce two more slow-roll parameters which are described by the following expressions:
\bea \xi^2_V=M^{4}_{\rm pl}\left(\frac{V^{'}(\phi)V^{'''}(\phi)}{V^2(\phi)}\right), \quad \quad \sigma^3_V=M^{6}_{\rm pl}\left(\frac{V^{'2}(\phi)V^{''''}(\phi)}{V^3(\phi)}\right).\eea
Using the above-mentioned two newly introduced slow-roll parameters and with the help of slow-roll hierarchy one can further derive the following expressions for $V^{'''}(\phi_*)$  and $V^{''''}(\phi_*)$, which are given by:
 \bea \textcolor{black}{V^{'''}(\phi_*)}&=&\textcolor{black}{\frac{3}{2}\bigg(\frac{\pi^2 \Delta^2_{\zeta}(k_*)}{{\cal J}}\bigg)\times\frac{r}{{\cal F}}\Bigg[\sqrt{\frac{r}{2{\cal F}}}\bigg(\frac{3}{8}\frac{r}{{\cal F}}+n_s-1\bigg)-3{\cal F}\left(\frac{r}{8{\cal F}}\right)^{3/2}-\frac{\alpha_s}{2}\sqrt{\frac{8{\cal F}}{r}}\Bigg]\;M_{\rm pl}}\nonumber\\
 &=&\textcolor{black}{\frac{3}{2}\bigg(\frac{\pi^2 r\Delta^2_{\zeta}(k_*)}{{\cal Y}}\bigg)\Bigg[\sqrt{\frac{r}{2{\cal F}}}\bigg(\frac{3}{8}\frac{r}{{\cal F}}+n_s-1\bigg)-3{\cal F}\left(\frac{r}{8{\cal F}}\right)^{3/2}-\frac{\alpha_s}{2}\sqrt{\frac{8{\cal F}}{r}}\Bigg]\;M_{\rm pl}},\\
\textcolor{black}{V^{''''}(\phi_*)}&=&\textcolor{black}{12\pi^2 \frac{r}{{\cal F}}\Delta^2_{\zeta}(k_*)\bigg(\frac{{\cal F}}{{\cal J}}\bigg)\Bigg\{\frac{r}{4{\cal F}}\bigg(\frac{3}{8}\frac{r}{{\cal F}}+n_s-1\bigg)^2+96\left(\frac{r}{8{\cal F}}\right)^2\bigg[1-\frac{1}{2}\bigg(\frac{3}{8}\frac{r}{{\cal F}}+n_s-1\bigg)\bigg]}\nonumber\\
&&\quad\quad\quad\quad\quad\quad\quad\quad\quad\textcolor{black}{-\frac{1}{2}\bigg(n_s-1-\frac{9}{8}\frac{r}{{\cal F}}\bigg)\Bigg[\frac{r}{4{\cal F}}\bigg(\frac{3}{8}\frac{r}{{\cal F}}+n_s-1\bigg)-12\left(\frac{r}{8{\cal F}}\right)^{2}-\frac{\alpha_s}{2}\Bigg]+\frac{\kappa_s}{2}\Bigg\}}\nonumber\\
&=&\textcolor{black}{12\pi^2 r\Delta^2_{\zeta}(k_*)\bigg(\frac{1}{{\cal J}}\bigg)\Bigg\{\frac{r}{4{\cal F}}\bigg(\frac{3}{8}\frac{r}{{\cal F}}+n_s-1\bigg)^2+96\left(\frac{r}{8{\cal F}}\right)^2\bigg[1-\frac{1}{2}\bigg(\frac{3}{8}\frac{r}{{\cal F}}+n_s-1\bigg)\bigg]}\nonumber\\
&&\quad\quad\quad\quad\quad\quad\quad\quad\textcolor{black}{-\frac{1}{2}\bigg(n_s-1-\frac{9}{8}\frac{r}{{\cal F}}\bigg)\Bigg[\frac{r}{4{\cal F}}\bigg(\frac{3}{8}\frac{r}{{\cal F}}+n_s-1\bigg)-12\left(\frac{r}{8{\cal F}}\right)^{2}-\frac{\alpha_s}{2}\Bigg]+\frac{\kappa_s}{2}\Bigg\}}\;.\eea

Here we have introduced two quantities $\alpha_s$ and $\kappa_s$, which represent the running and running of the running of scalar spectral tilt and are described by the following expressions:
\bea \alpha_s=\left(\frac{dn_s}{d\ln k}\right)_*,\quad\quad  \kappa_s=\left(\frac{d^2n_s}{d\ln k^2}\right)_*.\eea
Due to the present construction in the presence of NBD initial condition, one can immediately expect that the power law parametrization of the scalar spectrum can be slightly modified. However, contributions from both of these inflationary parameters are expected to be small and for this reason one can able able to safely ignore such additional effect in the parametrization. We have computed these values at the pivot scale for a given set of benchmark points from the reconstructed potential in the next section for completeness. Finally, here it is important to mention that, using our analysis one can clearly observe that $V(\phi_*)$, $V^{'}(\phi_*)$, $V^{''}(\phi_*)$, $V^{'''}(\phi_*)$ and $V^{''''}(\phi_*)$ can be fixed completely at the CMB pivot scale, $k=k_*=0.05{\rm Mpc}^{-1}$, corresponding to the field value $\phi=\phi_*$ using constraints from CMB observation from Planck and the recently observed NANOGrav 15 signal.

\subsection{Fixing the potential from observation}

Let us start our discussion with the following parametrization of the inflationary potential:
 \bea V(\phi)&=&V_0+A\phi^p-B\phi^q+C\phi^s,\eea
where the term $V_0$ mimics the role of Cosmological Constant and quantified by, $V_0\approx 3H^2_{\rm inf}M^2_{\rm pl}$, where $H_{\rm inf}$ represents the Hubble parameter during the epoch of inflation. Apart from this Cosmological Constant term $V_0$ the above mentioned generic single field potential is parametrized by $A$, $B$, $C$, $p$, $q$ and $s$. By imposing the observational constraints at the CMB pivot scale our job is to compute and determine all of the above-mentioned parameters for different benchmark points, where we need to supply the value of tensor-to-scalar ratio $r$, the amplitude of the scalar power spectrum $\Delta^2_{\zeta}$, and spectral index $n_s$. Once the tensor-to-scalar ratio is fixed from the benchmark criteria, automatically the value of the tenor spectral tilt $n_t$ will be fixed because of having a new consistency relation from the NBD initial condition. Apart from the above-mentioned seven unknown parameters using the previously mentioned observational constraints, one can further fix three more parameters ${\cal F}$ and ${\cal J}$, and hence ${\cal Y}$ which captures the detailed information regarding the structure of the Bogoliubov coefficients describing the scalar and tensor sectors of the cosmological perturbation characterizing the corresponding NBD initial condition.

Now, from the mentioned generic form of the inflationary potential one can further compute its various derivatives with respect to the inflaton field which will prove to be extremely useful while performing the rest of the analysis in this paper:
\bea V^{'}(\phi)&=&Ap\phi^{p-1}-Bq\phi^{q-1}+Cs\phi^{s-1},\\
V^{''}(\phi)&=&Ap(p-1)\phi^{p-2}-Bq(q-1)\phi^{q-2}+Cs(s-1)\phi^{s-2},\\
V^{'''}(\phi)&=&Ap(p-1)(p-2)\phi^{p-3}-Bq(q-1)(q-2)\phi^{q-3}+Cs(s-1)(s-2)\phi^{s-3},\\
V^{''''}(\phi)&=&Ap(p-1)(p-2)(p-3)\phi^{p-4}-Bq(q-1)(q-2)(q-3)\phi^{q-4}+Cs(s-1)(s-2)(s-3)\phi^{s-4}.\eea

Using the above-mentioned expressions of the various derivatives of inflationary potential evaluated at the CMB pivot scale, particularly at the inflaton field value $\phi=\phi_*$, which we fix below the Planck scale to validate the EFT prescription, we can write the following matrix equation in the present context:
\bea \begin{pmatrix}
V(\phi_*)-V_0\\
V^{'}(\phi_*) \\
V^{''}(\phi_*) \\
\end{pmatrix}=\underbrace{\begin{pmatrix}
\phi^p_* & -\phi^q_* & \phi^s_*\\
p\phi^{p-1}_* & -q\phi^{q-1}_* & s\phi^{s-1}_*\\
p(p-1)\phi^{p-2}_* & -q(q-1)\phi^{q-2}_* & s(s-1)\phi^{s-2}_*\\
\end{pmatrix}}_{\equiv K}\begin{pmatrix}
A\\
B \\
C \\
\end{pmatrix}\eea
Here for the transfer matrix $K$ we find ${\rm det}(K)\neq 0$, which implies the $K^{-1}$ exists in the present context. As a consequence the factors $A$, $B$ and $C$ can be determined through the following inverse matrix equation:
\bea \begin{pmatrix}
A\\
B \\
C \\
\end{pmatrix}=\underbrace{\begin{pmatrix}
\phi^p_* & -\phi^q_* & \phi^s_*\\
p\phi^{p-1}_* & -q\phi^{q-1}_* & s\phi^{s-1}_*\\
p(p-1)\phi^{p-2}_* & -q(q-1)\phi^{q-2}_* & s(s-1)\phi^{s-2}_*\\
\end{pmatrix}^{-1}}_{\equiv K^{-1}}\begin{pmatrix}
V(\phi_*)-V_0\\
V^{'}(\phi_*) \\
V^{''}(\phi_*) \\
\end{pmatrix}\eea
After doing algebraic manipulation and performing the matrix inversion we get the following simplified expressions for the unknown coefficients, $A$, $B$, and $C$ by the following expressions:
\begin{eqnarray} A&=&\frac{1}{\Delta}\displaystyle\Bigg\{ qs\left(q-s\right)\left(V(\phi_*)-V_0\right)
+\left(s(s-1)-q(q-1)\right)V^{'}(\phi_*)\phi_*+\left(q-s\right)V^{''}(\phi_*)\phi^2_*\Bigg\} ,\\
     B&=&\frac{1}{\Delta}\displaystyle\Bigg\{ ps\left(p-s\right)\left(V(\phi_*)-V_0\right)
+\left(s(s-1)-p(p-1)\right)V^{'}(\phi_*)\phi_*+\left(p-s\right)V^{''}(\phi_*)\phi^2_*\Bigg\} ,\\
     C&=&\frac{1}{\Delta}\displaystyle\Bigg\{ pq\left(p-q\right)\left(V(\phi_*)-V_0\right)
+\left(q(q-1)-p(p-1)\right)V^{'}(\phi_*)\phi_*+\left(p-q\right)V^{''}(\phi_*)\phi^2_*\Bigg\}.
\end{eqnarray}

where the factor $\Delta$ appearing in the denominator is defined as:
\bea \Delta=\displaystyle\bigg(pq(p-q)-ps(p-s)+qs(q-s) \bigg)\phi^p_*.\eea
Using the above-mentioned relation one can fix the mathematical form of the coefficients $A$, $B$, and $C$ in terms of the index $p$, $q$, $s$, Cosmological Constant $V_0$, potential and its derivatives $V(\phi_*)$, $V^{'}(\phi_*)$, $V^{''}(\phi_*)$ at the CMB pivot scale where the inflaton field value is $\phi=\phi_*<M_{\rm pl}$.

\subsection{Beta functions for non-standard cosmology}

In this section we talk about the Cosmological Beta functions which are obtained from the flow equations by maintaining the slow-roll hierarchy of the reconstructed inflationary potential. To serve this purpose we will consider the following steps:
\begin{enumerate}
    \item Using the determined structure of reconstructed inflationary potential first of all we compute the expressions for the slow-roll parameters $\epsilon_V$, $\eta_V$, $\xi^2_V$ and $\sigma^3_V$ with respect to inflaton field. To connect it with observational constraint we further compute their values at the CMB pivot scale $\phi_*$ considering all the modifications appearing from NBD initial condition.

    \item From the field dependent behaviour of either slow-roll parameters, $\epsilon_V$ or $\eta_V$, one can further determine the field value at the end of inflation by demanding that from the present prescription one should be able to get at least the necessary amount of e-foldings $\Delta {\cal N}\sim 60$. Here the end of inflation can be achieved from the violation of the slow-roll condition, which can be achieved either from $\epsilon_V(\phi_{\rm end})=1$ or from $|\eta_{V}(\phi_{\rm end})|=1$. Here $\phi_{\rm end}$ corresponds to the inflaton field value at the end of inflation. Since we have just mentioned the required number of e-foldings, it is better to provide the definition of this quantity in terms of the reconstructed inflationary potential which is given by the following expression:
    \bea \Delta{\cal N}=\frac{1}{M_{\rm pl}}\int^{\phi_*}_{\phi_{\rm end}}\frac{d\phi}{\sqrt{2\epsilon_V(\phi)}}\sim 60,\eea
    which we have to fix throughout our analysis. In this connection another important fact which we must remind ourselves is that the obtained inflaton field value at the end of inflation, $\phi_{\rm end}$, has to be in the sub-Planckian regime which further helps us to satisfy the necessary and sufficient condition for the new field excursion formula as a result of the NBD initial condition. This implies that throughout our analysis not only we have to maintain the sufficient number of e-foldings to pursue inflation successfully, but also one need to assure that the conditions, $|\Delta\phi|<M_{\rm pl}$ with $\phi_*<M_{\rm pl}$ and $\phi_{\rm end}<M_{\rm pl}$ always maintained.

    \item Further utilizing the obtained structure of the reconstructed inflationary potential and considering the crucial fact that inflation is initiated through the NBD initial condition, we obtain the following expressions for the Cosmological Beta functions from the scalar perturbations:
\bea \alpha_s&=&\left(\frac{dn_s}{d\ln k^2}\right)=\left(16\epsilon_V\eta_V-24\epsilon^2_V-2\xi^2_V\right),\\
   \kappa_s&=&\left(\frac{d^2n_s}{d\ln k^2}\right)=\left(192\epsilon^2_V\eta_V-192\epsilon^3_V+2\sigma^3_V-24\epsilon_V\xi^2_V+2\eta_V\xi^2_V-32\eta^2_V\epsilon_V\right).\eea

   During the derivations of the above-mentioned flow equations from the slow-roll hierarchy, we have utilized the following facts for the NBD initial condition-driven Bogoliubov coefficients describing the scalar perturbation sector:
   \bea &&\left(\frac{d^2{\cal J}}{d\ln k^2}\right),\left(\frac{d^3{\cal J}}{d\ln k^3}\right) \ll \left(\epsilon_V,\eta_V,\xi^2_V,\sigma^3_V\right)_*.
   \eea
   This implies that one can safely neglect the scale-dependent running behaviour of this modification factor at the pivot scale even in the presence of NBD initial condition. The correction factor appearing through this analysis becomes extremely small compared to all slow-roll parameters introduced in this work before. Now from the above structure, it is very obvious that the running and running of the scalar spectral tilt exactly map to the obtained result from BD initial condition. However, this statement is completely wrong as due to having correct reconstruction of inflationary potential all the slow-roll parameters will be corrected in the presence of two modification factors ${\cal F}$ and ${\cal J}$ due to having extremely crucial NBD initial condition in the present context of the discussion.

   \item \textcolor{black}{Finally, we will talk about the Cosmological Beta Functions from tensor perturbations:}
   \bea \textcolor{black}{\alpha_t}&=&\textcolor{black}{\left(\frac{dn_t}{d\ln k}\right)=\left(4\epsilon_V\eta_V-8\epsilon^2_V\right)+\left(\frac{d{\cal G}}{d\ln k}\right)},\\
   \textcolor{black}{\kappa_t}&=&\textcolor{black}{\left(\frac{d^2n_t}{d\ln k^2}\right)=\left(56\epsilon^2_V\eta_V-64\epsilon^3_V-8\eta^2_V\epsilon_V-4\epsilon_V\xi^2_V\right)+\left(\frac{d^2{\cal G}}{d\ln k^2}\right)}.\eea
  \textcolor{black}{ Here we have to use the following parametrization of the modification factor ${\cal G}$ as appearing from the NBD initial condition:}
 \bea \textcolor{black}{ {\cal G}}
&=& \textcolor{black}{\displaystyle
\displaystyle\left\{
	\begin{array}{ll}
		\displaystyle 0& \mbox{for}\quad {\bf Phase\;I:}\;10^{-20}{\rm Hz}< f < 10^{-17}{\rm Hz} \;  \\  
			\displaystyle 
			\displaystyle 10^{28}\exp(-22n_t)+n_t \quad\quad\quad\quad\quad\quad\quad\quad& \mbox{for}\quad  {\bf Phase\;II:}\;10^{-17}{\rm Hz}< f < 10^{-7}{\rm Hz} \; \\
   \displaystyle 
			\displaystyle 10^{28+6.89n_t}\exp(22n_t)-n_t \quad\quad\quad\quad\quad\quad\quad\quad& \mbox{for}\quad  {\bf Phase\;III:}\;10^{-7}{\rm Hz}< f < 1{\rm Hz} \;. 
	\end{array}
\right.} \eea
   \textcolor{black}{Now using this relation, we obtain the following results for the running and running of the running of modification factor ${\cal G}$:}
   \bea  \textcolor{black}{\left(\frac{d{\cal G}}{d\ln k}\right)}
&=&\textcolor{black}{ \displaystyle
\displaystyle\left\{
	\begin{array}{ll}
		\displaystyle 0& \mbox{for}\quad {\bf Phase\;I:}\;10^{-20}{\rm Hz}< f < 10^{-17}{\rm Hz} \;  \\  
			\displaystyle 
			\displaystyle \alpha_t\bigg(1-22\times 10^{28}\exp(-22n_t)\bigg) \quad\quad& \mbox{for}\quad  {\bf Phase\;II:}\;10^{-17}{\rm Hz}< f < 10^{-7}{\rm Hz} \; \\
   \displaystyle 
			\displaystyle \alpha_t \bigg(37.86\times10^{28+6.89n_t}\exp(22n_t)-1\bigg) \quad\quad& \mbox{for}\quad  {\bf Phase\;III:}\;10^{-7}{\rm Hz}< f < 1{\rm Hz} \;. 
	\end{array}
\right. }\eea
\bea  \textcolor{black}{\left(\frac{d^2{\cal G}}{d\ln k^2}\right)}
&=& \textcolor{black}{\displaystyle
\displaystyle\left\{
	\begin{array}{ll}
		\displaystyle 0& \mbox{for}\quad {\bf Phase\;I:}\;10^{-20}{\rm Hz}< f < 10^{-17}{\rm Hz} \;  \\  
			\displaystyle 
			\displaystyle \bigg\{\kappa_t\bigg(1-22\times 10^{28}\exp(-22n_t)\bigg)\\
   +4.84\times10^{30}\alpha^2_t\exp(-22n_t)\bigg\} \quad\quad& \mbox{for}\quad  {\bf Phase\;II:}\;10^{-17}{\rm Hz}< f < 10^{-7}{\rm Hz} \; \\
   \displaystyle 
			\displaystyle \bigg\{\kappa_t \bigg(37.86\times10^{28+6.89n_t}\exp(22n_t)-1\bigg)\\
   +18924\alpha^2_t\times10^{28+6.89n_t}\exp(22n_t)\bigg\}\quad\quad& \mbox{for}\quad  {\bf Phase\;III:}\;10^{-7}{\rm Hz}< f < 1{\rm Hz} \;. 
	\end{array}
\right.} \eea
   \textcolor{black}{Using both of these expressions we get the following simplified expressions for the Cosmological Beta Functions from tensor perturbations:}
 \bea \textcolor{black}{\alpha_t}
&=& \textcolor{black}{\displaystyle
\displaystyle\left\{
	\begin{array}{ll}
		\displaystyle 0& \mbox{for}\quad {\bf Phase\;I:}\;10^{-20}{\rm Hz}< f < 10^{-17}{\rm Hz} \;  \\  
			\displaystyle 
			\displaystyle4.55\times 10^{-30}\exp(22n_t)\times\left(4\epsilon_V\eta_V-8\epsilon^2_V\right) \quad\quad& \mbox{for}\quad  {\bf Phase\;II:}\;10^{-17}{\rm Hz}< f < 10^{-7}{\rm Hz} \; \\
   \displaystyle 
			\displaystyle \frac{\left(4\epsilon_V\eta_V-8\epsilon^2_V\right)}{\bigg(2-37.86\times10^{28+6.89n_t}\exp(22n_t)\bigg)} \quad\quad& \mbox{for}\quad  {\bf Phase\;III:}\;10^{-7}{\rm Hz}< f < 1{\rm Hz} \;. 
	\end{array}
\right. }\eea

 \bea \textcolor{black}{\kappa_t}
&=& \textcolor{black}{\displaystyle
\displaystyle\left\{
	\begin{array}{ll}
		\displaystyle 0& \mbox{for}\quad {\bf Phase\;I:}\;10^{-20}{\rm Hz}< f < 10^{-17}{\rm Hz} \;  \\  
			\displaystyle 
			\displaystyle\bigg\{4.55\times 10^{-30}\exp(22n_t)\times\\ \left(56\epsilon^2_V\eta_V-64\epsilon^3_V-8\eta^2_V\epsilon_V-4\epsilon_V\xi^2_V\right)\\+22\alpha^2_t\bigg\} \quad\quad& \mbox{for}\quad  {\bf Phase\;II:}\;10^{-17}{\rm Hz}< f < 10^{-7}{\rm Hz} \; \\
   \displaystyle 
			\displaystyle \frac{1}{\bigg(2-37.86\times10^{28+6.89n_t}\exp(22n_t)\bigg)}\times\\
   \bigg\{\left(56\epsilon^2_V\eta_V-64\epsilon^3_V-8\eta^2_V\epsilon_V-4\epsilon_V\xi^2_V\right)\\
   +18924\alpha^2_t\times10^{28+6.89n_t}\exp(22n_t)\bigg\}
   \quad\quad& \mbox{for}\quad  {\bf Phase\;III:}\;10^{-7}{\rm Hz}< f < 1{\rm Hz} \;. 
	\end{array}
\right.} \eea

  \textcolor{black}{ Despite the fact that we are aware about these parameters not appearing in the single field inflationary potential's final reconstructed mathematical structure and that they cannot be directly constrained by cosmological observations, we are still considering the results of our computation for the sake of completeness. We anticipate that this extra examination of the rebuilt inflationary potential will yield novel results that will allow us to explore some as-yet-unexplored territory in this direction and result in a physically consistent projection from the current research.}
\end{enumerate}

\section{UV completion of non-standard scenario }\label{sec6}

  In this section, our prime focus is to connect the reconstructed form of the inflationary potential with a known physical framework to guarantee successful UV completion of the present scenario. 
  
  To this end, let us start with the framework of $N=1$ SUGRA theory where the generic structure of the holomorphic superpotential and K$\ddot{a}$hler potential describing the visible light inflaton sector and the heavy field hidden sector is described by the following expressions \footnote{Apart from the mentioned form of the superpotential describing the hidden sector one can consider another example, where $W(S,\Phi)=\Phi(S^2-M^2)$. See refs.\cite{Choudhury:2013jya,Choudhury:2014sxa,Choudhury:2014uxa} for more details.}:
  \bea &&W(\Phi,S)=W(\Phi)+W(S), \quad\quad {\rm where} \quad\quad  W(\Phi)=\lambda\bigg(\frac{\Phi^n}{M^{n-3}_{\rm pl}}\bigg),\quad\quad\quad W(S)=M^2S,\\
&&K(\Phi,S)=\Phi^{\dagger}\Phi+S^{\dagger}S+\delta K(\Phi,S)\quad\quad{\rm with}\quad\delta K(\Phi,S)=f_1(\Phi^{\dagger}\Phi,S^{\dagger}S),f_2(S^{\dagger}\Phi\Phi),f_3(S^{\dagger}S^{\dagger}\Phi\Phi),f_4(S\Phi^{\dagger}\Phi),\quad\quad\eea
  where $\Phi$ represents a superfield describing along a SUGRA flat direction, $\lambda$ characterizes the self-coupling parameter of the theory which helps to shift the VEV of the superfield along the same flat direction, $S$ is the heavy field which belongs to the hidden sector of the underlying theory, $M$ represents the energy scale corresponding to the same sector which in turn fixes the final form of the Cosmological Constant term which describes the initial energy density of the universe, $\delta K(\Phi,S)$ represent the higher order SUGRA corrections, which are extremely hard to compute in a generalized prescription. Within the framework of String Theory, in some very special cases with the MSSM field contents, one can compute some of the corrections explicitly. For more details on this computation see the ref \cite{Choudhury:2013jya,Choudhury:2014sxa,Choudhury:2014uxa}.  In this work as discussed earlier to validate the prescription of EFT we always maintain the VEV of the inflaton field has to be below the Planckian cut-off scale, $M_{\rm pl}$. Within the framework of  $N=1$ SUGRA theory to get a meaningful physically justifiable form of the effective potential the lower bound on the index $n$ is fixed at, $n\geq 3$. Then for $N=1$ SUGRA theory for the $D$-flat directions, the effective inflationary potential can be further written in terms of the generic form of the superpotential and K$\ddot{a}$hler potential as:
  \bea V&=&\exp\left(\frac{K(\Phi,S)}{M^2_{\rm pl}}\right)\Bigg[\Bigg\{\bigg(D_{\Phi}W(\Phi,S)\bigg)K^{\Phi\Phi^{\dagger}}\bigg(D_{\Phi^{\dagger}}W(\Phi,S)\bigg)+\bigg(D_{S}W(\Phi,S)\bigg)K^{SS^{\dagger}}\bigg(D_{S^{\dagger}}W(\Phi,S)\bigg)\nonumber\\
  &&\quad\quad\quad\quad\quad\quad\quad\quad\quad\quad+\bigg(D_{S^{\dagger}}W(\Phi,S)\bigg)K^{S^{\dagger}\Phi}\bigg(D_{\Phi}W(\Phi,S)\bigg)+\bigg(D_{S}W(\Phi,S)\bigg)K^{S\Phi^{\dagger}}\bigg(D_{\Phi^{\dagger}}W(\Phi,S)\bigg)\Bigg\}
  \nonumber\\
  &&\quad\quad\quad\quad\quad\quad\quad\quad\quad\quad\quad\quad\quad\quad\quad\quad\quad\quad\quad\quad\quad\quad\quad\quad\quad\quad\quad\quad\quad\quad\quad\quad\quad\quad\quad\quad-\frac{3}{M^2_{\rm pl}}|W(\Phi,S)|^2\Bigg].\eea

  Here it is important to note that, due to having underlying $N=1$ SUGRA theory structure the inflaton field gets a soft SUSY breaking mass contribution which is of the order of, $m_{\phi}\geq {\cal O}(1\;{\rm TeV})$.
Now considering the relevant operators from the renormalizable sector and also considering the irrelevant operators from the non-renormalizable sector of the $N=1$ SUGRA theory, the final form of the effective potential can be easily computed, which can be easily parameterized in terms of the index $n$ and the soft SUSY breaking mass term $m_{\phi}$. To fully write down the final form of the effective potential here it is additionally important to note that within the framework of $N=1$ SUGRA theory, the potential is dominated by the Cosmological Constant terms which can be expressed in terms of the Hubble-induced correction factors \footnote{In $N=1$ SUGRA theory the Hubble induced corrections are characterized in terms of the mass and the supersymmetric trilinear $A$ term.}. In principle, the presence of the Hubble-induced corrections can be described in terms of a scenario where the structure of the K$\ddot{a}$hler term is canonical for the inflaton which describes the visible sector of the theory and the hidden sector in this situation is described by the heavy fields which one needs to be static during inflationary evolution with respect to the underlying time scale of the theory. This is actually mimicking the role of performing path integral over the heavy degrees of freedom of the theory and constructing a fruitful EFT description out of this scenario.  See refs. \cite{Choudhury:2017glj,Banerjee:2021lqu} for details on this issue. Considering the fact that, in the visible sector of the theory the superfield $\Phi$ is described by the parameterization, $\Phi=\phi\exp(i\theta)$ the general structure of the $N=1$ SUGRA potential for all possible allowed SUGRA flat directions can be written as \cite{Choudhury:2013jya,Choudhury:2014sxa,Choudhury:2014uxa}:

\vspace{-.35cm}
\bea V(\phi)=V_0+\frac{(m^2_\phi+c_H H^2)}{2}\phi^2+\frac{\left(a_{\lambda}m_{\phi}+a_{H}H\right)}{nM^{n-3}_{\rm pl}}\cos\left(n\theta+\theta_{a_{\lambda}}+\theta_{a_{H}}\right)\phi^n+\lambda^2\frac{\phi^{2(n-1)}}{M^{2n-3}_{\rm pl}},\eea

Now we consider the high-scale inflationary region, which is governed by, $m_{\phi}\ll H$ \footnote{On the contrary, one can consider a situation where the mass term related to soft SUSY breaking is much larger in comparison to the underlying Hubble scale of the theory i.e. $m_{\phi}\gg H$, which is commonly identified as the low scale inflation. Here one can safely neglect the contribution from the vacuum energy due to the absence of the Hubble-dominated contributions in the potential, i.e. $V_0\sim 0$. For this reason for low-scale inflation one can write down the following expression for the effective potential after minimizing along the angular direction $\theta$  of the light superfield \cite{Choudhury:2013jya,Choudhury:2014sxa,Choudhury:2014uxa}:
\bea V(\phi)=\frac{m^2_\phi}{2}\phi^2-\frac{a_{\lambda}m_{\phi}}{nM^{n-3}_{\rm pl}}\phi^n+\lambda^2\frac{\phi^{2(n-1)}}{M^{2n-3}_{\rm pl}}.\eea} and minimization with respect to the angular direction $\theta$ of the light superfield for further simplification and to connect with the realistic scenario supported by cosmological observation. 
In the case of high-scale inflation once we freeze the hidden sector contribution in this description then, within the $N=1$ SUGRA framework, a generic structure of the inflationary effective potential can be expressed in terms of the following expression considering a large possibility of SUGRA flat directions \cite{Choudhury:2013jya,Choudhury:2014sxa,Choudhury:2014uxa}:
\vspace{-.35cm}
\bea V(\phi)=V_0+\frac{c_H}{2} H^2\phi^2-\frac{a_{H}H}{nM^{n-3}_{\rm pl}}\phi^n+\lambda^2\frac{\phi^{2(n-1)}}{M^{2n-3}_{\rm pl}},\eea

where the first term represents Vacuum energy contribution which can be further expressed in terms of the Hubble parameter during inflation as, $V_0=3H^2M^2_{\rm pl}$ and is the most dominating term of the potential for the high scale inflation. In this above-mentioned expression, the second term corresponds to the Hubble parameter-induced mass term contribution where the corresponding mass is characterized by $m^2_H=c_HH^2$. Here further the coefficient $c_H$ is fixed from the detailed structure of the K$\ddot{a}$hler potential and can be safely absorbed in the contributions coming from the higher order K$\ddot{a}$hler correction terms. Within the present description, the flatness and corresponding slow-roll approximation is violated due to having the strict constraint from the soft SUSY breaking mass term, $m^2_{\phi}\ll m^2_H$. Due to this fact, one can immediately compute the expression for the second slow-roll parameter which turns out to be $\eta_V\sim c_H$, and now if we have large coupling parameter $c_H\sim {\cal O}(1)$ then this will further lead to the well known SUGRA $\eta_V$ problem \cite{Hardeman:2010fh}, which is an important fact in the related theories. Now in the above mentioned for the effective potential the last term represents the supersymmetric trilinear $A$ term, where the overall coefficient $a_H$ is a dimensionless positive number, and in principle, one can consider $a_H\sim c_H$ in the present discussion. Apart from having so many constraints from the theoretical ground, one can use the mentioned specific structure of the potential to study inflation by imposing a saddle or inflection point condition, where the mentioned problems can be easily evaded.  

For example, the inflection point condition can be accommodated by imposing the constraint condition, $a^2_H\approx 8(n-1)c_H$ on the coupling parameters of the $N=1$ SUGRA originated high scale models of inflation considering all possible $D$-flat direction for $3\leq n\leq 6$ with the coupling $\lambda=1$.
The inflection point is actually accommodated in the presence of fine-tuning parameter $\sigma\ll 1$, which is given by the following equation:
\bea \frac{a^2_H}{8(n-1)c_H}=1-\left(\frac{n-2}{2}\right)^2\sigma^2.\eea
For the small values of the fine-tuning parameter $\sigma$ the inflation point condition can be imposed by demanding the following constraint on the inflationary potential at the inflection point $\phi=\phi_0$ i.e.$V^{''}(\phi_0)=0$,
where the inflation field value at the inflection point is evaluated as:
\bea \phi_0=\left(\sqrt{\frac{c_H}{(n-1)}}HM^{n-3}_{\rm pl}\right)^{1/(n-2)}+{\cal O}(\sigma^2).\eea
As a result, the $N=1$ SUGRA-originated inflationary potential can be expanded in a Taylor series around the vicinity of the inflection point $\phi_0$ as:
\vspace{-.35cm}
\bea V(\phi)=V(\phi_0)+\alpha_1(\phi-\phi_0)+\alpha_2(\phi-\phi_0)^3+\alpha_3(\phi-\phi_0)^4,\eea

where the Taylor expansion coefficients $\alpha_1$, $\alpha_2$, $\alpha_3$ and $\alpha_4$ are given by the following expressions:
\begin{eqnarray}
     \alpha_1&=&V(\phi_{0})=V_{0}+\left(\frac{(n-2)^{2}}{n(n-1)}+\frac{(n-2)^2}{n}\sigma^{2}\right)c_{H}H^{2}\phi^{2}_{0}+{\cal O}(\sigma^{4}),\\
 \alpha_2&=&V^{'}(\phi_{0})=2\left(\frac{n-2}{2}\right)^{2}\sigma^{2}c_{H}H^{2}\phi_{0}+{\cal O}(\sigma^{4}),\\
 \alpha_3&=&\frac{V^{'''}(\phi_{0})}{3!}=\frac{c_{H}H^{2}}{\phi_{0}}\left(4(n-2)^2-\frac{(n-1)(n-2)^3}{2}\sigma^{2}\right)+{\cal O}(\sigma^{4}),\\
\alpha_4&=&\frac{V^{''''}(\phi_{0})}{4!}=\frac{c_{H}H^{2}}{\phi^{2}_{0}}\left(12(n-2)^3-\frac{(n-1)(n-2)(n-3)
(7n^2-27n+26)}{2}\sigma^{2}\right)+{\cal O}(\sigma^{4}).
   \end{eqnarray}
See refs. \cite{Allahverdi:2006iq,Enqvist:2003gh,Allahverdi:2006we,Allahverdi:2006cx,Enqvist:2003mr,Enqvist:2002rf,Enqvist:2010vd,Wang:2013hva,Mazumdar:2011ih,Chatterjee:2011qr,Allahverdi:2007wh,Choudhury:2013jya,Choudhury:2014sxa,Choudhury:2014uxa,Choudhury:2011jt,Choi:2016eif,Ghoshal:2022jeo,Allahverdi:2008bt,Hotchkiss:2011am,Enqvist:2010vd,Hotchkiss:2011gz,Choudhury:2014kma,Choudhury:2013woa,Choudhury:2018glz,Choudhury:2016wlj,Choudhury:2015pqa,Chatterjee:2014hna} to know more about the saddle and inflection point approximations in detail.

However, we have performed the analysis of the reconstruction without imposing any additional saddle point or inflection point constraint and without using any further approximation. This makes our analysis robust from the theoretical perspective and we are going to explicitly prove the robustness of our analysis in the numerical analysis section through the use of existing observational constraints from the CMB observation, from Planck, and the recently observed NANOGrav 15-year GWs signal. For this reason, our reconstructed form of the potential exactly matches the high-scale scale models of inflation when we use the following identifications:
\bea && V_0\neq 0,\quad\quad p=2,\quad\quad q=n,\quad\quad s=2(n-1),\quad\quad A=\frac{c_H}{2}H^2,\quad\quad B=\frac{a_H}{nM^{n-3}_{\rm pl}},\quad\quad C=\frac{\lambda^2}{M^{2n-3}_{\rm pl}}.\eea
where $n\geq 3$ restrictions in the choices of $D$-flat directions have been always maintained very carefully. Further using the above-mentioned values we get the following information regarding the couplings of $N=1$ SUGRA applicable within the framework of high-scale inflation:
\begin{eqnarray} c_H&=&\frac{1}{\left(n-2\right)^2\phi^2_*H^2}\displaystyle\Bigg\{ 2n(n-1)\left(V(\phi_*)-V_0\right)
-3(n-1)V^{'}(\phi_*)\phi_*+V^{''}(\phi_*)\phi^2_*\Bigg\} ,\\
     a_H&=&\frac{nM^{n-3}_{\rm pl}}{2\left(n-2\right)^2\phi^2_*}\displaystyle\Bigg\{ 8(n-1)\left(V(\phi_*)-V_0\right)
-2(2n-1)V^{'}(\phi_*)\phi_*+2V^{''}(\phi_*)\phi^2_*\Bigg\} ,\\
     \lambda^2&=&\frac{M^{2n-3}_{\rm pl}}{2\left(n-2\right)^2\phi^2_*}\displaystyle\Bigg\{ 2n\left(V(\phi_*)-V_0\right)
-(n+1)V^{'}(\phi_*)\phi_*+V^{''}(\phi_*)\phi^2_*\Bigg\}.\end{eqnarray}

Now we will talk about the inflaton candidates for $N=1$ SUGRA which can be obtained by imposing the $D$-flat condition in the present context and are given by:
\begin{itemize}[label={$\bullet$}]
    \item \underline{\bf For $n=3$:}
    In this case, the $D$-flat direction is $NH_uL$ on which ${\bf SSM}\otimes {\bf U(1)_{B-L}}$ symmetry is imposed with gauged ${\bf U(1)_{B-L}}$. Here ${\bf SSM}$ stands for the gauge group of the SUSY standard model. It is also important to note here, that $N$ represents the right-handed neutrino, $H_u$ corresponds to one of the Higgses which provide masses to the leptons and up quarks, and finally $L$ represents slepton having left-handedness. Consequently, for this specific flat direction, the superpotential and the corresponding inflaton candidate can be expressed as:
    \bea W\sim \lambda NH_uL\quad\quad {\rm with}\quad\quad \phi=\frac{1}{\sqrt{3}}\left(N+H_u+L\right).\eea
    Here the $\sqrt{3}$ factor comes from the correct normalization of the wave function.

    \item \underline{\bf For $n=4$:}
    In this case, the $D$-flat direction is $H_uH_d$. Here $H_d$ corresponds to one of the Higgses which provide masses to the leptons and down quarks. Consequently, for this specific flat direction, the superpotential and its corresponding inflaton candidate can be expressed as:
    \bea W\sim \frac{1}{M_{\rm pl}}H_uH_d\quad\quad {\rm with}\quad\quad \phi=\frac{1}{\sqrt{2}}\left(H_u+H_d\right).\eea
    Here the $\sqrt{2}$ factor comes from the correct normalization of the wave function.

     \item \underline{\bf For $n=6$:}
    In this case, the $D$-flat direction is $udd$ and $LLe$. Here $e$ represents selectron in the right-handed sector, $u$, and $d$ represents up and down squark superfield. Consequently, for this specific flat direction, the superpotential and its corresponding inflaton candidate can be expressed as:
    \bea && W\sim \lambda udd\quad\quad {\rm with}\quad\quad \phi=\frac{1}{\sqrt{3}}\left(u+d+d\right),\\
    && W\sim \lambda LLe\quad\quad {\rm with}\quad\quad \phi=\frac{1}{\sqrt{3}}\left(L+L+e\right).\eea
    Here the $\sqrt{3}$ factor comes from the correct normalization of the wave function.

\end{itemize}
   
\section{Numerical analysis and its implications}
\label{sec7}

\begin{table}


\begin{tabular}{|c|c|c|c|c|c|c|c|}

 \hline\hline

 \multicolumn{8}{|c|}{
 Parameters of the reconstructed potential.} \\
 
 \hline\hline
 
 \bf{(p,q,s)}  & \bf{A}$(\times 10^{-8})$& \bf{B}$(\times 10^{-8})$ & \bf{C}$(\times 10^{-8})$ &${\cal F}$& $\textcolor{black}{{\cal Y}(\times 10^{-21})}$ & $\phi_{\text{end}}$\; (in \;$M_{\rm pl}$)&$|\Delta\phi|$\; (in \;$M_{\rm pl}$)\\
 
 \hline\hline
$(2,3,4)$& $-7.356$ & $-10.906$ & $-4.543$ & $10^{-15.080}$& $8.424$& $0.262$& $0.638$\\
 
\hline\hline
$(2,4,6)$& $-3.676$ & $-4.545$ & $-1.870$ & $10^{-15.105}$& $7.954$&$0.354$ & $0.546$\\

\hline\hline
$(2,6,10)$& $-2.295$ & $-2.338$ & $-1.069$ & $10^{-15.104}$& $7.982$&$0.467$ & $0.433$\\

\hline\hline
$(2,5,8)$& $-2.721$ & $-2.993$ & $-1.283$ & $10^{-15.100}$& $8.071$&$0.418$  & $0.481$\\

\hline\hline
$(2,6,8)$& $-2.448$ & $-3.739$ & $-2.308$ & $10^{-15.006}$& $9.982$&$0.446$ & $0.454$\\

\hline\hline
$(2,4,8)$& $-3.266$ & $-3.029$ & $-0.769$ & $10^{-15.009}$& $9.899$&$0.380$  & $0.520$\\

\hline\hline

\end{tabular}

\caption{The table represents set of values for the parameters of a class of possible potentials labelled using the values in the first column. These parameters are obtained for a specific case where: tensor-to-scalar ratio \textcolor{black}{$r = 2.1\times 10^{-19}$, ${\cal J} = 10^{-6}$}, \textcolor{black}{with the scale of inflation, $V^{1/4}_{inf}\sim 0.12~M_{\rm pl}\sim 0.12\times 10^{16}~{\rm GeV}$ (i.e. $H_{inf}\sim 1.4\times 10^{-3}~M_{\rm pl}\sim 3.43\times 10^{15}~{\rm GeV}$)}, spectral index $n_{s} = 0.96$ while also achieving the required number of e-foldings to be ${\Delta\cal N} = 60$ for each of the potential at their respective $\phi_{\rm end}$ values keeping the quantity $|\Delta\phi| < 1$. All the estimates of ${\rm A}, {\rm B}, {\rm C},  {\cal Y}, \phi_{\rm end}$, and $|\Delta\phi|$ are rounded off to $3$ decimal points. Here we have taken reduced Planck mass, $M_{\rm pl}\sim 2.43\times 10^{18}~{\rm GeV}$}

\label{tab:1}

\end{table}

\begin{table}


\begin{tabular}{|c|c|c|c|c|c|c|c|c|c|}

 \hline\hline

 \multicolumn{10}{|c|}{
 Parameters of the reconstructed potential.} \\
 
 \hline\hline
 
 \bf{(p,q,s)}  & \bf{A}$(\times 10^{-8})$ & \bf{B}$(\times 10^{-8})$ & \bf{C}$(\times 10^{-8})$ & ${\cal F}$& ${\cal Y}$& $\phi_{\text{end}}$\; (in \;$M_{\rm pl}$)& $|\Delta\phi|$\; (in \;$M_{\rm pl}$)&\bf{r} & ${\cal J}$\\
 
 \hline\hline
$(2,3,4)$& $-7.071$ & $-16.516$ & $-4.382$& $10^{-11.091}$&$9.506 \times 10^{-21}$ &$0.259$ &$0.641$ &$\textcolor{black}{4.1 \times 10^{-14}}$ & $10^{-9}$\\

\hline\hline
$(2,3,4)$& $-7.118$ & $-10.578$ & $-4.408$ & $10^{-8.041}$&$9.374 \times 10^{-15}$ &$0.259$  &$0.640$& $\textcolor{black}{2.7 \times 10^{-11}}$ & $10^{-6}$\\

\hline\hline
$(2,3,4)$& $-7.211$ & $-10.701$ & $-4.459$ & $10^{-5.042}$ & $9.194 \times 10^{-9}$ & $0.261$ & $0.639$ & $\textcolor{black}{6.9 \times 10^{-9}}$ & $10^{-3}$\\

\hline\hline
$(2,4,6)$& $-3.599$ & $-4.458$ & $-1.835$ & $10^{-5.039}$& $9.236 \times 10^{-9}$ & $0.353$ &$0.546$ & $\textcolor{black}{6.9 \times 10^{-9}}$ & $10^{-3}$\\

\hline\hline
$(2,4,8)$& $-3.199$ & $-2.972$ & $-0.755$ & $10^{-5.057}$& $8.869 \times 10^{-9}$ &$0.379$  & $0.521$ & $\textcolor{black}{6.9 \times 10^{-9}}$ & $10^{-3}$\\

\hline\hline
$(2,3,4)$& $-7.027$ & $-10.458$ & $-4.356$ & $10^{-2.044}$ & $9.592 \times 10^{-3}$ & $0.259$ & $0.641$ & $\textcolor{black}{7.4 \times 10^{-5}}$ & $1$\\

\hline\hline

\end{tabular}

\caption{The table represents set of values for the parameters of a class of allowed potentials labelled using the values in the first column.  These parameters are obtained based on the conditions that: the spectral index $n_{s} = 0.96$, the number of e-foldings to be achieved is ${\Delta\cal N} = 60$ for each of the potential at their respective $\phi_{\rm end}$ values, $|\Delta\phi| < 1$, and the constraint on the slow-roll parameter $\xi^{2}_{V} < 0.5$ is also satisfied. All the estimates of ${\rm A}, {\rm B}, {\rm C},  {\cal Y}, \phi_{\rm end}$, and $|\Delta\phi|$ are rounded off to $3$ decimal points.}

\label{tab:2}

\end{table}

In this section, our prime objective is to discuss the numerical outcomes and prove the robustness of our analysis of reconstructing inflationary potential. To study the outcomes we consider five benchmark points which are described by the following conditions:
\begin{eqnarray} &&{\bf Benchmark\; I:}\quad\quad \;\;\; (r,n_t) = (\textcolor{black}{2.1 \times 10^{-19}},2.2),\\
&&{\bf Benchmark\; II:}\quad\quad\;\; (r,n_t) = (\textcolor{black}{4.1 \times 10^{-14}},1.8),\\
&&{\bf Benchmark\; III:}\quad\quad (r,n_t) = (\textcolor{black}{2.7 \times 10^{-11}},1.5),\\
&&{\bf Benchmark\; IV:}\quad\quad (r,n_t) = (\textcolor{black}{6.9 \times 10^{-9}},1.2),\\
&&{\bf Benchmark\; V:}\quad\quad\;\; (r,n_t) = (\textcolor{black}{7.4 \times 10^{-5}},0.85).
\end{eqnarray}

Using the aforementioned benchmark points in the parameter space we have found the following crucial outcomes, which we believe will be quite helpful to understand the vast range of applicability as well as the robustness of the reconstruction technique of inflationary potential studied in this paper:
\begin{itemize}[label={$\bullet$}]
    \item \underline{\bf{Benchmark I outcomes:}} \\ The values of the parameters obtained for the specific benchmark point of the ratio $r$ taken in Table (\ref{tab:1}) corresponds to the following results. These values are evaluated at the scale of CMB which is taken to be $\phi_{\rm *} = 0.9\;M_{\rm pl}$. Here we consider a very small tensor-to-scalar ratio i.e., \textcolor{black}{$r = 2.1 \times 10^{-19}$, that follows from the relation in equation (\ref{tensortoscalar}), and correspondingly fix ${\cal J} = 10^{-6}$}.
    \begin{figure*}[htb!]
    	\centering
   {
      	\includegraphics[width=18cm,height=9cm] {
      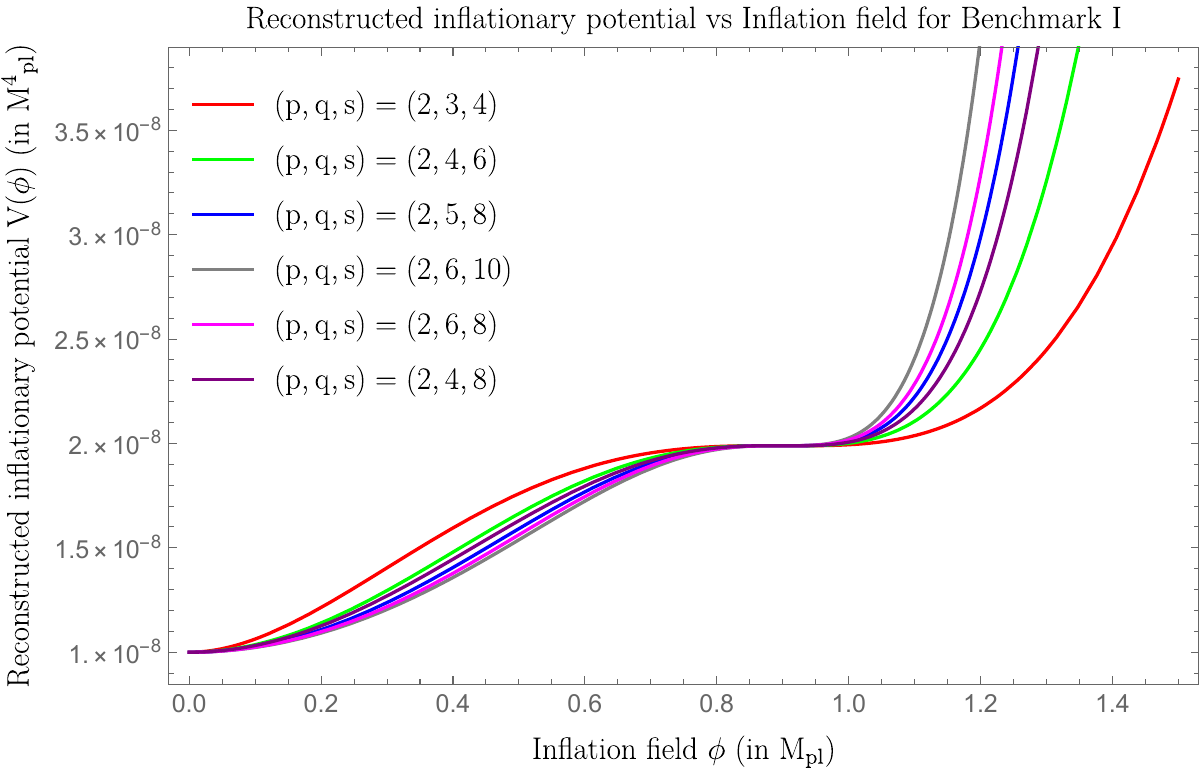}
        \label{fig1}
    }
    	\caption[Optional caption for list of figures]{The reconstructed potentials for the different allowed cases corresponding to Benchmark I is plotted against the inflaton field $\phi$.} 
        \label{potentialI}
    \end{figure*}
    \begin{figure*}[htb!]
    	\centering
   {
      	\includegraphics[width=18cm,height=9cm] {
      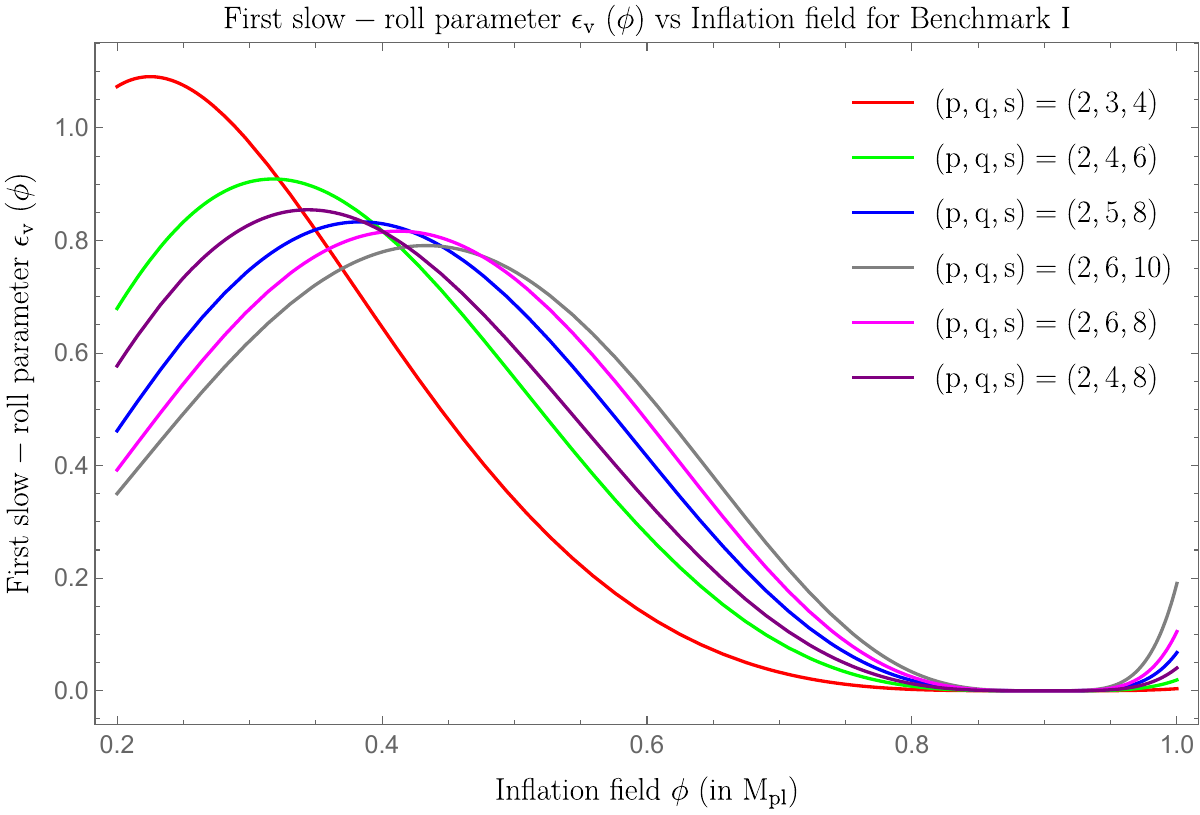}
        \label{fig2}
    }
    	\caption[Optional caption for list of figures]{The first slow-roll parameter $\epsilon_{V}$ for the same allowed cases corresponding to Benchmark I is plotted against the inflaton field $\phi$.} 
        \label{epsilonI}
    \end{figure*}
    \begin{figure*}[htb!]
    	\centering
   {
      	\includegraphics[width=18cm,height=9cm] {
      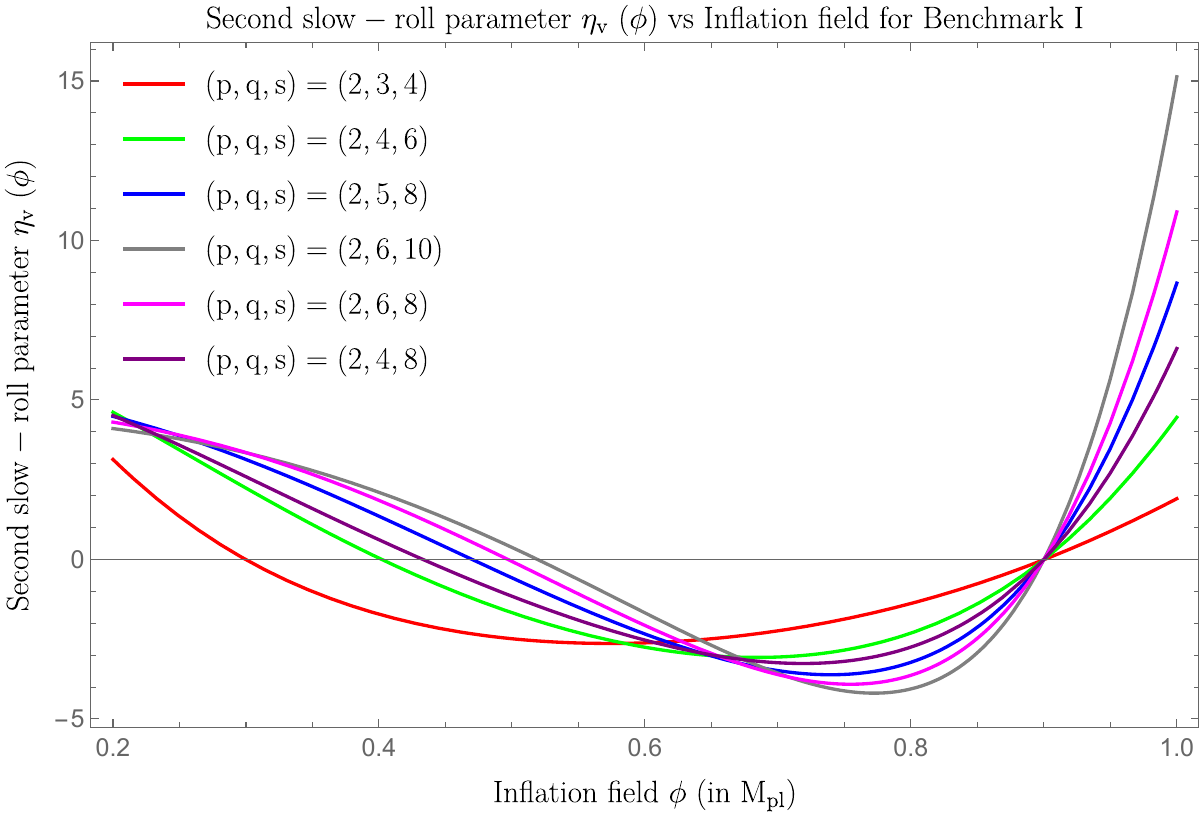}
        \label{fig3}
    }
    	\caption[Optional caption for list of figures]{The second slow-roll parameter $\eta_{V}$ for the same allowed cases corresponding to Benchmark I is plotted against the inflaton field $\phi$.} 
        \label{etaI}
    \end{figure*}
    \begin{enumerate}
        \item  Due to this we have the following values at the CMB scale for the inflationary observables related to the potential in the first row:
        \bea
        \Delta_{t}^{2} = 2.155 \times 10^{-28}, \quad \alpha_{s} =-0.17510, \quad \kappa_{s} =-0.00016, \quad \alpha_{t} = -2.736 \times 10^{-15},\quad \kappa_{t} = -2.750 \times 10^{-15}.\quad\quad
        \eea
        Also, the corresponding values of the slow-roll related parameters at the CMB scale for the same potential, in the first row, are written as follows:
        \bea
        \epsilon_{V} =0.000016196,\quad \eta_{V} =-0.0088356,\quad \xi^{2}_{V} =-0.093562,\quad \sigma^{3}_{V} =0.0017756,\quad \nu_{s} = 1.5088.\quad
        \eea
        \item The values at CMB scale for the observables from the potential described in the second row are as follows:
        \bea
        \Delta_{t}^{2} = 2.155 \times 10^{-28}, \quad \alpha_{s} =-0.074275,\quad \kappa_{s} =-0.0007104,\quad  \alpha_{t} = -3.079 \times 10^{-15},\quad \kappa_{t} = -1.220 \times 10^{-14}.\quad\quad
        \eea
        Similarly, the corresponding CMB values of the slow-roll related parameters for the same potential are written as follows:
        \bea
        \epsilon_{V} =0.000017167,\quad \eta_{V} =-0.0093707,\quad \xi^{2}_{V} =-0.19271,\quad \sigma^{3}_{V} =0.0075296,\quad \nu_{s} = 1.5094.\quad\quad
        \eea
        \item The values at CMB scale for the observables from the potential described in the third row are as follows:
        \bea
        \Delta_{t}^{2} = 2.155 \times 10^{-28},\quad \alpha_{s} =-0.46281,\quad \kappa_{s} =-0.0043105,\quad  \alpha_{t} = -3.064 \times 10^{-15},\quad \kappa_{t} = -3.302 \times 10^{-14}.\quad\quad
        \eea
        Similarly, the corresponding CMB values of the slow-roll related parameters for the same potential are written as follows:
        \bea
        \epsilon_{V} =0.000017106,\quad \eta_{V} =-0.0093368,\quad \xi^{2}_{V} =-0.48104,\quad \sigma^{3}_{V} =0.037521,\quad \nu_{s} = 1.5094.\quad\quad
        \eea
        \item The values at CMB scale for the observables from the potential described in the fourth row are as follows:
        \bea
        \Delta_{t}^{2} = 2.155 \times 10^{-28},\quad \alpha_{s} =-0.20396,\quad \kappa_{s} =-0.0019172,\quad  \alpha_{t} = -3.054 \times 10^{-15},\quad \kappa_{t} = -7.547 \times 10^{-14}.\quad\quad
        \eea
        Similarly, the corresponding CMB values of the slow-roll related parameters for the same potential are written as follows:
        \bea
        \epsilon_{V} =0.000016966,\quad \eta_{V} =-0.0092595,\quad \xi^{2}_{V} =-0.31934,\quad \sigma^{3}_{V} =0.018605,\quad \nu_{s} = 1.5093.\quad\quad
        \eea
        \item The values at CMB scale for the observables from the potential described in the fifth row are as follows:
        \bea
        \Delta_{t}^{2} = 2.155 \times 10^{-28},\quad \alpha_{s} =-0.23604,\quad \kappa_{s} =-0.0017750,\quad  \alpha_{t} = -1.936 \times 10^{-15},\quad \kappa_{t} = -3.069 \times 10^{-14}.\quad\quad
        \eea
        Similarly, the corresponding CMB values of the slow-roll related parameters for the same potential are written as follows:
        \bea
        \epsilon_{V} =0.000013641,\quad \eta_{V} =-0.0075235,\quad \xi^{2}_{V} =-0.34354,\quad \sigma^{3}_{V} =0.019939,\quad \nu_{s} = 1.5074.\quad\quad
        \eea
        \item The values at CMB scale for the observables from the potential described in the sixth row are as follows:
        \bea
        \Delta_{t}^{2} = 2.155 \times 10^{-28},\quad \alpha_{s} =-0.10576,\quad \kappa_{s} =-0.0008068,\quad  \alpha_{t} = -1.969 \times 10^{-15},\quad \kappa_{t} = -1.369 \times 10^{-14}.\quad\quad
        \eea
        Similarly, the corresponding CMB values of the slow-roll related parameters for the same potential are written as follows:
        \bea
        \epsilon_{V} =0.000013754,\quad \eta_{V} =-0.0074863,\quad \xi^{2}_{V} =-0.22996,\quad \sigma^{3}_{V} =0.010722,\quad \nu_{s} = 1.5075.\quad\quad
        \eea
    \end{enumerate}

    \item \underline{\bf{Benchmark II outcomes:}}\\ The values of the parameters obtained for the benchmark point represented by the first row in Table (\ref{tab:2}) corresponds to the following results. These values are evaluated at the scale of CMB which is kept the same as $\phi_{\rm *} = 0.9\;M_{\rm pl}$. Here we consider another value for tensor-to-scalar ratio \textcolor{black}{i.e., $r = 4.1 \times 10^{-14}$, that follows from the equation (\ref{tensortoscalar})}, and correspondingly fix ${\cal J} = 10^{-9}$.
     Due to this we have the values for the inflationary observables to be:
        \bea
        \Delta_{t}^{2} = 4.647 \times 10^{-23}, \quad \alpha_{s} =-0.33905,\quad \kappa_{s} =-0.01087,\quad  \alpha_{t} = -2.669 \times 10^{-17},\quad \kappa_{t} = -1.569 \times 10^{-16}.\quad\quad
        \eea
        Also, the corresponding values of the slow-roll related parameters at the CMB scale are as follows:
        \bea
        \epsilon_{V} =0.0003192,\quad \eta_{V} =-0.02847,\quad \xi_{V}^{2} =-0.4116,\quad \sigma_{V}^{3} =0.03467,\quad \nu_{s} = 1.5294.\quad\quad
        \eea

    \item \underline{\bf{Benchmark III outcomes:}}\\  The values of the parameters obtained for the benchmark point represented by the second row in Table (\ref{tab:2}) corresponds to the following  results. These values are evaluated at the scale of CMB, $\phi_{\rm *} = 0.9\;M_{\rm pl}$. Next we consider the value for tensor-to-scalar ratio i.e., \textcolor{black}{$r = 2.7 \times 10^{-11}$, that follows from equation (\ref{tensortoscalar}),} and correspondingly fix ${\cal J} = 10^{-6}$. Due to this we have the values for the inflationary observables to be:
        \bea
        \Delta_{t}^{2} = 2.698 \times 10^{-20},\quad \alpha_{s} =-0.20167,\quad \kappa_{s} =-0.001677,\quad  \alpha_{t} = -1.458 \times 10^{-20},\quad \kappa_{t} = -7.199 \times 10^{-20}.\quad\quad
        \eea
        Also, the corresponding values of the slow-roll related parameters at the CMB scale are as follows:
        \bea
        \epsilon_{V} =0.0001854,\quad \eta_{V} =-0.01976,\quad \xi_{V}^{2} =-0.3139,\quad \sigma_{V}^{3} =0.02014,\quad \nu_{s} = 1.5203.\quad\quad
        \eea

    \item \underline{\bf{Benchmark IV outcomes:}}\\ The values of the parameters obtained for the specific benchmark point represented by the third, fourth and fifth rows of Table (\ref{tab:2}) share the value of the ratio \textcolor{black}{$r = 6.9 \times 10^{-9}$}. The corresponding results from these parameters are evaluated at the scale of CMB, $\phi_{\rm *} = 0.9\;M_{\rm pl}$. \textcolor{black}{We move towards the upper bound and consider the value for tensor-to-scalar ratio to be $r = 6.9 \times 10^{-9}$, which follows from equation (\ref{tensortoscalar})} and correspondingly fix ${\cal J} = 10^{-3}$.
    \begin{figure*}[htb!]
    	\centering
   {
      	\includegraphics[width=18cm,height=9cm] {
      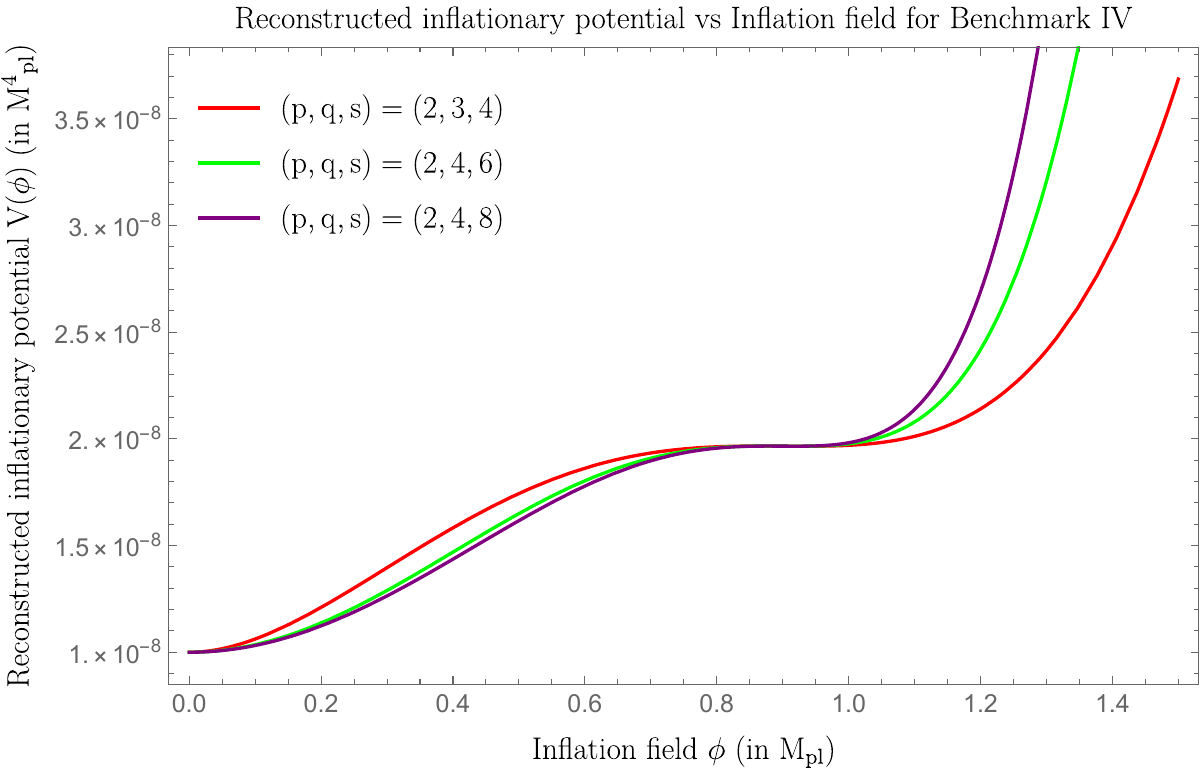}
        \label{fig4}
    }
    	\caption[Optional caption for list of figures]{The reconstructed potentials for the different allowed cases corresponding to Benchmark IV is plotted against the inflaton field $\phi$.} 
        \label{potentialIV}
    \end{figure*}
    \begin{figure*}[htb!]
    	\centering
   {
      	\includegraphics[width=18cm,height=9cm] {
      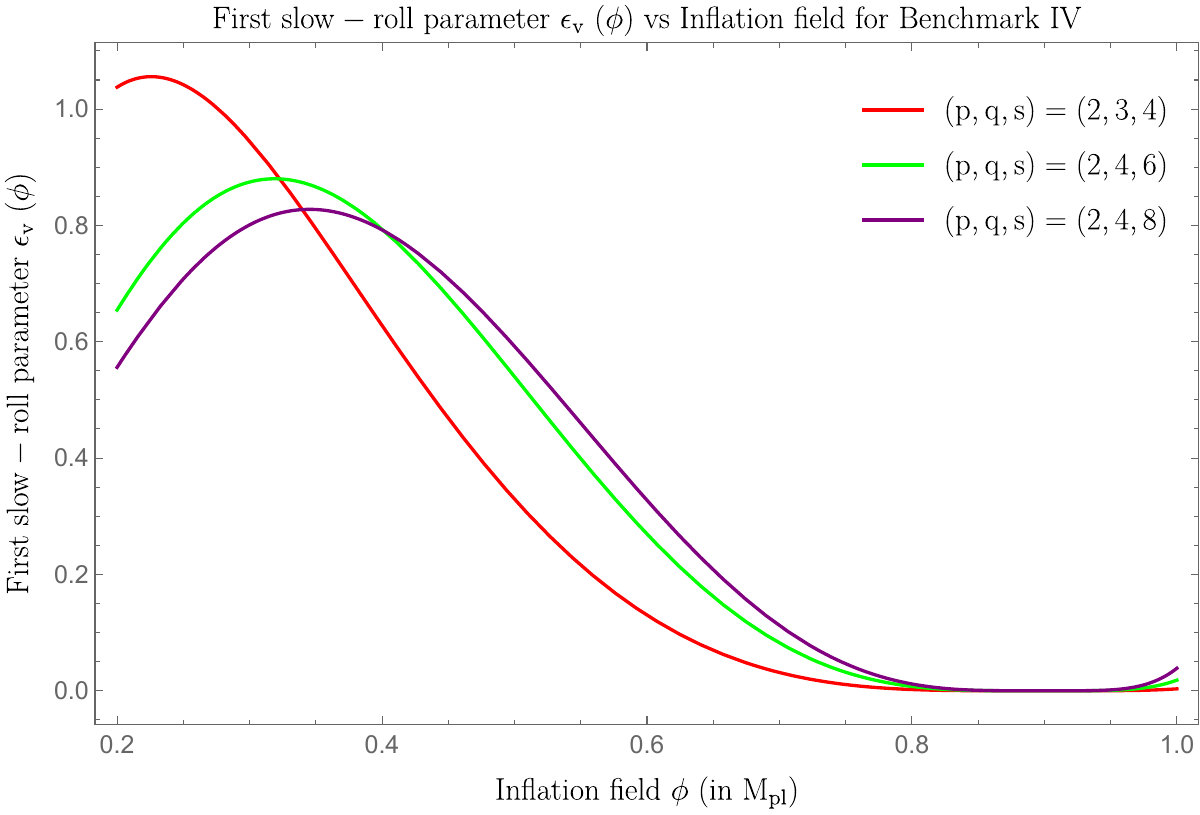}
        \label{fig5}
    }
    	\caption[Optional caption for list of figures]{The first slow-roll parameter $\eta_{V}$ for the same allowed cases corresponding to Benchmark IV is plotted against the inflaton field $\phi$.} 
        \label{epsilonIV}
    \end{figure*}
    \begin{figure*}[htb!]
    	\centering
   {
      	\includegraphics[width=18cm,height=9cm] {
      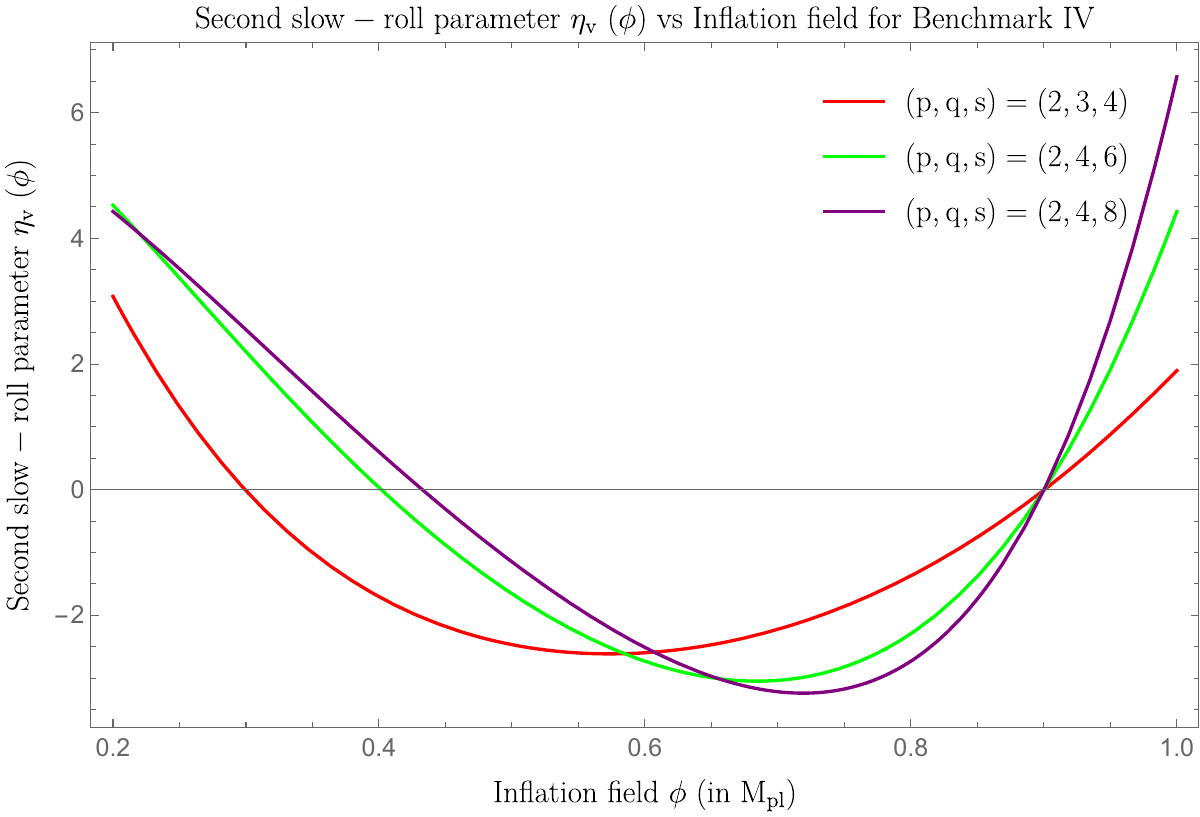}
        \label{fig6}
    }
    	\caption[Optional caption for list of figures]{The second slow-roll parameter $\eta_{V}$ for the same allowed cases corresponding to Benchmark IV is plotted against the inflaton field $\phi$.} 
        \label{etaIV}
    \end{figure*}
    \begin{enumerate}
        \item   Due to this we have the values for the inflationary observables corresponding to parameters in the third row as follows:
        \bea
        \Delta_{t}^{2} = 6.899 \times 10^{-18},\quad \alpha_{s} =-0.0507,\quad \kappa_{s} =-0.0004278,\quad  \alpha_{t} = -2.001 \times 10^{-24},\quad \kappa_{t} = -6.424 \times 10^{-24}.\quad\quad
        \eea
        Also, the corresponding values of the slow-roll related parameters at the CMB scale are as follows:
        \bea
        \epsilon_{V} =0.00004746,\quad \eta_{V} =-0.007574,\quad \xi_{V}^{2} =-0.1592,\quad \sigma_{V}^{3} =0.005168,\quad \nu_{s} = 1.5080.\quad\quad
        \eea
        \item For the same tensor-to-scalar ratio we have the values for the inflationary observables from the potential parameters in the fourth row as follows:
        \bea
        \Delta_{t}^{2} = 6.899 \times 10^{-18},\quad \alpha_{s} =-0.2017,\quad \kappa_{s} =-0.001677,\quad  \alpha_{t} = -1.991 \times 10^{-24},\quad \kappa_{t} = -2.534 \times 10^{-23}.\quad\quad
        \eea
        Also, the corresponding values of the slow-roll related parameters at the CMB scale are as follows:
        \bea
        \epsilon_{V} =0.00004724,\quad \eta_{V} =-0.001784,\quad \xi_{V}^{2} =-0.13178,\quad \sigma_{V}^{3} =0.001677,\quad \nu_{s} = 1.5079.\quad\quad
        \eea
        \item For the same tensor-to-scalar ratio we have the values for the inflationary observables from the potential parameters in the fifth row as follows:
        \bea
        \Delta_{t}^{2} = 6.899 \times 10^{-18},\quad \alpha_{s} =-0.3736,\quad \kappa_{s} =-0.003165,\quad  \alpha_{t} = -2.165 \times 10^{-24},\quad \kappa_{t} = -4.889 \times 10^{-23}.\quad\quad
        \eea
        Also, the corresponding values of the slow-roll related parameters at the CMB scale are as follows:
        \bea
        \epsilon_{V} =0.00004922,\quad \eta_{V} =-0.008178,\quad \xi_{V}^{2} =-0.4322,\quad \sigma_{V}^{3} =0.03811,\quad \nu_{s} = 1.5083.\quad\quad
        \eea
    \end{enumerate}

    \item \underline{\bf{Benchmark V outcomes:}}\\  The values of the parameters obtained for the benchmark point represented by the sixth row in Table (\ref{tab:2}) corresponds to the following results. These values are evaluated at the scale of CMB, $\phi_{\rm *} = 0.9\;M_{\rm pl}$. At last we consider the  \textcolor{black}{value for tensor-to-scalar ratio close to the upper bound i.e., $r = 7.4 \times 10^{-5}$, which follows from equation (\ref{tensortoscalar}),} and correspondingly fix ${\cal J} = 1$. Due to this we have the values for the inflationary observables to be:
        \bea
        \Delta_{t}^{2} = 7.386 \times 10^{-14},\quad \alpha_{s} =-0.542389,\quad \kappa_{s} =-0.02423,\quad  \alpha_{t} = -4.945 \times 10^{-26},\quad \kappa_{t} = -3.377 \times 10^{-25}.\quad\quad
        \eea
        Also, the corresponding values of the slow-roll related parameters at the CMB scale are as follows:
        \bea
      \epsilon_{V} =0.0005114,\quad
        \eta_{V} =-0.3915,\quad \xi_{V}^{2} =-0.5206,\quad \sigma_{V}^{3} =0.5528,\quad \nu_{s} = 1.5407.\quad\quad
        \eea
\end{itemize}
The reconstructed potential for different values of the parameters and the slow-roll variables derived from these potentials are examined through their behaviour upon plotting them against the inflation field value $\phi$ in Planck mass units. These parameters are listed in the Table \ref{tab:1} corresponding to the value of Benchmark I and in Table \ref{tab:2} for the other values of Benchmark II, III, IV, V. In fig.(\ref{potentialI}), we visualize the possible potentials for the same Benchmark I value, which are differentiated using the set of exponents of the field values. We see that for a very low value of the tensor-to-scalar ratio, there exist a large number of allowed potentials based on constraints on the slow-roll parameters and along with the need for maintaining the necessary number of e-foldings of expansion. For the following exponents: $(2,3,4), (2,4,6), (2,6,10)$ the respective potentials tends to become steeper as large field values are approached and when the field moves down to lower values, the respective potentials also falls down faster and similarly the end of inflation is reached earlier, which is also clear from the $\phi_{\rm end}$ values in the Table \ref{tab:1}. For the other $3$ values of the exponents, the potentials come closer to each other at large and small field values. The inflection point is produced roughly at the same field value for all the possible potentials. In figs.(\ref{epsilonI}, \ref{etaI}), we plot the behaviour of the first and second slow-roll parameters against the field values $\phi$. Here also a significant difference in the values of the first and second slow-roll parameter is observed for the same set of exponents, i.e., $(2,3,4), (2,4,6), (2,6,10)$, as their relative values are well-separated throughout the allowed field range values. For the rest of the exponents, their values tend to come closer to each other again within the range of field values. We must emphasize that our primary focus has been to concentrate on the sub-Planckian  behaviour for all the quantities and have only included the super-Planckian values to fully visualize the behaviour of the potentials.

For the Benchmarks II, III, V, we have not shown any behaviour of the potential and the first and second slow-roll parameters through plots as their results almost coincide for all these quantities throughout which would not be useful from an analysis perspective and hence only the specific values of the slow-roll parameters and the inflationary observables are listed explicitly in their respective outcomes as discussed above. 

In fig.(\ref{potentialIV}), we visualize the possible potentials for the Benchmark IV value which are differentiated using the set of exponents of the field. Here we get to consider only three possible cases. This possibility is again dependent on the constraint analysis performed for the values of the slow-roll parameters within the sub-Planckian region of field values. The behaviour of the potentials, as well as the slow-roll parameters plotted in figs.(\ref{epsilonIV}, \ref{etaIV}), is almost unchanged for all the examined exponents when compared to the similar cases in fig.(\ref{potentialI}) for Benchmark I. This is also reflected in the corresponding parameter values mentioned in the Tables (\ref{tab:1},\ref{tab:2}). However, there are noticeable changes in the values of the inflationary observables and the all the slow-roll parameters for Benchmark IV as is evident from their outcomes listed above.

\section{Conclusion}
\label{sec8}
 We finally conclude our discussion with the following key highlights from our analysis.
 We started this work with a discussion on the necessary theoretical inputs required for a observation based robust description of the inflationary potential by providing all the inflationary observables related to the scalar and tensor perturbations, which are computed in the presence of NBD initial condition. 
 Further, we have provided the details of the new consistency relation which can accommodate the blue-tilted tensor power spectrum feature in the presence of NBD initial condition. This information is one of the prime components of the reconstruction of the inflationary potential. This new consistency relation is manufactured in such a fashion that it changes the behaviour in three consecutive phases in the frequency (or wave number) domain. We have found that at very low frequencies ($f\leq 10^{-17}{\rm Hz}$) the behaviour matches with the known facts obtained from BD initial condition. In the intermediate frequencies ($10^{-17}{\rm Hz}\leq f\leq 10^{-7}{\rm Hz}$) and comparatively high frequencies ($10^{-7}{\rm Hz}\leq f\leq 1{\rm Hz}$), the consistency relation is strongly violated.  Next, we have discussed in detail the derivation of the new field excursion formula in presence of the NBD initial condition. Here we have shown a violation of the good old {\it Lyth bound} which was derived in the presence of BD initial condition. Also, we have shown that to validate the EFT prescription within the framework of the inflationary paradigm during the reconstruction, this newly derived bound plays an extremely significant role. Throughout performing the analysis we have maintained the necessary and sufficient conditions, which helps us to strictly validate EFT in the final form of the potential.
 Finally, we have discussed the theoretical framework for the inflationary potential reconstruction in the presence of NBD initial condition. With this designed method we have fixed the unknown coefficients, powers of the field, etc. in terms of the inflationary parameters which we directly probe in cosmological observation. We found after performing the analysis in a fully model-independent fashion, that in the end the generic feature of the reconstructed potential is suggesting towards having an inflection
 point potential which exactly matches with the inflection point models that can be obtained from N = 1 SUGRA MSSM inflationary framework. 
 This analysis is also highly favorable from the perspective of PBH formation. Since in most of cases, the effective potential during inflation is manufactured phenomenologically, but in this work, without the use of any additional features and with the help of present reconstructed potential, we strongly believe that one can mimic the SRI, USR and SRII phases necessarily required to produce large amplitude scalar fluctuations during the formation of PBHs \cite{Kristiano:2022maq,Kristiano:2023scm,Choudhury:2023vuj,Choudhury:2023jlt,Choudhury:2023rks,Choudhury:2023hvf,Choudhury:2023kdb,Riotto:2023hoz,Riotto:2023gpm,Firouzjahi:2023aum,Motohashi:2023syh,Firouzjahi:2023ahg,Franciolini:2023lgy,Tasinato:2023ukp,Cheng:2023ikq,Choudhury:2023hfm,Bhattacharya:2023ysp,Choudhury:2023fwk,Choudhury:2023fjs,Choudhury:2024one,Choudhury:2024ybk,Choudhury:2024jlz,Choudhury:2024dei,Choudhury:2024dzw,Choudhury:2024aji}. Though in refs. \cite{Choudhury:2023jlt,Choudhury:2023rks,Choudhury:2023hvf,Kristiano:2022maq,Kristiano:2023scm} we and some other authors have pointed out that one-loop effects in the scalar power spectrum rule out the formation of large mass PBHs at the scale, $k\sim {\cal O}(10^6-10^7) {\rm Mpc}^{-1}$ where the NANOGrav 15 signal appears. However, with the quantum loop effects, one can generate small mass PBHs, $M_{\rm PBH}\sim 10^2 {\rm gms}$, using which one can generate large amplitude gravitational waves in the frequency domain $f\sim {\cal O}(10^6-10^7){\rm Hz}$. 
 We also know that to accommodate PBH formation in the USR regime the parameter $\eta_V$ has to be large and approximately of the order of $\eta_V \sim -6$. From our analysis, $\eta_V$ reaches that limit, which means generically the PBH formation can be explained with the help of the reconstructed inflationary potential.

	\subsection*{Acknowledgements}
SC sincerely thanks Mohit Kumar Sharma for various useful discussions which helps to improve the analysis performed in this paper. SC would like to thank The North American Nanohertz Observatory for Gravitational Waves (NANOGrav) collaboration and the National Academy of Sciences (NASI), Prayagraj, India, for being elected as an associate member and the member of the academy respectively. SC would also like to thank all the members of Quantum Aspects of the Space-Time \& Matter
(QASTM) for elaborative discussions. Last but not least, we acknowledge our debt to the people
belonging to the various parts of the world for their generous and steady support for research in natural sciences. 


\bibliographystyle{utphys}

\begin{thebibliography}{100}

\bibitem{Kazanas:1980tx}
D.~Kazanas, ``{Dynamics of the Universe and Spontaneous Symmetry Breaking},''
  \href{http://dx.doi.org/10.1086/183361}{{\em Astrophys. J. Lett.} {\bfseries
  241} (1980) L59--L63}.

\bibitem{Starobinsky:1980te}
A.~A. Starobinsky, ``{A New Type of Isotropic Cosmological Models Without
  Singularity},'' \href{http://dx.doi.org/10.1016/0370-2693(80)90670-X}{{\em
  Phys. Lett. B} {\bfseries 91} (1980) 99--102}.

\bibitem{Sato:1981ds}
K.~Sato, ``{Cosmological Baryon Number Domain Structure and the First Order
  Phase Transition of a Vacuum},''
  \href{http://dx.doi.org/10.1016/0370-2693(81)90805-4}{{\em Phys. Lett. B}
  {\bfseries 99} (1981) 66--70}.

\bibitem{Guth:1980zm}
A.~H. Guth, ``{The Inflationary Universe: A Possible Solution to the Horizon
  and Flatness Problems},''
  \href{http://dx.doi.org/10.1103/PhysRevD.23.347}{{\em Phys. Rev. D}
  {\bfseries 23} (1981) 347--356}.

\bibitem{Mukhanov:1981xt}
V.~F. Mukhanov and G.~V. Chibisov, ``{Quantum Fluctuations and a Nonsingular
  Universe},'' {\em JETP Lett.} {\bfseries 33} (1981) 532--535.

\bibitem{Linde:1981mu}
A.~D. Linde, ``{A New Inflationary Universe Scenario: A Possible Solution of
  the Horizon, Flatness, Homogeneity, Isotropy and Primordial Monopole
  Problems},'' \href{http://dx.doi.org/10.1016/0370-2693(82)91219-9}{{\em Phys.
  Lett. B} {\bfseries 108} (1982) 389--393}.

\bibitem{Albrecht:1982wi}
A.~Albrecht and P.~J. Steinhardt, ``{Cosmology for Grand Unified Theories with
  Radiatively Induced Symmetry Breaking},''
  \href{http://dx.doi.org/10.1103/PhysRevLett.48.1220}{{\em Phys. Rev. Lett.}
  {\bfseries 48} (1982) 1220--1223}.

\bibitem{Baumann:2009ds}
D.~Baumann,
  \href{http://dx.doi.org/10.1142/9789814327183_0010}{``{Inflation},''} in {\em
  {Theoretical Advanced Study Institute in Elementary Particle Physics}:
  {Physics of the Large and the Small}}, pp.~523--686.
\newblock 2011.
\newblock \href{http://arxiv.org/abs/0907.5424}{{\ttfamily arXiv:0907.5424
  [hep-th]}}.

\bibitem{Baumann:2018muz}
D.~Baumann, ``{Primordial Cosmology},''
  \href{http://dx.doi.org/10.22323/1.305.0009}{{\em PoS} {\bfseries TASI2017}
  (2018) 009}, \href{http://arxiv.org/abs/1807.03098}{{\ttfamily
  arXiv:1807.03098 [hep-th]}}.

\bibitem{Senatore:2013roa}
L.~Senatore, \href{http://dx.doi.org/10.1142/9789814525220_0006}{``{TASI 2012
  Lectures on Inflation},''} in {\em {Theoretical Advanced Study Institute in
  Elementary Particle Physics}: {Searching for New Physics at Small and Large
  Scales}}, pp.~221--302.
\newblock 2013.

\bibitem{Choudhury:2011sq}
S.~Choudhury and S.~Pal, ``{Brane inflation in background supergravity},''
  \href{http://dx.doi.org/10.1103/PhysRevD.85.043529}{{\em Phys. Rev. D}
  {\bfseries 85} (2012) 043529},
  \href{http://arxiv.org/abs/1102.4206}{{\ttfamily arXiv:1102.4206 [hep-th]}}.

\bibitem{Choudhury:2011jt}
S.~Choudhury and S.~Pal, ``{Fourth level MSSM inflation from new flat
  directions},'' \href{http://dx.doi.org/10.1088/1475-7516/2012/04/018}{{\em
  JCAP} {\bfseries 04} (2012) 018},
  \href{http://arxiv.org/abs/1111.3441}{{\ttfamily arXiv:1111.3441 [hep-ph]}}.

\bibitem{Choudhury:2012yh}
S.~Choudhury and S.~Pal, ``{DBI Galileon inflation in background SUGRA},''
  \href{http://dx.doi.org/10.1016/j.nuclphysb.2013.05.010}{{\em Nucl. Phys. B}
  {\bfseries 874} (2013) 85--114},
  \href{http://arxiv.org/abs/1208.4433}{{\ttfamily arXiv:1208.4433 [hep-th]}}.

\bibitem{Choudhury:2012whm}
S.~Choudhury and S.~Pal, ``{Primordial non-Gaussian features from DBI Galileon
  inflation},'' \href{http://dx.doi.org/10.1140/epjc/s10052-015-3452-3}{{\em
  Eur. Phys. J. C} {\bfseries 75} no.~6, (2015) 241},
  \href{http://arxiv.org/abs/1210.4478}{{\ttfamily arXiv:1210.4478 [hep-th]}}.

\bibitem{Choudhury:2013zna}
S.~Choudhury, T.~Chakraborty, and S.~Pal, ``{Higgs inflation from new K\"ahler
  potential},'' \href{http://dx.doi.org/10.1016/j.nuclphysb.2014.01.002}{{\em
  Nucl. Phys. B} {\bfseries 880} (2014) 155--174},
  \href{http://arxiv.org/abs/1305.0981}{{\ttfamily arXiv:1305.0981 [hep-th]}}.

\bibitem{Choudhury:2013jya}
S.~Choudhury, A.~Mazumdar, and S.~Pal, ``{Low \& High scale MSSM inflation,
  gravitational waves and constraints from Planck},''
  \href{http://dx.doi.org/10.1088/1475-7516/2013/07/041}{{\em JCAP} {\bfseries
  07} (2013) 041}, \href{http://arxiv.org/abs/1305.6398}{{\ttfamily
  arXiv:1305.6398 [hep-ph]}}.

\bibitem{Choudhury:2013iaa}
S.~Choudhury and A.~Mazumdar, ``{An accurate bound on tensor-to-scalar ratio
  and the scale of inflation},''
  \href{http://dx.doi.org/10.1016/j.nuclphysb.2014.03.005}{{\em Nucl. Phys. B}
  {\bfseries 882} (2014) 386--396},
  \href{http://arxiv.org/abs/1306.4496}{{\ttfamily arXiv:1306.4496 [hep-ph]}}.

\bibitem{Choudhury:2013woa}
S.~Choudhury and A.~Mazumdar, ``{Primordial blackholes and gravitational waves
  for an inflection-point model of inflation},''
  \href{http://dx.doi.org/10.1016/j.physletb.2014.04.050}{{\em Phys. Lett. B}
  {\bfseries 733} (2014) 270--275},
  \href{http://arxiv.org/abs/1307.5119}{{\ttfamily arXiv:1307.5119
  [astro-ph.CO]}}.

\bibitem{Choudhury:2014sxa}
S.~Choudhury, A.~Mazumdar, and E.~Pukartas, ``{Constraining ${\cal N}=1$
  supergravity inflationary framework with non-minimal K\"ahler operators},''
  \href{http://dx.doi.org/10.1007/JHEP04(2014)077}{{\em JHEP} {\bfseries 04}
  (2014) 077}, \href{http://arxiv.org/abs/1402.1227}{{\ttfamily arXiv:1402.1227
  [hep-th]}}.

\bibitem{Choudhury:2014uxa}
S.~Choudhury, ``{Constraining N = 1 supergravity inflation with non-minimal
  Kaehler operators using $\delta$N formalism},''
  \href{http://dx.doi.org/10.1007/JHEP04(2014)105}{{\em JHEP} {\bfseries 04}
  (2014) 105}, \href{http://arxiv.org/abs/1402.1251}{{\ttfamily arXiv:1402.1251
  [hep-th]}}.

\bibitem{Choudhury:2014hua}
S.~Choudhury, ``{Inflamagnetogenesis redux: Unzipping sub-Planckian inflation
  via various cosmoparticle probes},''
  \href{http://dx.doi.org/10.1016/j.physletb.2014.06.029}{{\em Phys. Lett. B}
  {\bfseries 735} (2014) 138--145},
  \href{http://arxiv.org/abs/1403.0676}{{\ttfamily arXiv:1403.0676 [hep-th]}}.

\bibitem{Choudhury:2014kma}
S.~Choudhury and A.~Mazumdar, ``{Reconstructing inflationary potential from
  BICEP2 and running of tensor modes},''
  \href{http://arxiv.org/abs/1403.5549}{{\ttfamily arXiv:1403.5549 [hep-th]}}.

\bibitem{Choudhury:2014sua}
S.~Choudhury, ``{Can Effective Field Theory of inflation generate large
  tensor-to-scalar ratio within Randall\textendash{}Sundrum single
  braneworld?},'' \href{http://dx.doi.org/10.1016/j.nuclphysb.2015.02.024}{{\em
  Nucl. Phys. B} {\bfseries 894} (2015) 29--55},
  \href{http://arxiv.org/abs/1406.7618}{{\ttfamily arXiv:1406.7618 [hep-th]}}.

\bibitem{Choudhury:2014hja}
S.~Choudhury, B.~K. Pal, B.~Basu, and P.~Bandyopadhyay, ``{Quantum Gravity
  Effect in Torsion Driven Inflation and CP violation},''
  \href{http://dx.doi.org/10.1007/JHEP10(2015)194}{{\em JHEP} {\bfseries 10}
  (2015) 194}, \href{http://arxiv.org/abs/1409.6036}{{\ttfamily arXiv:1409.6036
  [hep-th]}}.

\bibitem{Choudhury:2015pqa}
S.~Choudhury, ``{Reconstructing inflationary paradigm within Effective Field
  Theory framework},'' \href{http://dx.doi.org/10.1016/j.dark.2015.11.003}{{\em
  Phys. Dark Univ.} {\bfseries 11} (2016) 16--48},
  \href{http://arxiv.org/abs/1508.00269}{{\ttfamily arXiv:1508.00269
  [astro-ph.CO]}}.

\bibitem{Choudhury:2015hvr}
S.~Choudhury and S.~Panda, ``{COSMOS-e\textquoteright{}-GTachyon from string
  theory},'' \href{http://dx.doi.org/10.1140/epjc/s10052-016-4072-2}{{\em Eur.
  Phys. J. C} {\bfseries 76} no.~5, (2016) 278},
  \href{http://arxiv.org/abs/1511.05734}{{\ttfamily arXiv:1511.05734
  [hep-th]}}.

\bibitem{Choudhury:2016wlj}
S.~Choudhury, {\em {Field Theoretic Approaches To Early Universe}}.
\newblock PhD thesis, Indian Statistical Inst., Calcutta, 2016.
\newblock \href{http://arxiv.org/abs/1603.08306}{{\ttfamily arXiv:1603.08306
  [hep-th]}}.

\bibitem{Choudhury:2016cso}
S.~Choudhury, S.~Panda, and R.~Singh, ``{Bell violation in the Sky},''
  \href{http://dx.doi.org/10.1140/epjc/s10052-016-4553-3}{{\em Eur. Phys. J. C}
  {\bfseries 77} no.~2, (2017) 60},
  \href{http://arxiv.org/abs/1607.00237}{{\ttfamily arXiv:1607.00237
  [hep-th]}}.

\bibitem{Choudhury:2017cos}
S.~Choudhury, ``{COSMOS-$e'$- soft Higgsotic attractors},''
  \href{http://dx.doi.org/10.1140/epjc/s10052-017-5001-8}{{\em Eur. Phys. J. C}
  {\bfseries 77} no.~7, (2017) 469},
  \href{http://arxiv.org/abs/1703.01750}{{\ttfamily arXiv:1703.01750
  [hep-th]}}.

\bibitem{Naskar:2017ekm}
A.~Naskar, S.~Choudhury, A.~Banerjee, and S.~Pal, ``{EFT of Inflation:
  Reflections on CMB and Forecasts on LSS Surveys},''
  \href{http://arxiv.org/abs/1706.08051}{{\ttfamily arXiv:1706.08051
  [astro-ph.CO]}}.

\bibitem{Choudhury:2017glj}
S.~Choudhury, ``{CMB from EFT},''
  \href{http://dx.doi.org/10.3390/universe5060155}{{\em Universe} {\bfseries 5}
  no.~6, (2019) 155}, \href{http://arxiv.org/abs/1712.04766}{{\ttfamily
  arXiv:1712.04766 [hep-th]}}.

\bibitem{Choudhury:2018glz}
S.~Choudhury, {\em {Quantum Field Theory approaches to Early Universe
  Cosmology}}.
\newblock LAP LAMBERT Academic Publishing, 5, 2018.

\bibitem{HosseiniMansoori:2023zop}
S.~A. Hosseini~Mansoori, F.~Felegary, M.~Roshan, O.~Akarsu, and M.~Sami,
  ``{$\mathbb{T}^{2}$- inflation: Sourced by energy-momentum squared
  gravity},'' \href{http://arxiv.org/abs/2306.09181}{{\ttfamily
  arXiv:2306.09181 [gr-qc]}}.

\bibitem{Geng:2015fla}
C.-Q. Geng, M.~W. Hossain, R.~Myrzakulov, M.~Sami, and E.~N. Saridakis,
  ``{Quintessential inflation with canonical and noncanonical scalar fields and
  Planck 2015 results},''
  \href{http://dx.doi.org/10.1103/PhysRevD.92.023522}{{\em Phys. Rev. D}
  {\bfseries 92} no.~2, (2015) 023522},
  \href{http://arxiv.org/abs/1502.03597}{{\ttfamily arXiv:1502.03597 [gr-qc]}}.

\bibitem{WaliHossain:2014usl}
M.~Wali~Hossain, R.~Myrzakulov, M.~Sami, and E.~N. Saridakis, ``{Unification of
  inflation and dark energy \`a la quintessential inflation},''
  \href{http://dx.doi.org/10.1142/S0218271815300141}{{\em Int. J. Mod. Phys. D}
  {\bfseries 24} no.~05, (2015) 1530014},
  \href{http://arxiv.org/abs/1410.6100}{{\ttfamily arXiv:1410.6100 [gr-qc]}}.

\bibitem{Hossain:2014coa}
M.~W. Hossain, R.~Myrzakulov, M.~Sami, and E.~N. Saridakis, ``{Class of
  quintessential inflation models with parameter space consistent with
  BICEP2},'' \href{http://dx.doi.org/10.1103/PhysRevD.89.123513}{{\em Phys.
  Rev. D} {\bfseries 89} no.~12, (2014) 123513},
  \href{http://arxiv.org/abs/1404.1445}{{\ttfamily arXiv:1404.1445 [gr-qc]}}.

\bibitem{Hossain:2014xha}
M.~W. Hossain, R.~Myrzakulov, M.~Sami, and E.~N. Saridakis, ``{Variable
  gravity: A suitable framework for quintessential inflation},''
  \href{http://dx.doi.org/10.1103/PhysRevD.90.023512}{{\em Phys. Rev. D}
  {\bfseries 90} no.~2, (2014) 023512},
  \href{http://arxiv.org/abs/1402.6661}{{\ttfamily arXiv:1402.6661 [gr-qc]}}.

\bibitem{Mazumdar:2010sa}
A.~Mazumdar and J.~Rocher, ``{Particle physics models of inflation and curvaton
  scenarios},'' \href{http://dx.doi.org/10.1016/j.physrep.2010.08.001}{{\em
  Phys. Rept.} {\bfseries 497} (2011) 85--215},
  \href{http://arxiv.org/abs/1001.0993}{{\ttfamily arXiv:1001.0993 [hep-ph]}}.

\bibitem{Pospelov:2010hj}
M.~Pospelov and J.~Pradler, ``{Big Bang Nucleosynthesis as a Probe of New
  Physics},'' \href{http://dx.doi.org/10.1146/annurev.nucl.012809.104521}{{\em
  Ann. Rev. Nucl. Part. Sci.} {\bfseries 60} (2010) 539--568},
  \href{http://arxiv.org/abs/1011.1054}{{\ttfamily arXiv:1011.1054 [hep-ph]}}.

\bibitem{Planck:2018vyg}
{\bfseries Planck} Collaboration, N.~Aghanim {\em et~al.}, ``{Planck 2018
  results. VI. Cosmological parameters},''
  \href{http://dx.doi.org/10.1051/0004-6361/201833910}{{\em Astron. Astrophys.}
  {\bfseries 641} (2020) A6}, \href{http://arxiv.org/abs/1807.06209}{{\ttfamily
  arXiv:1807.06209 [astro-ph.CO]}}. [Erratum: Astron.Astrophys. 652, C4
  (2021)].

\bibitem{Martin:2013tda}
J.~Martin, C.~Ringeval, and V.~Vennin, ``{Encyclop\ae{}dia Inflationaris},''
  \href{http://dx.doi.org/10.1016/j.dark.2014.01.003}{{\em Phys. Dark Univ.}
  {\bfseries 5-6} (2014) 75--235},
  \href{http://arxiv.org/abs/1303.3787}{{\ttfamily arXiv:1303.3787
  [astro-ph.CO]}}.

\bibitem{Benetti:2013cja}
M.~Benetti, ``{Updating constraints on inflationary features in the primordial
  power spectrum with the Planck data},''
  \href{http://dx.doi.org/10.1103/PhysRevD.88.087302}{{\em Phys. Rev. D}
  {\bfseries 88} (2013) 087302},
  \href{http://arxiv.org/abs/1308.6406}{{\ttfamily arXiv:1308.6406
  [astro-ph.CO]}}.

\bibitem{Martin:2013nzq}
J.~Martin, C.~Ringeval, R.~Trotta, and V.~Vennin, ``{The Best Inflationary
  Models After Planck},''
  \href{http://dx.doi.org/10.1088/1475-7516/2014/03/039}{{\em JCAP} {\bfseries
  03} (2014) 039}, \href{http://arxiv.org/abs/1312.3529}{{\ttfamily
  arXiv:1312.3529 [astro-ph.CO]}}.

\bibitem{Creminelli:2014oaa}
P.~Creminelli, D.~L\'opez~Nacir, M.~Simonovi\'c, G.~Trevisan, and
  M.~Zaldarriaga, ``{$\phi^2$ or Not $\phi^2$: Testing the Simplest
  Inflationary Potential},''
  \href{http://dx.doi.org/10.1103/PhysRevLett.112.241303}{{\em Phys. Rev.
  Lett.} {\bfseries 112} no.~24, (2014) 241303},
  \href{http://arxiv.org/abs/1404.1065}{{\ttfamily arXiv:1404.1065
  [astro-ph.CO]}}.

\bibitem{Dai:2014jja}
L.~Dai, M.~Kamionkowski, and J.~Wang, ``{Reheating constraints to inflationary
  models},'' \href{http://dx.doi.org/10.1103/PhysRevLett.113.041302}{{\em Phys.
  Rev. Lett.} {\bfseries 113} (2014) 041302},
  \href{http://arxiv.org/abs/1404.6704}{{\ttfamily arXiv:1404.6704
  [astro-ph.CO]}}.

\bibitem{Benetti:2016tvm}
M.~Benetti and J.~S. Alcaniz, ``{Bayesian analysis of inflationary features in
  Planck and SDSS data},''
  \href{http://dx.doi.org/10.1103/PhysRevD.94.023526}{{\em Phys. Rev. D}
  {\bfseries 94} no.~2, (2016) 023526},
  \href{http://arxiv.org/abs/1604.08156}{{\ttfamily arXiv:1604.08156
  [astro-ph.CO]}}.

\bibitem{Campista:2017ovq}
M.~Campista, M.~Benetti, and J.~Alcaniz, ``{Testing non-minimally coupled
  inflation with CMB data: a Bayesian analysis},''
  \href{http://dx.doi.org/10.1088/1475-7516/2017/09/010}{{\em JCAP} {\bfseries
  09} (2017) 010}, \href{http://arxiv.org/abs/1705.08877}{{\ttfamily
  arXiv:1705.08877 [astro-ph.CO]}}.

\bibitem{Keeley:2020rmo}
R.~E. Keeley, A.~Shafieloo, D.~K. Hazra, and T.~Souradeep, ``{Inflation Wars: A
  New Hope},'' \href{http://dx.doi.org/10.1088/1475-7516/2020/09/055}{{\em
  JCAP} {\bfseries 09} (2020) 055},
  \href{http://arxiv.org/abs/2006.12710}{{\ttfamily arXiv:2006.12710
  [astro-ph.CO]}}.

\bibitem{Vagnozzi:2020rcz}
S.~Vagnozzi, E.~Di~Valentino, S.~Gariazzo, A.~Melchiorri, O.~Mena, and J.~Silk,
  ``{The galaxy power spectrum take on spatial curvature and cosmic
  concordance},'' \href{http://dx.doi.org/10.1016/j.dark.2021.100851}{{\em
  Phys. Dark Univ.} {\bfseries 33} (2021) 100851},
  \href{http://arxiv.org/abs/2010.02230}{{\ttfamily arXiv:2010.02230
  [astro-ph.CO]}}.

\bibitem{Vagnozzi:2020dfn}
S.~Vagnozzi, A.~Loeb, and M.~Moresco, ``{Eppur \`e piatto? The Cosmic
  Chronometers Take on Spatial Curvature and Cosmic Concordance},''
  \href{http://dx.doi.org/10.3847/1538-4357/abd4df}{{\em Astrophys. J.}
  {\bfseries 908} no.~1, (2021) 84},
  \href{http://arxiv.org/abs/2011.11645}{{\ttfamily arXiv:2011.11645
  [astro-ph.CO]}}.

\bibitem{Vagnozzi:2023lwo}
S.~Vagnozzi, ``{Inflationary interpretation of the stochastic gravitational
  wave background signal detected by pulsar timing array experiments},''
  \href{http://arxiv.org/abs/2306.16912}{{\ttfamily arXiv:2306.16912
  [astro-ph.CO]}}.

\bibitem{Cabass:2022wjy}
G.~Cabass, M.~M. Ivanov, O.~H.~E. Philcox, M.~Simonovi\'c, and M.~Zaldarriaga,
  ``{Constraints on Single-Field Inflation from the BOSS Galaxy Survey},''
  \href{http://dx.doi.org/10.1103/PhysRevLett.129.021301}{{\em Phys. Rev.
  Lett.} {\bfseries 129} no.~2, (2022) 021301},
  \href{http://arxiv.org/abs/2201.07238}{{\ttfamily arXiv:2201.07238
  [astro-ph.CO]}}.

\bibitem{Cabass:2022ymb}
G.~Cabass, M.~M. Ivanov, O.~H.~E. Philcox, M.~Simonovi\'c, and M.~Zaldarriaga,
  ``{Constraints on multifield inflation from the BOSS galaxy survey},''
  \href{http://dx.doi.org/10.1103/PhysRevD.106.043506}{{\em Phys. Rev. D}
  {\bfseries 106} no.~4, (2022) 043506},
  \href{http://arxiv.org/abs/2204.01781}{{\ttfamily arXiv:2204.01781
  [astro-ph.CO]}}.

\bibitem{CMB-S4:2016ple}
{\bfseries CMB-S4} Collaboration, K.~N. Abazajian {\em et~al.}, ``{CMB-S4
  Science Book, First Edition},''
  \href{http://arxiv.org/abs/1610.02743}{{\ttfamily arXiv:1610.02743
  [astro-ph.CO]}}.

\bibitem{SimonsObservatory:2018koc}
{\bfseries Simons Observatory} Collaboration, P.~Ade {\em et~al.}, ``{The
  Simons Observatory: Science goals and forecasts},''
  \href{http://dx.doi.org/10.1088/1475-7516/2019/02/056}{{\em JCAP} {\bfseries
  02} (2019) 056}, \href{http://arxiv.org/abs/1808.07445}{{\ttfamily
  arXiv:1808.07445 [astro-ph.CO]}}.

\bibitem{SimonsObservatory:2019qwx}
{\bfseries Simons Observatory} Collaboration, M.~H. Abitbol {\em et~al.},
  ``{The Simons Observatory: Astro2020 Decadal Project Whitepaper},'' {\em
  Bull. Am. Astron. Soc.} {\bfseries 51} (2019) 147,
  \href{http://arxiv.org/abs/1907.08284}{{\ttfamily arXiv:1907.08284
  [astro-ph.IM]}}.

\bibitem{Kamionkowski:2015yta}
M.~Kamionkowski and E.~D. Kovetz, ``{The Quest for B Modes from Inflationary
  Gravitational Waves},''
  \href{http://dx.doi.org/10.1146/annurev-astro-081915-023433}{{\em Ann. Rev.
  Astron. Astrophys.} {\bfseries 54} (2016) 227--269},
  \href{http://arxiv.org/abs/1510.06042}{{\ttfamily arXiv:1510.06042
  [astro-ph.CO]}}.

\bibitem{NANOGrav:2023gor}
{\bfseries NANOGrav} Collaboration, G.~Agazie {\em et~al.}, ``{The NANOGrav 15
  yr Data Set: Evidence for a Gravitational-wave Background},''
  \href{http://dx.doi.org/10.3847/2041-8213/acdac6}{{\em Astrophys. J. Lett.}
  {\bfseries 951} no.~1, (2023) L8},
  \href{http://arxiv.org/abs/2306.16213}{{\ttfamily arXiv:2306.16213
  [astro-ph.HE]}}.

\bibitem{Antoniadis:2023ott}
J.~Antoniadis {\em et~al.}, ``{The second data release from the European Pulsar
  Timing Array III. Search for gravitational wave signals},''
  \href{http://arxiv.org/abs/2306.16214}{{\ttfamily arXiv:2306.16214
  [astro-ph.HE]}}.

\bibitem{Reardon:2023gzh}
D.~J. Reardon {\em et~al.}, ``{Search for an Isotropic Gravitational-wave
  Background with the Parkes Pulsar Timing Array},''
  \href{http://dx.doi.org/10.3847/2041-8213/acdd02}{{\em Astrophys. J. Lett.}
  {\bfseries 951} no.~1, (2023) L6},
  \href{http://arxiv.org/abs/2306.16215}{{\ttfamily arXiv:2306.16215
  [astro-ph.HE]}}.

\bibitem{Xu:2023wog}
H.~Xu {\em et~al.}, ``{Searching for the Nano-Hertz Stochastic Gravitational
  Wave Background with the Chinese Pulsar Timing Array Data Release I},''
  \href{http://dx.doi.org/10.1088/1674-4527/acdfa5}{{\em Res. Astron.
  Astrophys.} {\bfseries 23} no.~7, (2023) 075024},
  \href{http://arxiv.org/abs/2306.16216}{{\ttfamily arXiv:2306.16216
  [astro-ph.HE]}}.

\bibitem{NANOGrav:2023hde}
{\bfseries NANOGrav} Collaboration, G.~Agazie {\em et~al.}, ``{The NANOGrav 15
  yr Data Set: Observations and Timing of 68 Millisecond Pulsars},''
  \href{http://dx.doi.org/10.3847/2041-8213/acda9a}{{\em Astrophys. J. Lett.}
  {\bfseries 951} no.~1, (2023) L9},
  \href{http://arxiv.org/abs/2306.16217}{{\ttfamily arXiv:2306.16217
  [astro-ph.HE]}}.

\bibitem{NANOGrav:2023ctt}
{\bfseries NANOGrav} Collaboration, G.~Agazie {\em et~al.}, ``{The NANOGrav 15
  yr Data Set: Detector Characterization and Noise Budget},''
  \href{http://dx.doi.org/10.3847/2041-8213/acda88}{{\em Astrophys. J. Lett.}
  {\bfseries 951} no.~1, (2023) L10},
  \href{http://arxiv.org/abs/2306.16218}{{\ttfamily arXiv:2306.16218
  [astro-ph.HE]}}.

\bibitem{NANOGrav:2023hvm}
{\bfseries NANOGrav} Collaboration, A.~Afzal {\em et~al.}, ``{The NANOGrav 15
  yr Data Set: Search for Signals from New Physics},''
  \href{http://dx.doi.org/10.3847/2041-8213/acdc91}{{\em Astrophys. J. Lett.}
  {\bfseries 951} no.~1, (2023) L11},
  \href{http://arxiv.org/abs/2306.16219}{{\ttfamily arXiv:2306.16219
  [astro-ph.HE]}}.

\bibitem{NANOGrav:2023hfp}
{\bfseries NANOGrav} Collaboration, G.~Agazie {\em et~al.}, ``{The NANOGrav
  15-year Data Set: Constraints on Supermassive Black Hole Binaries from the
  Gravitational Wave Background},''
  \href{http://arxiv.org/abs/2306.16220}{{\ttfamily arXiv:2306.16220
  [astro-ph.HE]}}.

\bibitem{NANOGrav:2023tcn}
{\bfseries NANOGrav} Collaboration, G.~Agazie {\em et~al.}, ``{The NANOGrav
  15-year Data Set: Search for Anisotropy in the Gravitational-Wave
  Background},'' \href{http://arxiv.org/abs/2306.16221}{{\ttfamily
  arXiv:2306.16221 [astro-ph.HE]}}.

\bibitem{NANOGrav:2023pdq}
{\bfseries NANOGrav} Collaboration, G.~Agazie {\em et~al.}, ``{The NANOGrav
  15-year Data Set: Bayesian Limits on Gravitational Waves from Individual
  Supermassive Black Hole Binaries},''
  \href{http://arxiv.org/abs/2306.16222}{{\ttfamily arXiv:2306.16222
  [astro-ph.HE]}}.

\bibitem{NANOGrav:2023icp}
{\bfseries NANOGrav} Collaboration, A.~D. Johnson {\em et~al.}, ``{The NANOGrav
  15-year Gravitational-Wave Background Analysis Pipeline},''
  \href{http://arxiv.org/abs/2306.16223}{{\ttfamily arXiv:2306.16223
  [astro-ph.HE]}}.

\bibitem{Antoniadis:2023lym}
J.~Antoniadis {\em et~al.}, ``{The second data release from the European Pulsar
  Timing Array I. The dataset and timing analysis},''
  \href{http://arxiv.org/abs/2306.16224}{{\ttfamily arXiv:2306.16224
  [astro-ph.HE]}}.

\bibitem{Antoniadis:2023puu}
J.~Antoniadis {\em et~al.}, ``{The second data release from the European Pulsar
  Timing Array II. Customised pulsar noise models for spatially correlated
  gravitational waves},'' \href{http://arxiv.org/abs/2306.16225}{{\ttfamily
  arXiv:2306.16225 [astro-ph.HE]}}.

\bibitem{Antoniadis:2023aac}
J.~Antoniadis {\em et~al.}, ``{The second data release from the European Pulsar
  Timing Array IV. Search for continuous gravitational wave signals},''
  \href{http://arxiv.org/abs/2306.16226}{{\ttfamily arXiv:2306.16226
  [astro-ph.HE]}}.

\bibitem{Antoniadis:2023xlr}
J.~Antoniadis {\em et~al.}, ``{The second data release from the European Pulsar
  Timing Array: V. Implications for massive black holes, dark matter and the
  early Universe},'' \href{http://arxiv.org/abs/2306.16227}{{\ttfamily
  arXiv:2306.16227 [astro-ph.CO]}}.

\bibitem{Smarra:2023ljf}
C.~Smarra {\em et~al.}, ``{The second data release from the European Pulsar
  Timing Array: VI. Challenging the ultralight dark matter paradigm},''
  \href{http://arxiv.org/abs/2306.16228}{{\ttfamily arXiv:2306.16228
  [astro-ph.HE]}}.

\bibitem{Reardon:2023zen}
D.~J. Reardon {\em et~al.}, ``{The Gravitational-wave Background Null
  Hypothesis: Characterizing Noise in Millisecond Pulsar Arrival Times with the
  Parkes Pulsar Timing Array},''
  \href{http://dx.doi.org/10.3847/2041-8213/acdd03}{{\em Astrophys. J. Lett.}
  {\bfseries 951} no.~1, (2023) L7},
  \href{http://arxiv.org/abs/2306.16229}{{\ttfamily arXiv:2306.16229
  [astro-ph.HE]}}.

\bibitem{Zic:2023gta}
A.~Zic {\em et~al.}, ``{The Parkes Pulsar Timing Array Third Data Release},''
  \href{http://arxiv.org/abs/2306.16230}{{\ttfamily arXiv:2306.16230
  [astro-ph.HE]}}.

\bibitem{Siemens:2006yp}
X.~Siemens, V.~Mandic, and J.~Creighton, ``{Gravitational wave stochastic
  background from cosmic (super)strings},''
  \href{http://dx.doi.org/10.1103/PhysRevLett.98.111101}{{\em Phys. Rev. Lett.}
  {\bfseries 98} (2007) 111101},
  \href{http://arxiv.org/abs/astro-ph/0610920}{{\ttfamily
  arXiv:astro-ph/0610920}}.

\bibitem{Caprini:2010xv}
C.~Caprini, R.~Durrer, and X.~Siemens, ``{Detection of gravitational waves from
  the QCD phase transition with pulsar timing arrays},''
  \href{http://dx.doi.org/10.1103/PhysRevD.82.063511}{{\em Phys. Rev. D}
  {\bfseries 82} (2010) 063511},
  \href{http://arxiv.org/abs/1007.1218}{{\ttfamily arXiv:1007.1218
  [astro-ph.CO]}}.

\bibitem{Ramberg:2019dgi}
N.~Ramberg and L.~Visinelli, ``{Probing the Early Universe with Axion Physics
  and Gravitational Waves},''
  \href{http://dx.doi.org/10.1103/PhysRevD.99.123513}{{\em Phys. Rev. D}
  {\bfseries 99} no.~12, (2019) 123513},
  \href{http://arxiv.org/abs/1904.05707}{{\ttfamily arXiv:1904.05707
  [astro-ph.CO]}}.

\bibitem{Caprini:2019egz}
C.~Caprini {\em et~al.}, ``{Detecting gravitational waves from cosmological
  phase transitions with LISA: an update},''
  \href{http://dx.doi.org/10.1088/1475-7516/2020/03/024}{{\em JCAP} {\bfseries
  03} (2020) 024}, \href{http://arxiv.org/abs/1910.13125}{{\ttfamily
  arXiv:1910.13125 [astro-ph.CO]}}.

\bibitem{Ellis:2020awk}
J.~Ellis, M.~Lewicki, and J.~M. No, ``{Gravitational waves from first-order
  cosmological phase transitions: lifetime of the sound wave source},''
  \href{http://dx.doi.org/10.1088/1475-7516/2020/07/050}{{\em JCAP} {\bfseries
  07} (2020) 050}, \href{http://arxiv.org/abs/2003.07360}{{\ttfamily
  arXiv:2003.07360 [hep-ph]}}.

\bibitem{Rajagopal:1994zj}
M.~Rajagopal and R.~W. Romani, ``{Ultralow frequency gravitational radiation
  from massive black hole binaries},''
  \href{http://dx.doi.org/10.1086/175813}{{\em Astrophys. J.} {\bfseries 446}
  (1995) 543--549}, \href{http://arxiv.org/abs/astro-ph/9412038}{{\ttfamily
  arXiv:astro-ph/9412038}}.

\bibitem{Jaffe:2002rt}
A.~H. Jaffe and D.~C. Backer, ``{Gravitational waves probe the coalescence rate
  of massive black hole binaries},''
  \href{http://dx.doi.org/10.1086/345443}{{\em Astrophys. J.} {\bfseries 583}
  (2003) 616--631}, \href{http://arxiv.org/abs/astro-ph/0210148}{{\ttfamily
  arXiv:astro-ph/0210148}}.

\bibitem{Wyithe:2002ep}
J.~S.~B. Wyithe and A.~Loeb, ``{Low - frequency gravitational waves from
  massive black hole binaries: Predictions for LISA and pulsar timing
  arrays},'' \href{http://dx.doi.org/10.1086/375187}{{\em Astrophys. J.}
  {\bfseries 590} (2003) 691--706},
  \href{http://arxiv.org/abs/astro-ph/0211556}{{\ttfamily
  arXiv:astro-ph/0211556}}.

\bibitem{Sesana:2004sp}
A.~Sesana, F.~Haardt, P.~Madau, and M.~Volonteri, ``{Low - frequency
  gravitational radiation from coalescing massive black hole binaries in
  hierarchical cosmologies},'' \href{http://dx.doi.org/10.1086/422185}{{\em
  Astrophys. J.} {\bfseries 611} (2004) 623--632},
  \href{http://arxiv.org/abs/astro-ph/0401543}{{\ttfamily
  arXiv:astro-ph/0401543}}.

\bibitem{Burke-Spolaor:2018bvk}
S.~Burke-Spolaor {\em et~al.}, ``{The Astrophysics of Nanohertz Gravitational
  Waves},'' \href{http://dx.doi.org/10.1007/s00159-019-0115-7}{{\em Astron.
  Astrophys. Rev.} {\bfseries 27} no.~1, (2019) 5},
  \href{http://arxiv.org/abs/1811.08826}{{\ttfamily arXiv:1811.08826
  [astro-ph.HE]}}.

\bibitem{Athron:2023mer}
P.~Athron, A.~Fowlie, C.-T. Lu, L.~Morris, L.~Wu, Y.~Wu, and Z.~Xu, ``{Can
  Supercooled Phase Transitions explain the Gravitational Wave Background
  observed by Pulsar Timing Arrays?},''
  \href{http://arxiv.org/abs/2306.17239}{{\ttfamily arXiv:2306.17239
  [hep-ph]}}.

\bibitem{Kitajima:2023cek}
N.~Kitajima, J.~Lee, K.~Murai, F.~Takahashi, and W.~Yin, ``{Nanohertz
  Gravitational Waves from Axion Domain Walls Coupled to QCD},''
  \href{http://arxiv.org/abs/2306.17146}{{\ttfamily arXiv:2306.17146
  [hep-ph]}}.

\bibitem{Hossain:2014ova}
M.~W. Hossain, R.~Myrzakulov, M.~Sami, and E.~N. Saridakis, ``{Evading Lyth
  bound in models of quintessential inflation},''
  \href{http://dx.doi.org/10.1016/j.physletb.2014.08.051}{{\em Phys. Lett. B}
  {\bfseries 737} (2014) 191--195},
  \href{http://arxiv.org/abs/1405.7491}{{\ttfamily arXiv:1405.7491 [gr-qc]}}.

\bibitem{Piao:2004tq}
Y.-S. Piao and Y.-Z. Zhang, ``{Phantom inflation and primordial perturbation
  spectrum},'' \href{http://dx.doi.org/10.1103/PhysRevD.70.063513}{{\em Phys.
  Rev. D} {\bfseries 70} (2004) 063513},
  \href{http://arxiv.org/abs/astro-ph/0401231}{{\ttfamily
  arXiv:astro-ph/0401231}}.

\bibitem{Liu:2010dh}
Z.-G. Liu, J.~Zhang, and Y.-S. Piao, ``{Phantom Inflation with A Steplike
  Potential},'' \href{http://dx.doi.org/10.1016/j.physletb.2010.12.055}{{\em
  Phys. Lett. B} {\bfseries 697} (2011) 407--411},
  \href{http://arxiv.org/abs/1012.0673}{{\ttfamily arXiv:1012.0673 [gr-qc]}}.

\bibitem{Brandenberger:2008nx}
R.~H. Brandenberger, ``{String Gas Cosmology},''
\newblock 8, 2008.
\newblock \href{http://arxiv.org/abs/0808.0746}{{\ttfamily arXiv:0808.0746
  [hep-th]}}.

\bibitem{Brandenberger:2011et}
R.~H. Brandenberger, ``{String Gas Cosmology: Progress and Problems},''
  \href{http://dx.doi.org/10.1088/0264-9381/28/20/204005}{{\em Class. Quant.
  Grav.} {\bfseries 28} (2011) 204005},
  \href{http://arxiv.org/abs/1105.3247}{{\ttfamily arXiv:1105.3247 [hep-th]}}.

\bibitem{Khoury:2001wf}
J.~Khoury, B.~A. Ovrut, P.~J. Steinhardt, and N.~Turok, ``{The Ekpyrotic
  universe: Colliding branes and the origin of the hot big bang},''
  \href{http://dx.doi.org/10.1103/PhysRevD.64.123522}{{\em Phys. Rev. D}
  {\bfseries 64} (2001) 123522},
  \href{http://arxiv.org/abs/hep-th/0103239}{{\ttfamily arXiv:hep-th/0103239}}.

\bibitem{Lehners:2008vx}
J.-L. Lehners, ``{Ekpyrotic and Cyclic Cosmology},''
  \href{http://dx.doi.org/10.1016/j.physrep.2008.06.001}{{\em Phys. Rept.}
  {\bfseries 465} (2008) 223--263},
  \href{http://arxiv.org/abs/0806.1245}{{\ttfamily arXiv:0806.1245
  [astro-ph]}}.

\bibitem{Brandenberger:2016vhg}
R.~Brandenberger and P.~Peter, ``{Bouncing Cosmologies: Progress and
  Problems},'' \href{http://dx.doi.org/10.1007/s10701-016-0057-0}{{\em Found.
  Phys.} {\bfseries 47} no.~6, (2017) 797--850},
  \href{http://arxiv.org/abs/1603.05834}{{\ttfamily arXiv:1603.05834
  [hep-th]}}.

\bibitem{Brandenberger:2023wtd}
R.~Brandenberger and G.~A. Mitchell, ``{A bouncing cosmology from VECROs},''
  \href{http://dx.doi.org/10.1140/epjc/s10052-023-11501-2}{{\em Eur. Phys. J.
  C} {\bfseries 83} no.~4, (2023) 308},
  \href{http://arxiv.org/abs/2302.12924}{{\ttfamily arXiv:2302.12924
  [hep-th]}}.

\bibitem{Koehn:2015vvy}
M.~Koehn, J.-L. Lehners, and B.~Ovrut, ``{Nonsingular bouncing cosmology:
  Consistency of the effective description},''
  \href{http://dx.doi.org/10.1103/PhysRevD.93.103501}{{\em Phys. Rev. D}
  {\bfseries 93} no.~10, (2016) 103501},
  \href{http://arxiv.org/abs/1512.03807}{{\ttfamily arXiv:1512.03807
  [hep-th]}}.

\bibitem{Lehners:2015mra}
J.-L. Lehners and E.~Wilson-Ewing, ``{Running of the scalar spectral index in
  bouncing cosmologies},''
  \href{http://dx.doi.org/10.1088/1475-7516/2015/10/038}{{\em JCAP} {\bfseries
  10} (2015) 038}, \href{http://arxiv.org/abs/1507.08112}{{\ttfamily
  arXiv:1507.08112 [astro-ph.CO]}}.

\bibitem{Ijjas:2018qbo}
A.~Ijjas and P.~J. Steinhardt, ``{Bouncing Cosmology made simple},''
  \href{http://dx.doi.org/10.1088/1361-6382/aac482}{{\em Class. Quant. Grav.}
  {\bfseries 35} no.~13, (2018) 135004},
  \href{http://arxiv.org/abs/1803.01961}{{\ttfamily arXiv:1803.01961
  [astro-ph.CO]}}.

\bibitem{Bhargava:2020fhl}
P.~Bhargava, S.~Choudhury, S.~Chowdhury, A.~Mishara, S.~P. Selvam, S.~Panda,
  and G.~D. Pasquino, ``{Quantum aspects of chaos and complexity from bouncing
  cosmology: A study with two-mode single field squeezed state formalism},''
  \href{http://dx.doi.org/10.21468/SciPostPhysCore.4.4.026}{{\em SciPost Phys.
  Core} {\bfseries 4} (2021) 026},
  \href{http://arxiv.org/abs/2009.03893}{{\ttfamily arXiv:2009.03893
  [hep-th]}}.

\bibitem{Agullo:2016tjh}
I.~Agullo and P.~Singh, {\em {Loop Quantum Cosmology}},
  \href{http://dx.doi.org/10.1142/9789813220003_0007}{pp.~183--240}.
\newblock WSP, 2017.
\newblock \href{http://arxiv.org/abs/1612.01236}{{\ttfamily arXiv:1612.01236
  [gr-qc]}}.

\bibitem{Bojowald:2005epg}
M.~Bojowald, ``{Loop quantum cosmology},''
  \href{http://dx.doi.org/10.12942/lrr-2005-11}{{\em Living Rev. Rel.}
  {\bfseries 8} (2005) 11},
  \href{http://arxiv.org/abs/gr-qc/0601085}{{\ttfamily arXiv:gr-qc/0601085}}.

\bibitem{Bojowald:1999tr}
M.~Bojowald, ``{Loop quantum cosmology. I. Kinematics},''
  \href{http://dx.doi.org/10.1088/0264-9381/17/6/312}{{\em Class. Quant. Grav.}
  {\bfseries 17} (2000) 1489--1508},
  \href{http://arxiv.org/abs/gr-qc/9910103}{{\ttfamily arXiv:gr-qc/9910103}}.

\bibitem{Biswas:2005qr}
T.~Biswas, A.~Mazumdar, and W.~Siegel, ``{Bouncing universes in string-inspired
  gravity},'' \href{http://dx.doi.org/10.1088/1475-7516/2006/03/009}{{\em JCAP}
  {\bfseries 03} (2006) 009},
  \href{http://arxiv.org/abs/hep-th/0508194}{{\ttfamily arXiv:hep-th/0508194}}.

\bibitem{Choudhury:2023kam}
S.~Choudhury, ``{Single field inflation in the light of NANOGrav 15-year Data:
  Quintessential interpretation of blue tilted tensor spectrum through
  Non-Bunch Davies initial condition},''
  \href{http://arxiv.org/abs/2307.03249}{{\ttfamily arXiv:2307.03249
  [astro-ph.CO]}}.

\bibitem{Choudhury:2020yaa}
S.~Choudhury, ``{The Cosmological OTOC: Formulating new cosmological
  micro-canonical correlation functions for random chaotic fluctuations in
  Out-of-Equilibrium Quantum Statistical Field Theory},''
  \href{http://dx.doi.org/10.3390/sym12091527}{{\em Symmetry} {\bfseries 12}
  no.~9, (2020) 1527}, \href{http://arxiv.org/abs/2005.11750}{{\ttfamily
  arXiv:2005.11750 [hep-th]}}.

\bibitem{Choudhury:2021tuu}
S.~Choudhury, ``{The Cosmological OTOC: A New Proposal for Quantifying
  Auto-correlated Random Non-chaotic Primordial Fluctuations},''
  \href{http://dx.doi.org/10.20944/preprints202102.0616.v1}{{\em Symmetry}
  {\bfseries 13} no.~4, (2021) 599},
  \href{http://arxiv.org/abs/2106.01305}{{\ttfamily arXiv:2106.01305
  [physics.gen-ph]}}.

\bibitem{Adhikari:2021ked}
K.~Adhikari, S.~Choudhury, H.~N. Pandya, and R.~Srivastava, ``{PGW Circuit
  Complexity},'' \href{http://arxiv.org/abs/2108.10334}{{\ttfamily
  arXiv:2108.10334 [gr-qc]}}.

\bibitem{Akama:2023jsb}
S.~Akama and H.~W.~H. Tahara, ``{Imprints of primordial gravitational waves
  with non-Bunch-Davies initial states on CMB bispectra},''
  \href{http://arxiv.org/abs/2306.17752}{{\ttfamily arXiv:2306.17752 [gr-qc]}}.

\bibitem{Albayrak:2023hie}
S.~Albayrak, P.~Benincasa, and C.~D. Pueyo, ``{Perturbative Unitarity and the
  Wavefunction of the Universe},''
  \href{http://arxiv.org/abs/2305.19686}{{\ttfamily arXiv:2305.19686
  [hep-th]}}.

\bibitem{Choudhury:2022mch}
S.~Choudhury, ``{Entanglement negativity in de Sitter biverse from Stringy
  Axionic Bell pair: An analysis using Bunch-Davies vacuum},''
  \href{http://arxiv.org/abs/2301.05203}{{\ttfamily arXiv:2301.05203
  [hep-th]}}.

\bibitem{Colas:2022kfu}
T.~Colas, J.~Grain, and V.~Vennin, ``{Quantum recoherence in the early
  universe},'' \href{http://arxiv.org/abs/2212.09486}{{\ttfamily
  arXiv:2212.09486 [gr-qc]}}.

\bibitem{Aalsma:2022eru}
L.~Aalsma, M.~M. Faruk, J.~P. van~der Schaar, M.~Visser, and J.~de~Witte,
  ``{Late-Time Correlators and Complex Geodesics in de Sitter Space},''
  \href{http://arxiv.org/abs/2212.01394}{{\ttfamily arXiv:2212.01394
  [hep-th]}}.

\bibitem{Chapman:2022mqd}
S.~Chapman, D.~A. Galante, E.~Harris, S.~U. Sheorey, and D.~Vegh, ``{Complex
  geodesics in de Sitter space},''
  \href{http://dx.doi.org/10.1007/JHEP03(2023)006}{{\em JHEP} {\bfseries 03}
  (2023) 006}, \href{http://arxiv.org/abs/2212.01398}{{\ttfamily
  arXiv:2212.01398 [hep-th]}}.

\bibitem{Letey:2022hdp}
M.~I. Letey, Z.~Shumaylov, F.~J. Agocs, W.~J. Handley, M.~P. Hobson, and A.~N.
  Lasenby, ``{Quantum Initial Conditions for Curved Inflating Universes},''
  \href{http://arxiv.org/abs/2211.17248}{{\ttfamily arXiv:2211.17248 [gr-qc]}}.

\bibitem{Penna-Lima:2022dmx}
M.~Penna-Lima, N.~Pinto-Neto, and S.~D.~P. Vitenti, ``{New formalism to define
  vacuum states for scalar fields in curved spacetimes},''
  \href{http://dx.doi.org/10.1103/PhysRevD.107.065019}{{\em Phys. Rev. D}
  {\bfseries 107} no.~6, (2023) 065019},
  \href{http://arxiv.org/abs/2207.08270}{{\ttfamily arXiv:2207.08270 [gr-qc]}}.

\bibitem{Kanno:2022mkx}
S.~Kanno and M.~Sasaki, ``{Graviton non-gaussianity in
  \ensuremath{\alpha}-vacuum},''
  \href{http://dx.doi.org/10.1007/JHEP08(2022)210}{{\em JHEP} {\bfseries 08}
  (2022) 210}, \href{http://arxiv.org/abs/2206.03667}{{\ttfamily
  arXiv:2206.03667 [hep-th]}}.

\bibitem{Fumagalli:2021mpc}
J.~Fumagalli, G.~A. Palma, S.~Renaux-Petel, S.~Sypsas, L.~T. Witkowski, and
  C.~Zenteno, ``{Primordial gravitational waves from excited states},''
  \href{http://dx.doi.org/10.1007/JHEP03(2022)196}{{\em JHEP} {\bfseries 03}
  (2022) 196}, \href{http://arxiv.org/abs/2111.14664}{{\ttfamily
  arXiv:2111.14664 [astro-ph.CO]}}.

\bibitem{Sleight:2021plv}
C.~Sleight and M.~Taronna, ``{From dS to AdS and back},''
  \href{http://dx.doi.org/10.1007/JHEP12(2021)074}{{\em JHEP} {\bfseries 12}
  (2021) 074}, \href{http://arxiv.org/abs/2109.02725}{{\ttfamily
  arXiv:2109.02725 [hep-th]}}.

\bibitem{Chen:2010bka}
X.~Chen, ``{Folded Resonant Non-Gaussianity in General Single Field
  Inflation},'' \href{http://dx.doi.org/10.1088/1475-7516/2010/12/003}{{\em
  JCAP} {\bfseries 12} (2010) 003},
  \href{http://arxiv.org/abs/1008.2485}{{\ttfamily arXiv:1008.2485 [hep-th]}}.

\bibitem{Wang:2014kqa}
Y.~Wang and W.~Xue, ``{Inflation and Alternatives with Blue Tensor Spectra},''
  \href{http://dx.doi.org/10.1088/1475-7516/2014/10/075}{{\em JCAP} {\bfseries
  10} (2014) 075}, \href{http://arxiv.org/abs/1403.5817}{{\ttfamily
  arXiv:1403.5817 [astro-ph.CO]}}.

\bibitem{Ashoorioon:2014nta}
A.~Ashoorioon, K.~Dimopoulos, M.~M. Sheikh-Jabbari, and G.~Shiu,
  ``{Non-Bunch\textendash{}Davis initial state reconciles chaotic models with
  BICEP and Planck},''
  \href{http://dx.doi.org/10.1016/j.physletb.2014.08.038}{{\em Phys. Lett. B}
  {\bfseries 737} (2014) 98--102},
  \href{http://arxiv.org/abs/1403.6099}{{\ttfamily arXiv:1403.6099 [hep-th]}}.

\bibitem{Ashoorioon:2013eia}
A.~Ashoorioon, K.~Dimopoulos, M.~M. Sheikh-Jabbari, and G.~Shiu,
  ``{Reconciliation of High Energy Scale Models of Inflation with Planck},''
  \href{http://dx.doi.org/10.1088/1475-7516/2014/02/025}{{\em JCAP} {\bfseries
  02} (2014) 025}, \href{http://arxiv.org/abs/1306.4914}{{\ttfamily
  arXiv:1306.4914 [hep-th]}}.

\bibitem{Guendelman:2024ezk}
E.~Guendelman, ``{Holomorphic gravity and its regularization of Locally Signed
  Coordinate Invariance},'' \href{http://arxiv.org/abs/2402.00140}{{\ttfamily
  arXiv:2402.00140 [gr-qc]}}.

\bibitem{Guendelman:2023spd}
E.~Guendelman, ``{Holomorphic general coordinate invariant modified measure
  gravitational theory},''
  \href{http://dx.doi.org/10.1016/j.aop.2023.169466}{{\em Annals Phys.}
  {\bfseries 458} (2023) 169466},
  \href{http://arxiv.org/abs/2308.09246}{{\ttfamily arXiv:2308.09246 [gr-qc]}}.

\bibitem{Guendelman:2023vso}
E.~Guendelman, ``{Signed Coordinate Invariance, invariant lagrangians and
  manifolds, the time problem in quantum cosmology, quantum space time,
  spacetimes and antispacetimes},''
  \href{http://arxiv.org/abs/2304.04056}{{\ttfamily arXiv:2304.04056 [gr-qc]}}.

\bibitem{Hui:2001ce}
L.~Hui and W.~H. Kinney, ``{Short distance physics and the consistency relation
  for scalar and tensor fluctuations in the inflationary universe},''
  \href{http://dx.doi.org/10.1103/PhysRevD.65.103507}{{\em Phys. Rev. D}
  {\bfseries 65} (2002) 103507},
  \href{http://arxiv.org/abs/astro-ph/0109107}{{\ttfamily
  arXiv:astro-ph/0109107}}.

\bibitem{Ganc:2011dy}
J.~Ganc, ``{Calculating the local-type fNL for slow-roll inflation with a
  non-vacuum initial state},''
  \href{http://dx.doi.org/10.1103/PhysRevD.84.063514}{{\em Phys. Rev. D}
  {\bfseries 84} (2011) 063514},
  \href{http://arxiv.org/abs/1104.0244}{{\ttfamily arXiv:1104.0244
  [astro-ph.CO]}}.

\bibitem{Brahma:2018hrd}
S.~Brahma and M.~Wali~Hossain, ``{Avoiding the string swampland in single-field
  inflation: Excited initial states},''
  \href{http://dx.doi.org/10.1007/JHEP03(2019)006}{{\em JHEP} {\bfseries 03}
  (2019) 006}, \href{http://arxiv.org/abs/1809.01277}{{\ttfamily
  arXiv:1809.01277 [hep-th]}}.

\bibitem{Allahverdi:2010xz}
R.~Allahverdi, R.~Brandenberger, F.-Y. Cyr-Racine, and A.~Mazumdar,
  ``{Reheating in Inflationary Cosmology: Theory and Applications},''
  \href{http://dx.doi.org/10.1146/annurev.nucl.012809.104511}{{\em Ann. Rev.
  Nucl. Part. Sci.} {\bfseries 60} (2010) 27--51},
  \href{http://arxiv.org/abs/1001.2600}{{\ttfamily arXiv:1001.2600 [hep-th]}}.

\bibitem{Kohri:2009ka}
K.~Kohri, A.~Mazumdar, and N.~Sahu, ``{Inflation, baryogenesis and gravitino
  dark matter at ultra low reheat temperatures},''
  \href{http://dx.doi.org/10.1103/PhysRevD.80.103504}{{\em Phys. Rev. D}
  {\bfseries 80} (2009) 103504},
  \href{http://arxiv.org/abs/0905.1625}{{\ttfamily arXiv:0905.1625 [hep-ph]}}.

\bibitem{Mazumdar:2003bs}
A.~Mazumdar and A.~Perez-Lorenzana, ``{Sneutrino condensate source for density
  perturbations, leptogenesis and low reheat temperature},''
  \href{http://dx.doi.org/10.1103/PhysRevLett.92.251301}{{\em Phys. Rev. Lett.}
  {\bfseries 92} (2004) 251301},
  \href{http://arxiv.org/abs/hep-ph/0311106}{{\ttfamily arXiv:hep-ph/0311106}}.

\bibitem{Choudhury:2018rjl}
S.~Choudhury, A.~Mukherjee, P.~Chauhan, and S.~Bhattacherjee, ``{Quantum
  Out-of-Equilibrium Cosmology},''
  \href{http://dx.doi.org/10.1140/epjc/s10052-019-6751-2}{{\em Eur. Phys. J. C}
  {\bfseries 79} no.~4, (2019) 320},
  \href{http://arxiv.org/abs/1809.02732}{{\ttfamily arXiv:1809.02732
  [hep-th]}}.

\bibitem{Choudhury:2018bcf}
S.~Choudhury and A.~Mukherjee, ``{Quantum randomness in the Sky},''
  \href{http://dx.doi.org/10.1140/epjc/s10052-019-7072-1}{{\em Eur. Phys. J. C}
  {\bfseries 79} no.~7, (2019) 554},
  \href{http://arxiv.org/abs/1812.04107}{{\ttfamily arXiv:1812.04107
  [physics.gen-ph]}}.

\bibitem{Planck:2018jri}
{\bfseries Planck} Collaboration, Y.~Akrami {\em et~al.}, ``{Planck 2018
  results. X. Constraints on inflation},''
  \href{http://dx.doi.org/10.1051/0004-6361/201833887}{{\em Astron. Astrophys.}
  {\bfseries 641} (2020) A10},
  \href{http://arxiv.org/abs/1807.06211}{{\ttfamily arXiv:1807.06211
  [astro-ph.CO]}}.

\bibitem{Ashoorioon:2018sqb}
A.~Ashoorioon, ``{Rescuing Single Field Inflation from the Swampland},''
  \href{http://dx.doi.org/10.1016/j.physletb.2019.02.009}{{\em Phys. Lett. B}
  {\bfseries 790} (2019) 568--573},
  \href{http://arxiv.org/abs/1810.04001}{{\ttfamily arXiv:1810.04001
  [hep-th]}}.

\bibitem{Banerjee:2021lqu}
S.~Banerjee, S.~Choudhury, S.~Chowdhury, J.~Knaute, S.~Panda, and K.~Shirish,
  ``{Thermalization in Quenched De Sitter Space},''
  \href{http://arxiv.org/abs/2104.10692}{{\ttfamily arXiv:2104.10692
  [hep-th]}}.

\bibitem{Hardeman:2010fh}
S.~Hardeman, J.~M. Oberreuter, G.~A. Palma, K.~Schalm, and T.~van~der Aalst,
  ``{The everpresent eta-problem: knowledge of all hidden sectors required},''
  \href{http://dx.doi.org/10.1007/JHEP04(2011)009}{{\em JHEP} {\bfseries 04}
  (2011) 009}, \href{http://arxiv.org/abs/1012.5966}{{\ttfamily arXiv:1012.5966
  [hep-ph]}}.

\bibitem{Allahverdi:2006iq}
R.~Allahverdi, K.~Enqvist, J.~Garcia-Bellido, and A.~Mazumdar, ``{Gauge
  invariant MSSM inflaton},''
  \href{http://dx.doi.org/10.1103/PhysRevLett.97.191304}{{\em Phys. Rev. Lett.}
  {\bfseries 97} (2006) 191304},
  \href{http://arxiv.org/abs/hep-ph/0605035}{{\ttfamily arXiv:hep-ph/0605035}}.

\bibitem{Enqvist:2003gh}
K.~Enqvist and A.~Mazumdar, ``{Cosmological consequences of MSSM flat
  directions},'' \href{http://dx.doi.org/10.1016/S0370-1573(03)00119-4}{{\em
  Phys. Rept.} {\bfseries 380} (2003) 99--234},
  \href{http://arxiv.org/abs/hep-ph/0209244}{{\ttfamily arXiv:hep-ph/0209244}}.

\bibitem{Allahverdi:2006we}
R.~Allahverdi, K.~Enqvist, J.~Garcia-Bellido, A.~Jokinen, and A.~Mazumdar,
  ``{MSSM flat direction inflation: Slow roll, stability, fine tunning and
  reheating},'' \href{http://dx.doi.org/10.1088/1475-7516/2007/06/019}{{\em
  JCAP} {\bfseries 06} (2007) 019},
  \href{http://arxiv.org/abs/hep-ph/0610134}{{\ttfamily arXiv:hep-ph/0610134}}.

\bibitem{Allahverdi:2006cx}
R.~Allahverdi, A.~Kusenko, and A.~Mazumdar, ``{A-term inflation and the
  smallness of neutrino masses},''
  \href{http://dx.doi.org/10.1088/1475-7516/2007/07/018}{{\em JCAP} {\bfseries
  07} (2007) 018}, \href{http://arxiv.org/abs/hep-ph/0608138}{{\ttfamily
  arXiv:hep-ph/0608138}}.

\bibitem{Enqvist:2003mr}
K.~Enqvist, A.~Jokinen, S.~Kasuya, and A.~Mazumdar, ``{MSSM flat direction as a
  curvaton},'' \href{http://dx.doi.org/10.1103/PhysRevD.68.103507}{{\em Phys.
  Rev. D} {\bfseries 68} (2003) 103507},
  \href{http://arxiv.org/abs/hep-ph/0303165}{{\ttfamily arXiv:hep-ph/0303165}}.

\bibitem{Enqvist:2002rf}
K.~Enqvist, S.~Kasuya, and A.~Mazumdar, ``{Adiabatic density perturbations and
  matter generation from the MSSM},''
  \href{http://dx.doi.org/10.1103/PhysRevLett.90.091302}{{\em Phys. Rev. Lett.}
  {\bfseries 90} (2003) 091302},
  \href{http://arxiv.org/abs/hep-ph/0211147}{{\ttfamily arXiv:hep-ph/0211147}}.

\bibitem{Enqvist:2010vd}
K.~Enqvist, A.~Mazumdar, and P.~Stephens, ``{Inflection point inflation within
  supersymmetry},'' \href{http://dx.doi.org/10.1088/1475-7516/2010/06/020}{{\em
  JCAP} {\bfseries 06} (2010) 020},
  \href{http://arxiv.org/abs/1004.3724}{{\ttfamily arXiv:1004.3724 [hep-ph]}}.

\bibitem{Wang:2013hva}
L.~Wang, E.~Pukartas, and A.~Mazumdar, ``{Visible sector inflation and the
  right thermal history in light of Planck data},''
  \href{http://dx.doi.org/10.1088/1475-7516/2013/07/019}{{\em JCAP} {\bfseries
  07} (2013) 019}, \href{http://arxiv.org/abs/1303.5351}{{\ttfamily
  arXiv:1303.5351 [hep-ph]}}.

\bibitem{Mazumdar:2011ih}
A.~Mazumdar, S.~Nadathur, and P.~Stephens, ``{Inflation with large supergravity
  corrections},'' \href{http://dx.doi.org/10.1103/PhysRevD.85.045001}{{\em
  Phys. Rev. D} {\bfseries 85} (2012) 045001},
  \href{http://arxiv.org/abs/1105.0430}{{\ttfamily arXiv:1105.0430 [hep-th]}}.

\bibitem{Chatterjee:2011qr}
A.~Chatterjee and A.~Mazumdar, ``{Tuned MSSM Higgses as an inflaton},''
  \href{http://dx.doi.org/10.1088/1475-7516/2011/09/009}{{\em JCAP} {\bfseries
  09} (2011) 009}, \href{http://arxiv.org/abs/1103.5758}{{\ttfamily
  arXiv:1103.5758 [hep-ph]}}.

\bibitem{Allahverdi:2007wh}
R.~Allahverdi, A.~R. Frey, and A.~Mazumdar, ``{Graceful exit from a stringy
  landscape via MSSM inflation},''
  \href{http://dx.doi.org/10.1103/PhysRevD.76.026001}{{\em Phys. Rev. D}
  {\bfseries 76} (2007) 026001},
  \href{http://arxiv.org/abs/hep-th/0701233}{{\ttfamily arXiv:hep-th/0701233}}.

\bibitem{Choi:2016eif}
S.-M. Choi and H.~M. Lee, ``{Inflection point inflation and reheating},''
  \href{http://dx.doi.org/10.1140/epjc/s10052-016-4150-5}{{\em Eur. Phys. J. C}
  {\bfseries 76} no.~6, (2016) 303},
  \href{http://arxiv.org/abs/1601.05979}{{\ttfamily arXiv:1601.05979
  [hep-ph]}}.

\bibitem{Ghoshal:2022jeo}
A.~Ghoshal, G.~Lambiase, S.~Pal, A.~Paul, and S.~Porey, ``{Inflection-point
  inflation and dark matter redux},''
  \href{http://dx.doi.org/10.1007/JHEP09(2022)231}{{\em JHEP} {\bfseries 09}
  (2022) 231}, \href{http://arxiv.org/abs/2206.10648}{{\ttfamily
  arXiv:2206.10648 [hep-ph]}}.

\bibitem{Allahverdi:2008bt}
R.~Allahverdi, B.~Dutta, and A.~Mazumdar, ``{Attraction towards an inflection
  point inflation},'' \href{http://dx.doi.org/10.1103/PhysRevD.78.063507}{{\em
  Phys. Rev. D} {\bfseries 78} (2008) 063507},
  \href{http://arxiv.org/abs/0806.4557}{{\ttfamily arXiv:0806.4557 [hep-ph]}}.

\bibitem{Hotchkiss:2011am}
S.~Hotchkiss, A.~Mazumdar, and S.~Nadathur, ``{Inflection point inflation: WMAP
  constraints and a solution to the fine-tuning problem},''
  \href{http://dx.doi.org/10.1088/1475-7516/2011/06/002}{{\em JCAP} {\bfseries
  06} (2011) 002}, \href{http://arxiv.org/abs/1101.6046}{{\ttfamily
  arXiv:1101.6046 [astro-ph.CO]}}.

\bibitem{Hotchkiss:2011gz}
S.~Hotchkiss, A.~Mazumdar, and S.~Nadathur, ``{Observable gravitational waves
  from inflation with small field excursions},''
  \href{http://dx.doi.org/10.1088/1475-7516/2012/02/008}{{\em JCAP} {\bfseries
  02} (2012) 008}, \href{http://arxiv.org/abs/1110.5389}{{\ttfamily
  arXiv:1110.5389 [astro-ph.CO]}}.

\bibitem{Chatterjee:2014hna}
A.~Chatterjee and A.~Mazumdar, ``{Bound on largest $r\lesssim 0.1$ from
  sub-Planckian excursions of inflaton},''
  \href{http://dx.doi.org/10.1088/1475-7516/2015/01/031}{{\em JCAP} {\bfseries
  01} (2015) 031}, \href{http://arxiv.org/abs/1409.4442}{{\ttfamily
  arXiv:1409.4442 [astro-ph.CO]}}.

\bibitem{Kristiano:2022maq}
J.~Kristiano and J.~Yokoyama, ``{Ruling Out Primordial Black Hole Formation
  From Single-Field Inflation},''
  \href{http://arxiv.org/abs/2211.03395}{{\ttfamily arXiv:2211.03395
  [hep-th]}}.

\bibitem{Kristiano:2023scm}
J.~Kristiano and J.~Yokoyama, ``{Response to criticism on ''Ruling Out
  Primordial Black Hole Formation From Single-Field Inflation'': A note on
  bispectrum and one-loop correction in single-field inflation with primordial
  black hole formation},'' \href{http://arxiv.org/abs/2303.00341}{{\ttfamily
  arXiv:2303.00341 [hep-th]}}.

\bibitem{Choudhury:2023vuj}
S.~Choudhury, M.~R. Gangopadhyay, and M.~Sami, ``{No-go for the formation of
  heavy mass Primordial Black Holes in Single Field Inflation},''
  \href{http://arxiv.org/abs/2301.10000}{{\ttfamily arXiv:2301.10000
  [astro-ph.CO]}}.

\bibitem{Choudhury:2023jlt}
S.~Choudhury, S.~Panda, and M.~Sami, ``{No-go for PBH formation in EFT of
  single field inflation},'' \href{http://arxiv.org/abs/2302.05655}{{\ttfamily
  arXiv:2302.05655 [astro-ph.CO]}}.

\bibitem{Choudhury:2023rks}
S.~Choudhury, S.~Panda, and M.~Sami, ``{Quantum loop effects on the power
  spectrum and constraints on primordial black holes},''
  \href{http://arxiv.org/abs/2303.06066}{{\ttfamily arXiv:2303.06066
  [astro-ph.CO]}}.

\bibitem{Choudhury:2023hvf}
S.~Choudhury, S.~Panda, and M.~Sami, ``{Galileon inflation evades the no-go for
  PBH formation in the single-field framework},''
  \href{http://arxiv.org/abs/2304.04065}{{\ttfamily arXiv:2304.04065
  [astro-ph.CO]}}.

\bibitem{Choudhury:2023kdb}
S.~Choudhury, A.~Karde, S.~Panda, and M.~Sami, ``{Primordial non-Gaussianity
  from ultra slow-roll Galileon inflation},''
  \href{http://arxiv.org/abs/2306.12334}{{\ttfamily arXiv:2306.12334
  [astro-ph.CO]}}.

\bibitem{Riotto:2023hoz}
A.~Riotto, ``{The Primordial Black Hole Formation from Single-Field Inflation
  is Not Ruled Out},'' \href{http://arxiv.org/abs/2301.00599}{{\ttfamily
  arXiv:2301.00599 [astro-ph.CO]}}.

\bibitem{Riotto:2023gpm}
A.~Riotto, ``{The Primordial Black Hole Formation from Single-Field Inflation
  is Still Not Ruled Out},'' \href{http://arxiv.org/abs/2303.01727}{{\ttfamily
  arXiv:2303.01727 [astro-ph.CO]}}.

\bibitem{Firouzjahi:2023aum}
H.~Firouzjahi, ``{One-loop Corrections in Power Spectrum in Single Field
  Inflation},'' \href{http://arxiv.org/abs/2303.12025}{{\ttfamily
  arXiv:2303.12025 [astro-ph.CO]}}.

\bibitem{Motohashi:2023syh}
H.~Motohashi and Y.~Tada, ``{Squeezed bispectrum and one-loop corrections in
  transient constant-roll inflation},''
  \href{http://arxiv.org/abs/2303.16035}{{\ttfamily arXiv:2303.16035
  [astro-ph.CO]}}.

\bibitem{Firouzjahi:2023ahg}
H.~Firouzjahi and A.~Riotto, ``{Primordial Black Holes and Loops in
  Single-Field Inflation},'' \href{http://arxiv.org/abs/2304.07801}{{\ttfamily
  arXiv:2304.07801 [astro-ph.CO]}}.

\bibitem{Franciolini:2023lgy}
G.~Franciolini, A.~Iovino, Junior., M.~Taoso, and A.~Urbano, ``{One loop to
  rule them all: Perturbativity in the presence of ultra slow-roll dynamics},''
  \href{http://arxiv.org/abs/2305.03491}{{\ttfamily arXiv:2305.03491
  [astro-ph.CO]}}.

\bibitem{Tasinato:2023ukp}
G.~Tasinato, ``{A large $|\eta|$ approach to single field inflation},''
  \href{http://arxiv.org/abs/2305.11568}{{\ttfamily arXiv:2305.11568
  [hep-th]}}.

\bibitem{Cheng:2023ikq}
S.-L. Cheng, D.-S. Lee, and K.-W. Ng, ``{Primordial perturbations from
  ultra-slow-roll single-field inflation with quantum loop effects},''
  \href{http://arxiv.org/abs/2305.16810}{{\ttfamily arXiv:2305.16810
  [astro-ph.CO]}}.

\bibitem{Choudhury:2023hfm}
S.~Choudhury, A.~Karde, S.~Panda, and M.~Sami, ``{Scalar induced gravity waves
  from ultra slow-roll Galileon inflation},''
  \href{http://arxiv.org/abs/2308.09273}{{\ttfamily arXiv:2308.09273
  [astro-ph.CO]}}.

\bibitem{Bhattacharya:2023ysp}
G.~Bhattacharya, S.~Choudhury, K.~Dey, S.~Ghosh, A.~Karde, and N.~S. Mishra,
  ``{Evading no-go for PBH formation and production of SIGWs using Multiple
  Sharp Transitions in EFT of single field inflation},''
  \href{http://arxiv.org/abs/2309.00973}{{\ttfamily arXiv:2309.00973
  [astro-ph.CO]}}.

\bibitem{Choudhury:2023fwk}
S.~Choudhury, K.~Dey, A.~Karde, S.~Panda, and M.~Sami, ``{Primordial
  non-Gaussianity as a saviour for PBH overproduction in SIGWs generated by
  Pulsar Timing Arrays for Galileon inflation},''
  \href{http://arxiv.org/abs/2310.11034}{{\ttfamily arXiv:2310.11034
  [astro-ph.CO]}}.

\bibitem{Choudhury:2023fjs}
S.~Choudhury, K.~Dey, and A.~Karde, ``{Untangling PBH overproduction in
  $w$-SIGWs generated by Pulsar Timing Arrays for MST-EFT of single field
  inflation},'' \href{http://arxiv.org/abs/2311.15065}{{\ttfamily
  arXiv:2311.15065 [astro-ph.CO]}}.

\bibitem{Choudhury:2024one}
S.~Choudhury, A.~Karde, S.~Panda, and M.~Sami, ``{Realisation of the ultra-slow
  roll phase in Galileon inflation and PBH overproduction},''
  \href{http://arxiv.org/abs/2401.10925}{{\ttfamily arXiv:2401.10925
  [astro-ph.CO]}}.

\bibitem{Choudhury:2024ybk}
S.~Choudhury, ``{Large fluctuations in the Sky},''
  \href{http://arxiv.org/abs/2403.07343}{{\ttfamily arXiv:2403.07343
  [astro-ph.CO]}}.

\bibitem{Choudhury:2024jlz}
S.~Choudhury, A.~Karde, P.~Padiyar, and M.~Sami, ``{Primordial Black Holes from
  Effective Field Theory of Stochastic Single Field Inflation at NNNLO},''
  \href{http://arxiv.org/abs/2403.13484}{{\ttfamily arXiv:2403.13484
  [astro-ph.CO]}}.

\bibitem{Choudhury:2024dei}
S.~Choudhury, A.~Karde, S.~Panda, and S.~SenGupta,
  ``{Regularized-Renormalized-Resummed loop corrected power spectrum of
  non-singular bounce with Primordial Black Hole formation},''
  \href{http://arxiv.org/abs/2405.06882}{{\ttfamily arXiv:2405.06882
  [astro-ph.CO]}}.

\bibitem{Choudhury:2024dzw}
S.~Choudhury, S.~Ganguly, S.~Panda, S.~SenGupta, and P.~Tiwari, ``{Obviating
  PBH overproduction for SIGWs generated by pulsar timing arrays in loop
  corrected EFT of bounce},''
  \href{http://dx.doi.org/10.1088/1475-7516/2024/09/013}{{\em JCAP} {\bfseries
  09} (2024) 013}, \href{http://arxiv.org/abs/2407.18976}{{\ttfamily
  arXiv:2407.18976 [astro-ph.CO]}}.

\bibitem{Choudhury:2024aji}
S.~Choudhury and M.~Sami, ``{Large fluctuations and Primordial Black Holes},''
  \href{http://arxiv.org/abs/2407.17006}{{\ttfamily arXiv:2407.17006 [gr-qc]}}.

\end{thebibliography}

\end{document}